# The Habitability of Proxima Centauri b: II: Environmental States and Observational Discriminants

Victoria S. Meadows[1,2,3], Giada N. Arney[1,2], Edward W. Schwieterman[1,2], Jacob Lustig-Yaeger[1,2], Andrew P. Lincowski[1,2], Tyler Robinson[4,2], Shawn D. Domagal-Goldman[5,2], Rory K. Barnes[1,2], David P. Fleming[1,2], Russell Deitrick[1,2], Rodrigo Luger[1,2], Peter E. Driscoll[6,2], Thomas R. Quinn[1,2], David Crisp[7,2]

[1]Astronomy Department, University of Washington, Box 951580, Seattle, WA 98195

[2]NASA Astrobiology Institute – Virtual Planetary Laboratory Lead Team, USA

[3]E-mail: vsm@astro.washington.edu

[4]Department of Astronomy and Astrophysics, University of California, Santa Cruz, CA 95064,

[5]Planetary Environments Laboratory, NASA Goddard Space Flight Center, 8800 Greenbelt Road, Greenbelt, MD 20771

[6]Department of Terrestrial Magnetism, Carnegie Institution for Science, Washington, DC

[7]Jet Propulsion Laboratory, California Institute of Technology, M/S 183-501, 4800 Oak Grove Drive, Pasadena, CA 91109

## Abstract

Proxima Centauri b provides an unprecedented opportunity to understand the evolution and nature of terrestrial planets orbiting M dwarfs. Although Proxima Cen b orbits within its star's habitable zone, multiple plausible evolutionary paths could have generated different environments that may or may not be habitable. Here we use 1D coupled climate-photochemical models to generate self-consistent atmospheres for several of the evolutionary scenarios predicted in our companion paper (Barnes et al., 2016). These include high-$O_2$, high-$CO_2$, and more Earth-like atmospheres, with either oxidizing or reducing compositions. We show that these modeled environments can be habitable or uninhabitable at Proxima Cen b's position in the habitable zone. We use radiative transfer models to generate synthetic spectra and thermal phase curves for these simulated environments, and use instrument models to explore our ability to discriminate between possible planetary states. These results are applicable not only to Proxima Cen b, but to other terrestrial planets orbiting M dwarfs. Thermal phase curves may provide the first constraint on the existence of an atmosphere. We find that JWST observations longward of 7 μm could characterize atmospheric heat transport and molecular composition. Detection of ocean glint is unlikely with JWST, but may be within the reach of larger aperture telescopes. Direct imaging spectra may detect $O_4$ absorption, which is diagnostic of massive water loss and $O_2$ retention, rather than a photosynthetic biosphere. Similarly, strong $CO_2$ and CO bands at wavelengths shortward of 2.5 μm would indicate a $CO_2$-dominated atmosphere. If the planet is habitable and volatile-rich, direct imaging will be the best means of detecting habitability. Earth-like planets with microbial biospheres may be identified by the presence of $CH_4$ - which has a longer atmospheric lifetime under Proxima Centauri's incident UV - and either photosynthetically produced $O_2$ or a hydrocarbon haze layer.

## 1.0 Introduction.

The discovery of a possibly terrestrial mass planet sitting squarely in the habitable zone of the Solar System's nearest neighbor (Anglada-Escudé et al., 2016) is a remarkable opportunity to further our understanding of the evolution of terrestrial planets and the distribution of life in the Universe.  If confirmed, Proxima Cen b is arguably the most explorable potentially habitable planet orbiting an M dwarf host.  M dwarfs comprise 70% of all stars, and habitable planets orbiting M dwarfs may be the most common environment for life in the Universe.   However, although Proxima Cen b has several promising attributes, including its possibly small size (minimum mass of 1.3 $M_{\oplus}$) and its position in the conservative habitable zone (Kopparapu et al., 2013) with an effective insolation 65% that of Earth's (Anglada-Escudé et al., 2016), its habitability is not guaranteed. A determination of the degree to which this planet is habitable, or any other insights into the fate of terrestrial planets orbiting M dwarfs, awaits follow-up information on its orbit, the planet's phase- and time-dependent photometry, and on spectra of the atmosphere and surface. Conversely, Proxima Cen b will ultimately provide an excellent observational laboratory for our current understanding of terrestrial planet evolution, which will in turn inform the prevalence of habitability and life in our Galaxy.

In advance of these data, we can synthesize what has been learned over the past several decades on star-planet and planetary environmental interactions for terrestrial planets to make a broad, preliminary assessment of plausible evolutionary processes and current environmental states for Proxima Cen b.  In our companion paper (Barnes et al., 2016) we explored likely evolutionary paths for Proxima Cen b, as it coevolved with the Galaxy, companion stars and planets, and its host star.  In this paper, we model several of these possible planetary environmental outcomes with coupled climate-photochemical models that are self-consistently forced by the spectrum of Proxima Cen. We then use radiative transfer and instrument simulation models to generate synthetic phase-dependent lightcurves, and transmission and direct imaging spectra, which are relevant to missions such as the James Webb Space Telescope (JWST[1]; e.g., Cowan et al. 2015), ground-based extremely large telescopes (ELTs, e.g. Kasper et al. 2008), the Wide-Field Infrared Survey Telescope (WFIRST[2]; Spergel et al. 2015), the Habitable Exoplanet Imaging Mission (HabEx; e.g., Menneson et al., 2016)[3], and the Large UltraViolet Optical Infrared Surveyor Telescope (LUVOIR[4]; e.g., Kouveliotou et al. 2014,  Dalcanton et al. 2015).  These spectra can be used to predict possible attributes and necessary measurements to identify discriminants for habitable and uninhabitable environments, and to explore the potential detectability of environmental signs of habitability and life.

---

[1] http://www.jwst.nasa.gov/
[2] http://wfirst.gsfc.nasa.gov/
[3] http://www.jpl.nasa.gov/habex/
[4] http://cor.gsfc.nasa.gov/studies/luvoir.php

In particular, we consider the scenarios for atmospheric composition from Barnes et al., (2016), and use our models to explore the photochemical and climatic outcomes, and observable attributes. Our first scenario is an oxygen-dominated atmosphere, generated by the loss of oceans of water during Proxima Cen's pre-main sequence phase (Luger & Barnes, 2015; Barnes et al., 2016). In this scenario, the planet can be either desiccated or have retained some of its initial water inventory. In the second class of scenario, the atmosphere is $CO_2$-dominated and progressively desiccated, and is formed when the majority of $O_2$ from ocean loss is either lost to space or sequestered in the planetary crust or interior (Schaefer et al., 2016; Barnes et al., 2016), while $CO_2$ is outgassed from the mantle. In this case, $O_2/CO_2$ dominated atmospheres could form, and potentially evolve to more $CO_2$-rich Venus-like states as the $O_2$ is lost or sequestered. Ultimately, with extreme desiccation, $CO_2$-dominated atmospheres could photolyze to stable $CO_2/CO/O_2$ atmospheres (Gao et al., 2015). In the final set of scenarios explored here, the planet is potentially habitable, having evolved as a "Habitable Evaporated Core" (Luger et al., 2015), where an early hydrogen envelope protected a terrestrial or volatile-rich core from water loss. Alternatively, a terrestrial planet initially orbiting farther from the star could have moved to the current orbit of Proxima Cen b via orbital instabilities, possibly triggered by close a passage of Proxima Cen to α Cen A and B Two cases are presented for this habitable scenario: an oxidizing, modern Earth-like atmosphere, and a more reducing Archean (3.8-2.5Gya) early Earth-like atmosphere.

In Section 2 we review the possible evolutionary scenarios for Proxima Centauri b, and briefly discuss observable parameters for identification of habitability and biosignatures for terrestrial planets. In Section 3 we describe the radiative transfer, instrument, climate and photochemical models, along with their model inputs. In Section 4 we present our results. In Section 5 we discuss the implications of our modeling, including an assessment the scenarios that might lead to habitability for Proxima Cen b. We also identify future observations that will help discriminate between the proposed evolutionary paths, and potentially identify signs of habitability and life. These results are summarized in our conclusions in Section 6.

## 2.0 Review of Habitability and Detectability.

In this section we briefly review the factors that can affect terrestrial planet habitability, and the plausible evolutionary paths for Proxima Cen b as explored in our companion paper, Barnes et al., (2016). We describe how various aspects of the planet and planetary system could affect its current climate. We also briefly review relevant knowledge on observations to identify signs of habitability and life for terrestrial planets orbiting M dwarfs - to motivate the detectability simulations and discussion that follow.

## 2.1 Evolutionary Processes and the Possible Habitability of Proxima Cen b.

Many factors other than planetary mass and star-planet distance affect a planet's volatile inventory, atmosphere, and surface environment. These factors, in conjunction with the evolution of the star and planet can maintain or destroy habitability. Classically, exoplanet habitability is defined as the ability to maintain liquid water on the surface of a rocky, terrestrial planet, and the habitable zone is that range of distances from the star over which a planet can maintain surface liquid water (Kasting et al., 1993). Proxima Cen b orbits in the habitable zone of an M5.5V star, and models of *in situ* planet formation suggest that terrestrial planets that form close to the star may be water-poor (Raymond et al., 2007; Lissauer et al., 2007). Proxima Cen b's close-in orbit ($a = 0.0485$ AU; Anglada-Escudé et al., 2016) also leaves it particularly vulnerable to the effects of stellar evolution and longer-term radiative and gravitational interaction between star and planet. These interactions could drive volatile and atmospheric loss processes and orbital changes that could preclude habitability for this planet. These processes are examined in more detail in our companion paper and an overview of the scope of plausible scenarios is given here to motivate the climate and spectral modeling work that follows.

Perhaps the greatest threat to Proxima Cen b's habitability is the luminosity evolution of its M dwarf host (Luger & Barnes, 2015). Lower-mass M dwarfs can experience extended pre-main sequence phases of up to a billion years, in which they are more luminous than they will ultimately be when they settle on to the main sequence (Baraffe et al., 1998; 2015). During this super-luminous pre-main sequence phase, any terrestrial planet that forms in what will become the main sequence habitable zone is subjected to extremely high levels of radiation, which can vaporize and photolyze oceans, and strip lighter elements from the atmosphere (Luger & Barnes, 2015). Simulations performed in our companion paper indicate that, had Proxima Cen b formed at its current location, then its host star would have taken as long as 160 million years after formation to dim sufficiently to allow the planet to enter the habitable zone for the first time. Consequently, if Proxima Cen b is a terrestrial planet, it may have been in a runaway greenhouse state for the first 160 million years. The vaporized ocean would have been photolyzed by UV from the host star, with subsequent loss of H to space, and between ~3 - 10 terrestrial ocean equivalents of water could have been lost, depending on the efficiency of oxygen sinks, photochemical shielding, and the planet's initial water endowment (Barnes et al., 2016).

Since loss of one Earth ocean equivalent of water can potentially produce up to 240 bars of atmospheric oxygen (Kasting, 1997), the remaining atmosphere may be strongly oxygen dominated (Luger & Barnes, 2015), but may transition to being more $CO_2$ dominated with time as the free oxygen reacts with the surface. The final amount of oxygen will depend on the initial water inventory, the stellar XUV flux, atmospheric losses through hydrodynamic escape and other top-of-atmosphere processes (Collinson et al., 2016), and the efficiency of planetary sequestration processes (e.g. Schaefer et al., 2016). However, if oxygen loss and sequestration is efficient—and $CO_2$ outgassing proceeds via terrestrial geological activity—then the atmosphere may transition from $O_2$-dominated, through $O_2/CO_2$-dominated, to a more Venus-like, $CO_2$-dominated atmosphere, if large quantities of outgassed $CO_2$ are present (Chassefière, 1996a; 1996b). Note that Barnes et al. (2016) found that the mantle temperature of Proxima Cen b could be maintained at high values due to either tidal heating or increased radiogenic heating, and so high outgassing rate may be likely. If the planet also outgasses $SO_2$, photochemical processes can result in the formation of a planet-wide $H_2SO_4$ cloud and haze deck. If the atmosphere is desiccated, with $H_2O$ abundances of the order of tens of parts per million, HDO may be enhanced as an indicator of early water loss, as is also the case for Venus (de Bergh et al., 1991). Should such a $CO_2$-dominated planet become significantly more desiccated than Venus—with atmospheric hydrogen inventories < 1ppm —then photochemical processes can split $CO_2$ while the lack of hydrogen-bearing species inhibit its recombination, resulting in a stable equilibrium mixture of $CO_2$, CO, $O_2$, and $O_3$ (Gao et al., 2015).

If, however, Proxima Cen b formed at its current position with a dense hydrogen envelope, or initially formed in a region of the solar nebula rich in ices—and then migrated inwards—then this scenario could lead to a habitable planet. In this case, rather than stripping water from the planet, the super-luminous pre-main-sequence phase of the star may have removed the primordial hydrogen atmosphere which would otherwise preclude habitability (Owen & Mohanty, 2016) to reveal a habitable evaporated core—a potentially volatile-rich planet without a dense hydrogen envelope (Luger et al., 2015). Our calculations suggest that if Proxima Cen b started with 0.1-1% of its planetary mass in hydrogen, then it could have survived the pre-MS phase and remained habitable (Barnes et al., 2016). The extreme ultraviolet radiation from Proxima Cen would have caused $H_2$ loss—protecting the water vapor beneath—for the 160 Myrs required for the star to dim and the planet to enter the habitable zone. At this point, the planet would have been safe from further $H_2O$ loss if atmospheric water vapor were cold trapped in the troposphere by a sufficient inventory of non-condensable gases, such as $N_2$ (Wordsworth & Pierrehumbert, 2014). Even if sufficient $N_2$ were not initially available, $O_2$ from the loss of $H_2O$, if not rapidly sequestered at the surface, could potentially reestablish the cold trap and inhibit subsequent water loss (Wordsworth & Pierrehumbert, 2014) Another potential path to habitability of Proxima Cen b is a later, large scale dynamical instability of its planetary system possibly caused by Proxima Centauri passing close to α Cen A and B. If the planet formed beyond the habitable zone, orbital disruption could have allowed it to arrive in its current orbit after the pre-main sequence phase. In this scenario, the planet would not need an initial thick hydrogen envelope to protect it from desiccation and could start off as a terrestrial body.

Over the 3.5–6 Gyr age of the system (Bazot et al., 2007; Barnes et al., 2016), Proxima Cen's activity levels may have also affected the planet's habitability. Despite its relatively long rotation period (83.5 days, Benedict et al., 1998), which often correlates with lower activity levels, Proxima Cen is a moderately active star with many strong flares per year (Davenport et al., 2016), and the stellar magnetic field which drives stellar activity is ~600 times stronger than that of our Sun (Reiners & Basri, 2008). Earlier modeling suggested that Earth-mass planets in the habitable zones of M stars would suffer continuous exposure to strong stellar winds originating from coronal mass ejection events with a subsequent rapid loss of planetary atmosphere (Lammer et al., 2008). This loss could be exacerbated by the shutdown of magnetic dynamo production due to tidal heating of the planet (Driscoll & Barnes, 2015), or higher initial radiogenic abundances than Earth (Barnes et al., 2016). However, other processes that may inhibit atmospheric loss still need to be taken into account, such as cooling to space from upper atmosphere $CO_2$ (Tian, 2009) or shielding via formation of ozone from high-$O_2$ atmospheres. Ribas et al., (2016) also argue that in the extreme case, with ion pickup losses, up to 100 bars of $N_2$ could be lost from the planet, resulting in 16–21 Earth ocean equivalents of hydrogen being lost.

If, however, the atmosphere survived and by the age of the system (3.5–6 Gy; Bazot et al., 2007) was ultimately Earth-like, the protons released with repeatedly flaring of this magnitude would destroy any incipient ozone layer, resulting in high surface UV fluxes during flare activity (Segura et al., 2010; M. Tilley, private communication). However, water shields from UV light, and so life may persist in the oceans, as long as there is sufficient atmosphere remaining to keep liquid water stable and liquid at the planetary surface. For even the strongest flares exhibited by Proxima Cen b, which are comparable to the great AD Leo flare (Hawley & Pettersen, 1991; Hawley et al., 2003), previous calculations suggest that UV damage to life can simply be avoided by remaining at a water depth of 9m or more, while still allowing photosynthesis (Kiang et al., 2007b). The resultant flux of photosynthetically active radiation would still be several orders of magnitude above the lower limit for useful light levels set by red algae, but the productivity of such a biosphere would be significantly lower than on Earth, with an estimated 4% of Earth's productivity for a star with AD Leo-type flares (Kiang et al., 2007b).

Proxima Cen b's close-in orbit also makes the planet more vulnerable to gravitational tidal interaction with the star (Jackson et al., 2008), which falls off rapidly with semi-major axis, as a$^{-7.5}$. Over time, the star could have circularized Proxima Cen b's orbit, trapped it into synchronous rotation, reduced the semi-major axis, and set the obliquity to zero (Barnes et al., 2008; Barnes et al., 2009; Heller et al., 2011). However, if Proxima Cen b's orbit has even a small eccentricity, possibly due to a companion planet or a recent perturbation due to a stellar encounter, then the gravitational interaction with the star can induce "tidal heating" due to friction as the body of the planet is flexed due to differential gravitational fields at different points in its orbit (Barnes et al., 2009). Sufficiently high levels of tidal heating could result in surface heat fluxes on the planet that could evaporate oceans of water on the planetary surface (Barnes et al., 2013). Our initial simulations suggest that by 3 Gy, for a starting eccentricity close to 0.1 then the orbit should have evolved to be currently close to circular (unless it is being perturbed by another planet) in which case the planet is experiencing very little internal heating from tidal forces. However, these forces would have been much more significant in the past, exceeding that of the volcanically active Io for the first billion years of the planet's existence (Barnes et al., 2016). However, these heat fluxes fall far short of those required to trigger a runaway greenhouse (Barnes et al., 2013). Instead, the heat deposited into the mantle of the planet may have reduced circulation in the planet's interior, and shut down the generation of a protective magnetic field (Driscoll & Barnes, 2015) which could have left the atmosphere more vulnerable to erosion.

Additional considerations for factors that could affect the habitability of Proxima Cen b include the fact that the orbit of Proxima Cen, in relation to α Cen A and B, is very poorly constrained (c.f. Wertheimer & Laughlin, 2006; Matvienko & Orlov, 2014). If this orbit takes Proxima Cen closer to its two companions, then the orbit of Proxima Cen b could have been significantly perturbed and induced large changes in eccentricity. As discussed above, these would in turn have made the planet vulnerable to climatic swings (Williams & Pollard, 2002; Williams & Kasting, 1997; Dressing et al., 2010), catastrophic tidal heating (Barnes et al., 2013), and the shutdown of the magnetic field (Driscoll & Barnes, 2015), which could have made atmospheric escape more likely. This disruption of Proxima Cen b's orbit could have occurred at any time in the past and is difficult to predict from the current position of the three stars. On the other hand, as discussed above, orbital instabilities caused by close passage to α Cen A and B is also a mechanism to aid this planet against having experienced extreme water loss during its star's pre main-sequence phase if these instabilities transferred the planet into its current orbit from a more distant one after the pre-main sequence phase ended. Perturbations from stellar encounters could also enhance habitability via impacts, which may generate atmospheric blow off of a dense $H_2$ envelope, or deliver volatiles to the planet after formation or after the pre-main sequence phase.

In summary, based on our current knowledge of M dwarf-planet interactions, there are several plausible scenarios for the environmental state of Proxima Cen b, including: an abiotic $O_2$-rich atmosphere, a $CO_2$-rich atmosphere, and a habitable terrestrial environment. In the abiotic $O_2$-rich scenarios, Proxima Cen b formed at or close to its current position and suffered catastrophic water loss during the star's super-luminous pre-main sequence phase. The resulting steam atmosphere was photolyzed, and H was lost to space. $O_2$, and possibly remnant water, was left behind, and so there could be two cases from this scenario: $O_2$-rich without water, and $O_2$-rich with water (Luger & Barnes., 2016). Similarly, if massive water loss occurs, and $O_2$ is lost either via hydrodynamic escape or sequestration in the planet's crust or mantle, or via a magma ocean, then $CO_2$ may be the dominant gas that persists in the atmosphere. In this case, the atmosphere may consist of remnant $O_2$ and outgassed $CO_2$, $CO_2$-rich and largely desiccated (Venus-like) or a $CO_2$-dominated, highly-desiccated planet (H < 1ppm) which produces a stable $CO/CO_2/O_2$ atmosphere (Gao et al., 2015). Finally, in the habitable terrestrial environment scenario, Proxima Cen b was a terrestrial body that migrated to its current orbit after the pre main-sequence phase through instability processes, or formed with a protective $H_2$ layer of no more than 1% of the solid mass of the object, either because a terrestrial planet formed *in situ* with that envelope, or because a more volatile and $H_2$-rich planet migrated inwards from beyond the snowline. That $H_2$ envelope would have been sufficiently thick to protect the volatile rich planet underneath during the super-luminous pre-main sequence phase, but not thick enough to remain and compromise the planet's habitability (Owen & Mohanty, 2016). For these cases, the resultant planet could have had a strongly oxidizing atmosphere, or one that was more reducing, depending on where in the planetary system it formed, and how it evolved.

*2.2 Impact of planetary characteristics and star-planet interactions on climate.*

Each of these scenarios will generate an environment that is currently being acted upon by the host star via its incident spectrum, activity levels, and orbital and tidal interactions. Each of these agents, when interacting with the planet's environment, can strongly impact the atmospheric composition, climate and potential habitability. For example, the planet's current climate and potentially-enhanced ability to maintain surface liquid water is caused by the interaction of the spectral energy distribution of the M dwarf star with the planet's atmospheric and surface composition (Shields et al., 2013), any clouds or hazes (Arney et al., 2016b), and the planet's orbital parameters and obliquity (Armstrong et al., 2014; Barnes et al., 2013). In particular, the UV spectrum of the star is critically important for understanding the planet's photochemistry - and therefore the atmospheric composition and climate (Segura et al., 2005). It is also the key to interpretation of any spectra obtained from the planet. The stellar UV also affects whether or not a UV-absorbing haze will form in a reducing atmosphere (Arney et al., 2016b) or an ozone layer in an oxidizing atmosphere (Segura et al., 2005; Domagal-Goldman et al., 2014). The presence and strength of these UV shields will affect the resultant surface UV flux, which could in turn strongly impact habitability. Sufficiently high UV flux could potentially sterilize the land surfaces, although life may still be adequately shielded in an ocean as previously mentioned (Kiang et al., 2007). Stellar flaring activity can also greatly increase stellar UV flux and eject protons, which collide with the planet's atmosphere and drive $NO_x$ chemistry in the stratosphere, potentially damaging or destroying an ozone layer (Segura et al., 2010; M. Tilley, private communication). Consequently, to assess the current environmental state of the planet, and to interpret any spectra obtained of this object, one of the first steps in planet characterization will be to observe and monitor the UV characteristics and activity of the host star.

Due to the gravitational tidal interactions described above, a terrestrial planet in the habitable zone around an M star should be tidally locked, a situation that can lead to synchronous rotation for planets with circular orbits, where one side of the planet is constantly facing the star. Although this was originally hypothesized to preclude planetary habitability, as the planet's atmosphere would eventually freeze out on the eternal night side of the planet (Kasting et al., 1993), subsequent modeling showed that the presence of a planetary atmosphere of sufficient density would protect against atmospheric collapse onto the dark side (Joshi et al. 1997; Goldblatt, 2016). Leconte et al., (2015) discussed how thermal tides in the planetary atmosphere may cause asynchronous rotation of tidally locked planets. However, Proxima Cen b is still more likely to be synchronously rotating under most assumptions for orbital position and atmospheric mass (Barnes et al., 2016), although existence in an asynchronous 3:2 spin-orbit resonance (similar to Mercury's) is also possible (Ribas et al., 2016; Turbet et al., 2016). Nonetheless, 3D General Circulation Models (GCMs) of terrestrial exoplanets now routinely show that atmospheric rotation can transfer heat from the day side to night side of the planet, and habitable conditions are possible for planets with as little $CO_2$ as 1 ppm near the inner edge (Kopparapu et al., 2016). For planets like Proxima Cen b with an orbital period of 11 days, 3D GCMs suggest that the relatively rapid rotation for a synchronously rotating planet can in fact result in cloud banding that reduces the overall albedo of the planet, further warming it (Kopparapu et al., 2016). However, although there is a high probability that Proxima Cen b is tidally locked, if companion planets exist in the system they could drive an eccentricity in Proxima Cen b's orbit that would cause the host star to librate in the sky, and make the planet susceptible to tidal heating. Therefore observations and constraints on the current orbit of Proxima Cen b, and the presence and orbital parameters of any companions, are extremely important for our understanding of Proxima Cen b's current climate.

*2.3 Identifying Planetary Habitability for Planets Orbiting M Dwarfs.*

Given the diversity of plausible evolutionary scenarios discussed above, and the star-planet interactions that may sculpt the current environment of the planet, one of the biggest questions posed for Proxima Cen b is, "how do we determine whether or not this planet is habitable?" Habitability can be assessed most straightforwardly by detecting liquid water on the planetary surface. This could be done using photometric measurements of the distant planet at visible or near-infrared wavelengths to search for signs of enhanced reflectivity near crescent phase due to the presence of ocean "glint" (Williams & Gaidos, 2008; Robinson et al., 2010; Robinson et al., 2014). Glint is specular reflectance at glancing angles from a smooth surface (Cox & Munk, 1954), which on a terrestrial planet is most likely to come from a liquid—since rock, snow, and snow-covered ice tend to have non-specular scattering properties.

Robinson et al., (2010) used a sophisticated 3D spectral model of the Earth (Robinson et al., 2011), validated against the EPOXI mission (Livengood et al., 2011) and Earthshine (Pallé et al., 2003) observations of the disk-averaged Earth, to show that the Earth deviates strongly from Lambertian (i.e. isotropic) scattering behavior at phases crescent-ward of quadrature. While a similar deviation from Lambertian behavior can occur due to forward scattering from clouds, Robinson et al., (2010) were able to show that an Earth-like planet with realistic (~50% coverage) forward-scattering water clouds and a Lambertian ocean produces a difference of as much as 100% when compared to an Earth-like planet with forward-scattering clouds and a specularly reflecting ocean, was as much as 100%. This was for measurements between phase angles of 90 and near 130 degrees, and wavelengths between 0.8–0.9 μm (Robinson et al., 2010; Robinson et al., 2014). Consequently, glint from the Earth's ocean is potentially detectable as a deviation in the observed reflectivity of the planet near crescent phase, even in the presence of forward-scattering clouds. A potential false positive for ocean glint may occur if the observer is preferentially sounding ice- and/or cloud-covered portions of the planet near crescent phases (Cowan et al., 2015), although this effect may be distinguishable with spectroscopic measurements. Glint may also polarize incident starlight, which could potentially induce a signal in the planet's polarimetric light curve. Williams & Gaidos (2008) used idealized models to show that cloud-free, ocean planets with non-polarizing atmospheres may exhibit a strong (30–70%) polarization signal over an orbit. However, Rayleigh scattering and clouds are also a source of polarization, possibly overwhelming the signal due to a surface ocean (Zugger et al., 2010; 2011) and making it difficult to use polarization to identify water under an atmosphere.

Note that glint is far less ambiguous for surface water detection than the presence of water vapor in the planetary atmosphere, as a planet may maintain a steam atmosphere without being habitable. "Anti-ocean" signatures may also be present in the form of highly soluble gases, e.g. $SO_2$, that would not accumulate in the atmosphere unless an ocean was not present (or the ocean was saturated with that gas). Conversely, false negatives for the detection of weater can occur, especially in transmission observations, if the water is cold-trapped in the troposphere of a habitable planet, resulting in an appropriately drier stratosphere. Water that is kept near the surface is most valuable for habitability, and less susceptible to loss processes. In this case, clouds and refraction in transmission (Misra et al., 2014a,b, Bétrémieux & Kaltenegger, 2014) limit our ability to probe into the relatively water-rich troposphere, where water is more likely to be detected. Drier stratospheres are less likely to affect direct imaging observations, however, except in the presence of high-altitude, planet-wide clouds. Even in the presence of broken cloud cover, or some planet-wide hazes, direct imaging observations can still sample the lower atmosphere and surface to detect water vapor (Arney et al., 2016a).

Another means to assess habitability is to constrain the surface temperature and pressure to determine if surface liquid water is feasible. This is best assessed with spectra of the planet in the visible and near-infrared and/or photometry or spectra in the mid-infrared. While a Rayleigh scattering slope from atmospheric molecules has been proposed as a means of determining atmospheric pressure for a terrestrial planet (Arnold et al., 2002; Woolf et al., 2002; Benneke & Seager, 2012) this method—like all methods for atmospheric pressure assessment—is not robust in the presence of cloud or haze cover, either complete or partial. At best it will return the pressure at the top of the clouds in the former, and an average of the cloud top altitudes and surface in the latter. This is graphically demonstrated in our Solar System by Venus, which has a 93 bar atmosphere, but exhibits extremely weak Rayleigh scattering, because the line of sight into the atmosphere is truncated by sulfuric acid haze at an altitude of 70 km and a pressure near 30 mbar. In direct imaging, which looks at planetary reflectivity, attempts to measure Rayleigh scattering are additionally vulnerable to the spectral properties of the underlying aerosols and surfaces. On Mars, strong absorption by surface iron oxide absorbs Rayleigh scattering in the blue. Similarly, for a $CO_2$- and $CH_4$-rich early-Earth-like atmosphere, the formation of a hydrocarbon haze results in strong absorption in the UV and blue, strongly altering the Rayleigh signature (Arney et al., 2016a,b).

A potentially more promising means for determining atmospheric pressure comes from measurements of collisional absorption from molecules such as $N_2$ (Schwieterman et al., 2015b) and $O_2$ (Tinetti et al., 2006b; Pallé et al., 2009; Misra et al., 2014a), which are both seen in the Earth's disk-averaged spectrum. Unlike Rayleigh scattering, which is degenerate in terms of characterizing the mix of gases that are producing the scattering, observations of collisional absorption can provide a direct measurement of the atmosphere's bulk constituents. However, this method is also subject to path length truncation due to clouds, and so will return a composite pressure depending on the available path lengths to clouds and the surface. In the case of $O_2$, observations of the $O_2$ molecule can be compared to the strength of the absorption band produced by the $O_2$-$O_2$ collisional pair—which is sensitive to density squared—to quantify the partial pressure of oxygen in the atmosphere (Misra et al., 2014a), and in the case of massive oxygen atmospheres, the presence of strong $O_2$-$O_2$ bands in the visible (0.35–0.65 μm) is diagnostic of massive $O_2$ atmospheres (Schwieterman et al., 2016).

In addition to searching for oceans, and determining surface temperature and pressure, habitability could be assessed by undertaking a spectroscopic survey of greenhouse gases and other planetary characteristics that affect climate. In particular, retrieving abundances for greenhouse gases such as $CO_2$, $CH_4$, $H_2O$, $SO_2$, $O_3$, and $N_2O$, obtaining pressure estimates using observations of $O_4$ and $N_4$, and searching for and characterizing hazes and clouds, will provide important constraints on planetary climate. These observations could then be used as input to coupled climate and photochemistry models (e.g. Segura et al., 2005) to understand the composition of the atmosphere, and the surface temperature and pressure.

It is important to note that for terrestrial planets with $CO_2$ (distinct from Titan whose atmosphere is extremely reducing), methane and organic hazes could also be a sign of either habitability or life (Arney et al., 2016a). Methane in a planet's atmosphere can produce organic haze if the $CH_4$/$CO_2$ ratio exceeds 0.1, and these hazes dramatically impact their planet's spectrum. Higher carbon dioxide levels makes haze formation more difficult, so larger fluxes of methane are needed to produce haze in the atmospheres of planets with larger $CO_2$ inventories compared to very reducing atmospheres like Titan. Hazes in Earthlike atmospheres containing $CO_2$ can therefore be a sign of a high methane production rate. Methane can be sourced from either biological or abiotic processes, but even abiotic methane is a potential habitability marker as its dominant source on an Earthlike planet is serpentinization—which are liquid water/rock reactions (Kelley et al., 2005). Serpentinization requires freshly exposed seafloor minerals to react, and while a limited area could be obtained from cracking, new seafloor crust would likely be needed for the maintenance of robust serpentinization on long timescales. Consequently, abundant methane on a terrestrial planet may indicate both liquid water and plate tectonics, two hallmarks of habitability. More intriguingly, existing measurements and models suggest that the presence of an organic haze on an Earthlike exoplanet with > 1% $CO_2$ in the atmosphere requires a higher methane production rate than is thought to be plausible from abiotic processes alone (Etiope & Sherwood-Lollar, 2013; Guzman-Marmolejo et al., 2013; Kharecha et al., 2005), so organic haze in an Earth-like atmosphere could also be suggestive of life if the $CO_2$ abundance can also be constrained (Arney et al., 2016a).

*2.4 Biosignature Considerations for Planets Orbiting M Dwarfs.*

Exoplanet biosignatures are biological modifications of a planet's global environment that are potentially observable over interstellar distances. On Earth, biosignatures can be classified into three major groups: atmospheric gases that are produced by life, such as the Earth's abundant photosynthetically-generated $O_2$ (Hitchcock & Lovelock, 1967); surface reflectivity signatures like the enhanced "red edge" reflectivity at wavelengths longward of 0.7 μm from vegetation (Gates et al.,1965) or non-photosynthetic pigments from other organisms (Schwieterman et al., 2015a); and time-dependent phenomena, such as seasonal changes in surface coverage or atmospheric gases (Meadows, 2006).

For planets orbiting M dwarfs, the UV spectrum and stellar activity of the star can work via photochemistry to either enhance or destroy the detectability of potential atmospheric biosignatures. Segura et al., (2005) showed that, for Earth-like surface fluxes of the biogenic gases $CH_4$ and $N_2O$, extremely large abundances of these gases can build up in a terrestrial planet atmosphere. In particular, the lifetime and abundance of atmospheric methane is increased from 10–12 years and 1.6 ppm for an Earth-like atmosphere orbiting a Sun-like star to over 200 years and over 300 ppm for a planet orbiting the M3.5V star AD Leo. This is due in large part to the slope of the M dwarf's UV spectrum which has smaller relative amounts of NUV radiation, and its subsequent relative inefficiency at photolyzing ozone to produce reactive $O(^1D)$ that generates the OH from water vapor that ultimately destroys $CH_4$ (Segura et al., 2005). On a related note, Segura et al. showed that for the same atmospheric $O_2$ abundance, ozone column density could be larger or smaller than Earth's by roughly a factor of two, again depending on the UV spectrum of the star. However, these calculations were performed for quiescent versions of M dwarf spectra. Segura et al., (2010) explored the effect on a planetary atmosphere of large flares, looking at both UV and proton flux on the thickness of an Earth-like planet's ozone layer. For the single flare that they studied, they found that the UV flux had negligible effect on the thickness of the ozone layer, dropping it by of order 1%. However, if the planet intercepted the proton beam from the coronal mass ejection, and did not have a protective magnetic field, the effect on the ozone was devastating, as proton-driven $NO_x$ chemistry from this single flare resulted in a 94% depletion of the ozone layer over a 2-year period (Segura et al., 2010). Consequently both the stellar spectrum and activity levels need to be well-characterized as an adjunct to interpreting potential biosignatures from planets orbiting M dwarfs.

Other important considerations when interpreting biosignatures of M dwarf planets are the likely detectability of the biosignature gas in question, as well as the likelihood that the planetary environment could produce the biosignature gas without life being present, and thereby exhibit a false positive for the biosignature. Oxygen is a particularly good gas for this discussion, as it has been well studied as a potential biosignature (see Meadows, 2016 for a comprehensive review). Oxygen is produced by photosynthetic organisms on Earth and is particularly attractive as a biosignature gas because it is present in high abundance, is evenly mixed throughout the atmosphere—and so can potentially be detected in the stratosphere by transmission observations— and has strong absorption bands at UV and NIR wavelengths where the first generation of exoplanet telescopes will observe. More exotic potential biosignatures, including organic sulfur compounds (e.g. Domagal-Goldman et al., 2011) are likely to be at lower abundances and be more sensitive to photolysis, so that they are confined closer to the planetary surface. These molecules also tend to absorb predominantly in the mid-infrared, which may be challenging to access for terrestrial planets with upcoming telescopes. However, to offset its advantages as a biosignature, oxygen is now known to have false positive production mechanisms – abiotic processes that can also produce $O_2$ in a planetary environment - and the majority of those mechanisms currently known are thought to be more likely to occur for planets orbiting M dwarfs (Domagal-Goldman & Meadows, 2010; Domagal-Goldman et al., 2014; Tian et al., 2014; Luger & Barnes, 2015; Gao et al., 2015, Harman et al., 2015.) For these mechanisms, carbon dioxide and water vapor serve as key sources of abiotic oxygen, and the spectrum of the host star and the resultant photochemistry plays a significant role, leaving tell-tale signs in the planetary atmosphere that can be sought to discriminate between biological and abiotic sources for oxygen (Meadows, 2016). In particular, two of the possible atmospheres for Proxima Cen b discussed above—including the massive $O_2$ atmospheres generated from water loss (Luger & Barnes, 2015) and the $CO_2$-rich, desiccated atmospheres that may contain large, stable $O_2$ fractions (Gao et al., 2015)—are both potential false positives for biologically produced oxygen. However, both may be discriminated by searching for either $O_4$ absorption, or both CO and $CO_2$ absorption in the planetary spectrum (Schwieterman et al., 2016). These studies show the importance of obtaining contextual information about the planetary environment, and provide a guide to increasing our confidence in biosignature detection by searching for additional gases and planetary characteristics.

### 3.0 Models.

We use a suite of planetary climate and photochemistry models to simulate the current environmental state for the evolutionary scenarios described in Sections 1 and 2. We then use a radiative transfer model to predict the potentially observable photometric and spectral parameters that would help discriminate between these environmental states. These models and the stellar and surface albedo inputs are described in the sections below.

*3.1 SMART—Direct Imaging Synthetic Spectra.*

The SMART (Spectral Mapping Atmospheric Radiative Transfer) code is a 1D line-by-line, multi-stream, fully multiple scattering radiative transfer model (Meadows & Crisp 1996; Crisp 1997) that computes accurate synthetic planetary spectra. It is used in this study to generate synthetic planetary spectra for both direct imaging, and through the SMART-T modification, for transmission spectroscopy. It also serves as the radiative transfer engine for phase curve generation and for the VPL climate model. We have validated SMART against observations of Solar System planets, including Mars, Earth, and Venus (Tinetti et al 2005; Robinson et al. 2011; Arney et al. 2014). SMART requires a number of user inputs, such as a pressure/altitude-temperature grid, gas mixing ratios, molecular absorption coefficients generated using LBLABC (see Section 3.4.7, below) from line lists (i.e. HITRAN, HITEMP), collision-induced absorption data, UV-visible absorption cross sections, a stellar spectrum, and wavelength-dependent surface albedo data. We can also incorporate aerosols such as hazes and clouds, which require input containing the altitude-dependent optical depths as well as the particle asymmetry parameter and the extinction, scattering, and absorption efficiencies ($Q_{ext}$, $Q_{scat}$, and $Q_{abs}$). SMART can generate spectra at any arbitrary spectral resolution.

*3.2 SMART-T—Transit Transmission spectra.*

Although Proxima Cen b may not transit, the simulations provided here for its plausible environments are also more generally applicable to planets orbiting late type M dwarfs. Consequently we simulate transmission as well as direct imaging observations for this object. To generate the transit transmission spectra, we use the SMART-T model (Misra et al. 2014a; Misra et al. 2014b). This version of SMART uses the normal-incidence optical depths of SMART to calculate the transmission through limb-traversing atmospheric paths. The model takes into account the effects of refraction using the path integration method of Werf (2008). Starlight refracted out of the beam of an observer sets a limit on the tangent altitudes at which the atmosphere can be probed (Sidis & Sari, 2010; García Muñoz et al., 2012; Misra et al., 2014a,b). SMART-T requires the same inputs as the standard SMART code, but to accurately calculate refraction additionally requires the radius of the host star and the index of refraction of the bottom layer of the atmosphere at standard temperature and pressure. SMART-T incorporates limb-darkening using the coefficients from Claret (2000). For all transmission spectra presented here, an impact parameter of 0 is assumed.

*3.3 SMART Phase Curves*

We use the SMART Phase Curve model (developed by T. Robinson and D. Crisp) to calculate the multi-band, orbital phase-variability of a planet in reflected and emitted light. SMART is used to compute a library of multi-stream, 1D radiative transfer calculations over a grid of solar and observer zenith and azimuth angles and this code then generates a phase-dependent, disk-integrated spectrum of the planet using the method of n-point Gaussian quadrature. In this formalism, the zenith angles for the n radiance streams computed by SMART for each set of solar and observer, zenith and azimuth angles form the abscissa for an integral over the surface of a sphere. The corresponding weights for the integration are determined based on the longitudes of the planetary disk that are visible and illuminated as seen by the observer. The wavelength-dependent radiances are interpolated on a sphere and integrated using the weights. We apply this disk-integration over a grid of observed planetary phase angles to simulate high-resolution, phase-dependent spectra. These high-resolution spectra are rebinned to a variety of lower spectral resolutions to determine the detectability of the planet's phase variability against the bright stellar background.

Our focus is on the phase dependence of the planet due to the vertically resolved atmospheric structure atop a reflective surface, and not on the spatially resolved features accessible with a GCM. Therefore, the distinguishing observables from one plausible planetary state to another are limited to bulk surface and atmospheric characteristics, such as deviations from Lambertian scattering due to the presence of forward-scattering clouds and hazes, or the presence of ocean glint (Robinson et al., 2010; Robinson et al., 2014), modulations in the peak amplitude of phase curves as a function of wavelength due to molecular features (Selsis et al., 2011; Stevenson et al., 2014), and thermal phase curve amplitudes due to a day-night temperature contrast. We are not using a GCM in this study and we have not made self-consistent predictions for the day-night temperature contrast. To simulate phase curves with a thermal emission contribution from the nightside of the planet we use our self-consistent dayside surface temperature, assuming no day-night temperature contrast and extremely efficient heat redistribution (e.g. Venus). We also simulate cases where we assume no thermal contribution from the nightside of the planet to simulate the maximum day-night temperature contrast, and severely inefficient heat redistribution more appropriate for an airless body (Maurin et al., 2012). In this way, we assess the extreme end-member cases for the phase curve amplitude. We also simulate an intermediate case where the nightside surface temperature and temperature-pressure profile is 20 K lower than the dayside to emphasize a plausible phase curve amplitude.

*3.4 Instrument Simulators and Noise Model*

We use a general coronagraph instrument noise model to simulate directly-imaged reflectivity spectra for the atmospheric states considered in this work. We refer the reader to Robinson et al. (2016) for a thorough description of the model. In brief, the coronagraph instrument noise model computes wavelength-dependent photon count rates on the detector due to the planet, zodiacal and exozodiacal light, dark current, read noise, speckles, and thermal emission from the mirror. The model also considers the CCD quantum efficiency of assumed detectors, the dependence of dark current on the near-IR detector due to detector temperature, and the limitations to achieving the photon counting noise limit due to the spacecraft roll maneuver required for background subtraction (see Brown, 2005). With the coronagraph noise model, we simulate observations using three future telescope concepts: a space-based 16 meter LUVOIR, a space-based 6.5 meter HabEx, and a ground-based 30 meter telescope assumed to be located in the Atacama Desert, Chile.

To model ground-based observations, we modified the coronagraph noise model to account for transmission through, and downwelling thermal emission from, Earth's atmosphere. Both of these quantities can be calculated using the SMART model. We assume a cold planetary surface temperature (269 K) and an observing zenith angle of 30°, which corresponds to the altitude of Proxima Centauri crossing the meridian at La Silla Observatory. The parameters used to simulate coronagraphic observations are given in Table 1; other model parameters are identical to those assumed in Robinson et al. (2016; their Table 3).

Table 1. Parameters used for the coronagraph noise model. See Robinson et al. (2016) for a complete list of baseline telescope and astrophysical parameters.

| Parameter Description | HabEx | LUVOIR | 30 m |
|---|---|---|---|
| Telescope diameter | 6.5 m | 16 m | 30 m |
| Mirror/System temperature | 200 K | 200 K | 269 K |
| Contrast | $10^{-10}$ | $10^{-10}$ | $10^{-10}$ |

*3.5 SMART Climate Model.*

Our new 1D radiative-convective equilibrium (RCE) climate model (Robinson & Crisp, in prep) incorporates SMART for full-physics radiative transfer as its core, which is coupled with a variety of convection and time-stepping methods to determine the equilibrium pressure-temperature structure of an atmosphere. As our radiative fluxes are computed using a multi-stream solver, we avoid inaccuracies that can impact models that use δ 2-stream approximations (Kitzmann et al., 2013). Note that the SMART climate model is, at present, not dynamically coupled to any atmospheric chemistry tools. Therefore, in this work we present results from this model only for cases for which our other 1D climate models lack key capabilities, such as for high-pressure atmospheres.

Net radiative fluxes in the SMART climate model are computed using a first-order linearized flux-adding approach, as described in detail by Robinson & Crisp (in prep). Full multi-stream, multi-scattering solar and thermal radiative fluxes are computed by SMART, as well as a set of Jacobians that describe the response of the radiative fluxes to changes in key elements in the atmospheric state vector (e.g., temperature, gas mixing ratios, aerosol optical depth). These Jacobians consist of derivatives of the layer-by-layer, wavelength-dependent stellar and thermal source terms, as well as derivatives of layer reflectivity, transmissivity, and absorptivity. This information is used in a linear flux-adding approach to determine the upwelling and downwelling solar and thermal flux profiles at each time-step. If the evolved atmospheric state is outside of the linear range of the Jacobians, additional iterations of computing Jacobians can be required. Since SMART computes fluxes at extremely high spectral resolution, spectral quantities in our flux-adding approach are degraded to a lower resolution (typically 10 cm$^{-1}$) by convolution with a slit function, which still provides thousands of wavelength intervals.

Our model can implement convection in a number of ways, using either convective adjustment (Manabe & Strickler, 1964), mixing length theory (e.g. Gierasch & Goody, 1968), or turbulent mixing theory (Mellor & Yamada, 1972). Convective adjustment is a simple method to reset the lapse rate between atmospheric layers to a prescribed lapse rate (e.g. the dry adiabatic lapse rate). This provides a quick and simple way to incorporate the effect of convection into the thermal profile, which is critical for calculating a realistic surface temperature. A more physically-informed method is to use mixing length or turbulent mixing theory.  Specifically, for the simulations computed using the SMART climate model, we first use convective adjustment to rapidly iterate from an isothermal initial profile. We then transition to mixing length theory, which is used to calculate the eddy diffusion rates, kinematic fluxes, and convective heat transport. The dry adiabatic lapse rate is calculated for each layer at each timestep to determine convective stability. The timescales at which we resolve our mixing length model are small, consistent with the time it takes for an air parcel to rise one layer. Since radiative flux calculations are more computationally expensive, we employ a timestep splitting approach where many convective timesteps are resolved within each radiative timestep. Here, we use adaptive methods to calculate the maximum radiative and convective time-steps to more rapidly iterate the temperature-pressure profile to equilibrium. The limit for the radiative timestep is simply an allowed change in temperature for any given layer per timestep (e.g. 0.1 K), while mixing length theory provides for a maximum stable convective timestep in a single layer:

$$dt_c = \frac{1}{2} \frac{dz^2}{K_{zz}}$$

where $dz$ is the change in altitude and $K_{zz}$ is the eddy diffusion coefficient. We take the smallest of these from all layers as the maximum convective timestep in the convective region.

The model atmosphere constructed for SMART Climate simulations generally consist of 64 plane-parallel layers in hydrostatic equilibrium, spanning from the surface pressure to 10$^{-7}$ bar, equally spaced in log-space, except that there is finer spacing at the surface and coarser spacing at the top of atmosphere. Fluxes are computed for the entire stellar SED range (~0.1–5.5 μm), divided into ~8600 intervals, and thermal flux from the planet is computed for the wavelength range ~2–200 μm, divided into ~500 intervals, all spaced at 10 cm$^{-1}$. Gas absorption is computed in all intervals for which cross sections or line data are available.

Validations of Earth, Mars, and Venus are detailed by Robinson & Crisp (in prep). These validations provide confidence that this model can be applied to a variety of planetary climates, including those we discuss for Proxima b. In Section 4.1 we also provide a cross-comparison between the SMART Climate model and the LMD model used by Turbet et al., for a subset of similar Proxima Cen b calculations.

*3.6 Atmos Coupled Climate-Photochemical Model.*

We use a 1D photochemical-climate model, *Atmos,* to simulate photochemistry and climate of terrestrial planet environments. With this coupled code, we are able to generate atmospheres that are both chemically and climatically self-consistent with the atmospheric composition, temperature profile, and incident stellar spectrum. To use *Atmos*, the photochemical model (which can include particle microphysics) is run first to generate an initial atmospheric state based on user-specified boundary conditions (i.e. gas mixing ratios or fluxes and deposition velocities, the stellar spectrum, the total atmospheric pressure, the initial temperature-pressure profile). Once the photochemical model reaches a converged state, the photochemical model feeds its outputs into the climate model. The climate model uses the photochemical model's final state as its initial condition, and the models iterate in this manner until global convergence is reached. This coupling and the criterion for convergence is described in more detail in Arney et al. (2016a).

The photochemical portion of *Atmos* is based on the 1D photochemical code originally developed by Kasting et al. (1979). The version we use here has been significantly modernized as described in Zahnle et al. (2006) and can simulate a wide range of planetary redox states ranging from extremely anoxic (pO$_2$ = 10$^{-16}$) to 100 bars of O$_2$ (Schwieterman et al., 2016). An organic haze-formation scheme is in place for reducing, methane-rich atmospheres as described in Pavlov et al. (2001), Zerkle et al. (2012) and Arney et al. (2016a). For the simulations presented here, the model atmosphere is divided into 200 plane-parallel layers up to 100 km in altitude with a layer spacing of 0.5 km. Hydrostatic equilibrium is assumed. A vertical transport scheme includes molecular and eddy diffusion. Boundary conditions can be set for each species at the top and bottom of the atmosphere, including gaseous mixing ratios and/or fluxes in or out of the atmosphere. Radiative transfer in the photochemical model is done via a δ 2-stream method (Toon et al. 1989).The primary equations solved by the model are the continuity and flux equations, which are, in order,

$$\frac{\partial n_i}{\partial t} = P_i - l_i n_i - \frac{\partial \Phi_i}{\partial z}$$

and

$$\Phi_i = -Kn\frac{\partial f_i}{\partial z} - D_i n_i \left(\frac{1}{n_i}\frac{\partial n_i}{\partial z} + \frac{1}{H_i} + \frac{1 + \alpha T_i}{T}\frac{\partial T}{\partial z}\right)$$

where $z$ is altitude (cm), $t$ is model time (s), $n_i$ is the number density (cm$^{-3}$) of species $i$, $P_i$ is chemical production rate (in molecules cm$^{-3}$ s$^{-1}$), $l_i$ is the chemical loss frequency (s$^{-1}$), $\Phi_i$ is the flux of species $i$ (cm$^2$ s$^{-1}$), $f_i$ is the mixing ratio of the species $i$ ($n_i/n$), K is the eddy diffusion coefficient (cm$^2$ s$^{-1}$), $n$ is the total density, $D_i$ is the diffusion coefficient between the background atmospheres and species $i$, and $\alpha_{Ti}$ is the thermal diffusion coefficient between species $i$ and the background atmosphere. $H_i$ is the scale height of species $i$ (note $H = kT/m_ig$). These equations are integrated using a variable timestep reverse Euler method appropriate for stiff systems. This method relaxes to the steady state solution when timesteps are large.

The climate portion of *Atmos* is based on the 1D climate model originally developed by Kasting et al. (1983), Kasting et al. (1984), Kasting & Ackerman (1986), but the version we use here has been significantly modernized and was recently used to simulate habitable zones around varied spectral types (Kopparapu et al., 2013). The model uses correlated-$k$ coefficients to compute absorption by spectrally active gases including $O_3$, $CO_2$, $H_2O$, $CH_4$, and $C_2H_6$. The KSPECTRUM[5] program was originally used to calculate the correlated-$k$ coefficients using the HITRAN 2008 line lists (Rothman et al., 2009) and updated in Kopparapu et al. (2013) with the HITEMP 2010 line lists (Rothman et al., 2010). As in the photochemical model, this climate model uses a δ2-stream multiple scattering algorithm (Toon et al., 1989). The shortwave (absorbed stellar radiation) wavelength grid spans from λ = 0.2 to 4.5 μm in 38 spectral intervals; to compute outgoing longwave IR radiation, there is a separate set of correlated-$k$ coefficients in 55 spectral intervals for each gas included in our scheme for wavenumbers 0–15,000 cm$^{-1}$ (λ > 0.67 μm). The density structure of the atmosphere is calculated assuming hydrostatic equilibrium. The tropospheric temperature profile is calculated by following a wet adiabatic lapse rate to the altitude at which the stratospheric temperature is reached (Kasting 1988), except in desiccated cases, for which a dry adiabat is assumed. The water vapor distribution with altitude is determined by a Manabe & Wetherald (1967) profile with a surface relative humidity of 80% (Kasting & Ackerman 1986). Gases in the upper atmosphere can have a heating or cooling effect on the temperature profile depending on the relative abundance of gases in the upper atmosphere and the extend of shortwave heating.

Note that we have been unable to run this climate model to a converged state using the same top-of-atmosphere pressure that the photochemical model has. The photochemical model grid extends to 100 km in altitude, but for 1 bar atmospheres, we are typically unable to run this climate model with pressures above 80 km in altitude at the top of the pressure grid due to model instabilities.

---

[5] http://code.google.com/p/kspectrum/

Therefore, when this climate model passes its temperature and water profiles to the photochemical model, they are fixed at their values at the top of the climate grid, and they become isoprofiles above the top of this grid. Previous tests (Arney et al. 2016a) suggest that this treatment does not strongly impact the resultant photochemistry.

Organic haze particles, which are relevant to some of our simulations, are treated as fractal (rather than spherical) in shape in both the photochemical and climate portions of the *Atmos* model (Arney et al., 2016a). Fractal particles are agglomerates of spherical particles, and studies of organic hazes in the laboratory (e.g. Trainer et al., 2006) and from observations of Titan (e.g. Rannou et al., 1997) suggest that fractal particles are more realistic for organic hazes compared to spherical particles. In particular, we treat our fractal particles as agglomerates of 0.05 μm spherical "monomers." This size of monomer was chosen because it is similar to the size of the monomers of Titan's haze particles (Rannou et al., 1997; Tomasko et al., 2008), and it was the size used in the first study to simulate fractal hazes in an Earth-like atmosphere (Wolf & Toon, 2010), which our haze input files are based on. Haze scattering and absorption properties are calculated with Mie scattering for our sub-monomer particles (r < 0.05 μm), and the fractal mean field approximation (Botet et al., 1997) for fractal particles (r > 0.05 μm).

*3.7 Model inputs.*

To simulate planetary environments and observable properties, our models require information on the planetary and stellar characteristics, including: stellar parameters and spectrum, planetary physical and orbital parameters, as well as environmental information such as the surface albedo, aerosol and atmospheric molecular absorption properties.

*3.7.1 Planetary and Stellar Parameters.*

We use the best-fit minimum mass of 1.3 $M_\oplus$ from Anglada-Escudé et al. (2016) and adopt the silicate planet scaling law from Sotin et al. (2007) to obtain a planetary radius of 1.074 $R_\oplus$ (6850 km). Note that this assumes an edge-on, or nearly edge-on, inclination for the system, which is currently unknown. We adopt the best-fit semi-major axis for Proxima Centauri b of 0.0485 AU. We also assume the stellar radius is 0.141 $R_\odot$ (Boyajian et al., 2012). These parameters are the same as those used in our companion paper Barnes et al. (2016).

*3.7.2 The Stellar Energy Distribution of Proxima Centauri.*

To self-consistently model the photochemistry, climate, and the expected reflectance and transmission spectra from Proxima Centauri b, an EUV to MIR input spectrum is required. Since Proxima Centauri is the nearest star to our Sun, many spectra have been recorded over the years, though no single source currently provides a calibrated, normalized spectrum for all necessary wavelengths. We have compiled a self-consistent, panchromatic stellar spectrum representative of Proxima Centauri (Figure 1) using publicly available data downloaded from the Mikulski Archive for Space Telescopes (MAST)[6] and using PHOENIX 2.0 spectral models[7] at wavelengths where calibrated observations are not yet publicly available (Husser et al. 2013).

For UV wavelengths, we combined calibrated HST Space Telescope Imaging Spectrograph (STIS) observations from $\lambda = 1153$–1691 Å, observations from the International Ultraviolet Explorer's (IUE) Short Wavelength prime (SWP) instrument from $\lambda = 1500$–1816 Å, and observations from the IUE Long Wavelength Prime (LWP) instrument from 1851–3347 Å. We preferentially used STIS data at all UV wavelengths that overlapped with IUE SWP data (1500–1691 Å) due to the higher spectral resolution of STIS. The final STIS contribution was calculated as the median point at each wavelength of six available calibrated observations (HST Proposal ID: 8040). There were multiple IUE observations available; we took the median of several calibrated observations (IUE Proposal ID: RPERG). For the visible spectrum ($\lambda = 4569$–8499 Å) we used observations from the HST Faint Object Spectrograph (FOD/RD) (HST Proposal ID: 6059). The infrared spectrum ($\lambda > 8499$ Å) and the wavelength gap between the IUE/LWP and HST/FOD data ($\lambda = 3349$–4569 Å), were computed using a linear interpolation of PHOENIX 2.0 spectral models (Husser et al., 2013) with $T_{\mathrm{eff}} = [3000 \text{ K}, 3100 \text{K}]$ and [Fe/H] = [0.0, 0.5]. The combined spectrum was normalized to match Proxima Centauri's bolometric luminosity of $L/L_\odot = 0.00155$ (Boyajian et al., 2012). Table 2 provides a summary of the sources for our combined spectrum. An important limitation to note for this spectrum is that it is a static representation of the star in its quiescent state and does not account for flare activity (i.e., Davenport et al., 2016).

---

[6] https://archive.stsci.edu/
[7] http://phoenix.astro.physik.uni-goettingen.de/

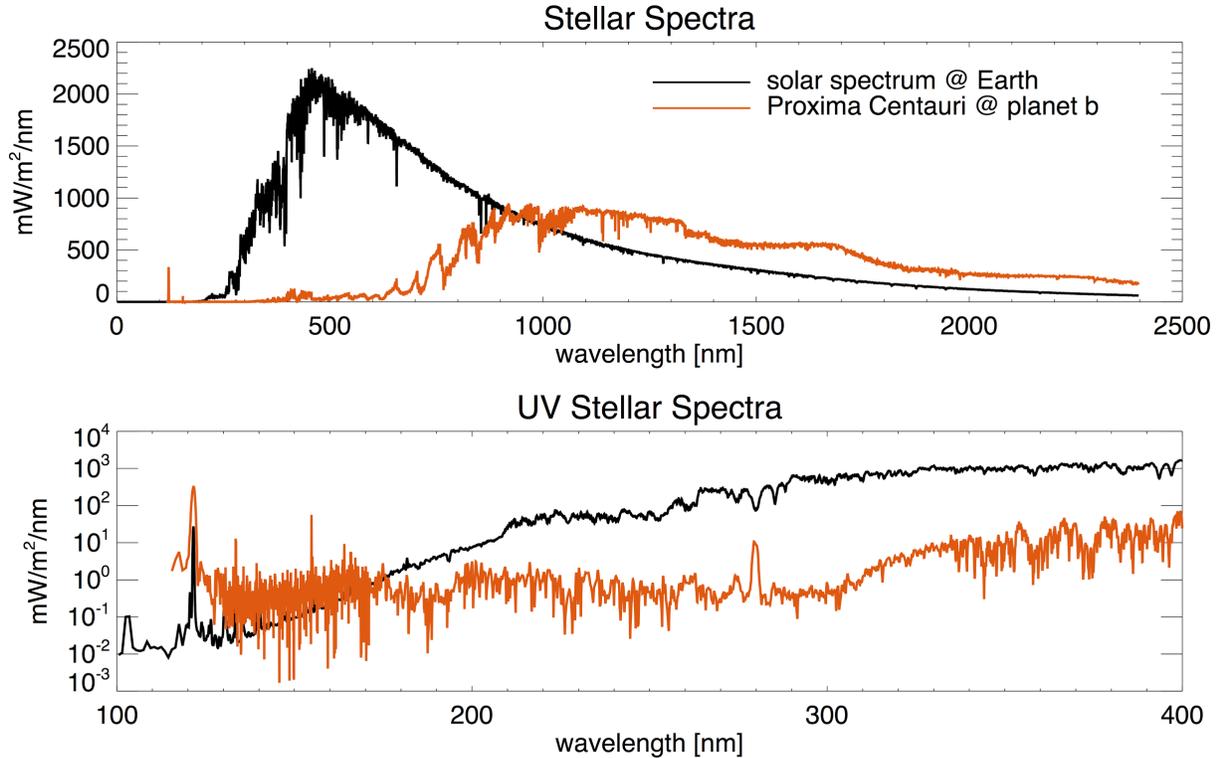

*Figure 1. The spectrum of Proxima Centauri (at the distance of Planet b) compared to the solar spectrum. Proxima Centauri b receives about 0.66 times the insolation Earth receives at 1 AU from the Sun.*

### 3.7.3 Input Line Lists.

Our models use a variety of sources to compute gas absorption from line lists, collision-induced absorption coefficients, and UV-visible cross sections. The line lists are from HITRAN (Rothman et al., 2013) for all species except for high-temperature applications for $CO_2$ and $H_2O$, for which we use HITEMP (Rothman et al., 2010). Collision-induced absorption (CIA) for $CO_2$-$CO_2$, $O_2$-$O_2$, and $N_2$-$N_2$ are from a number of sources. The $CO_2$-$CO_2$ absorption data, which is included in high-$CO_2$ models and cover a range of temperatures up to 700 K, are sourced from Moore (1971), Kasting et al. (1984), Gruszka & Borysow (1997), Baranov et al. (2004), and Wordsworth et al. (2010). $N_2$-$N_2$ CIA coefficients are calculated based on the empirical model from Lafferty et al. (1996) as described in Schwieterman et al. (2015b). The $O_2$-$O_2$ absorption coefficients are taken from C. Hermans[8] (Hermans et al. 1999) for the 0.333–0.666 μm range and from Greenblatt et al. (1990) and Maté et al. (1999) for the $O_2$-$O_2$ absorption at 1.06 μm and 1.27 μm, respectively. UV-visible cross section data is sourced from primary references available from the MPI-Mainz UV/VIS Spectral Atlas of Gaseous Molecules of Atmospheric Interest[9].

---

[8] http://spectrolab.aeronomie.be/o2.htm
[9] http://satellite.mpic.de/spectral_atlas

Table 2. Data Sources for Proxima Centauri Spectrum

| Wavelength Range [Å] | Platform/Instrument | Observation ID/references |
|---|---|---|
| 1153–1691 | HST/STIS[1] | o5eo01010[20,30], o5eo02010[20,40] |
| 1691–1816 | IUE/SWP[1] | swp54725, swp55040, swp55378 |
| 1851–3347 | IUE/LWP[1] | lwp30641, lwp30654, lwp30692, lwp30741, lwp30765, lwp30836, lwp30855, lwp30864, lwp30914, lwp30917, lwp30928, lwp30931, lwp30941, lwp30947, lwp30964, lwp31016, lwp31079, lwp31092, lwp31140, lwp31162, lwp31180, lwp31208, lwp31242, lwp31475 |
| 3347–4569 | PHOENIX 2.0 Model | Husser et al. (2013) |
| 4569–8499 | HST/FOD/RD[1] | Y2WY0705T, Y2WY0305T |
| $\lambda > 8499$ | PHOENIX 2.0 Model | Husser et al. (2013) |

[1]Data obtained from MAST [https://archive.stsci.edu/]

*3.7.4 LBLABC.*

A companion code to SMART, the Line-By-Line Absorption Coefficients (LBLABC) code, generates line-by-line temperature-dependent absorption coefficient profiles using input line lists (e.g. HITRAN) and temperature, pressure, and gas mixing ratio profiles (Meadows & Crisp 1996). LBLABC fully resolves the narrow line cores and wings as far as 1000 cm$^{-1}$ from the line center and was designed for a wide range of pressures ($10^{-5}$–100 bars) and temperatures (130–750 K). Unless otherwise specified, a line cutoff of 1000 cm$^{-1}$ is used here.

*3.7.5 Surface Spectral Albedo Inputs.*

In the case of Lambertian scattering, we chose surface spectral albedos to be consistent with the possible surface conditions for the type of atmosphere modeled. Depending on the type of atmosphere, we used a basalt, desert, iron oxide (goethite), ice, seawater, or composite surface albedo spectrum. For modern Earth-like cases, a weighted composite spectrum was used that consists of 65.6% seawater, 13.6% grassland/brush, 4% conifer forest, 5.5% soil/desert (kaolinite), and 11.3% snow/ice, based on a diurnally-averaged equatorial Earth view during spring equinox (composite 1; Robinson et al., 2011). For early Earth-like cases we used a similar composite spectrum with land vegetation removed: 65.6% seawater, 23.1% soil/desert, and 11.3% snow/ice (composite 2). The forest spectrum was taken from the ASTER spectral library (Baldridge et al., 2009), while all other surface spectra were taken from the USGS spectral library (Clark et al., 2007). In the subset of cases that included glint, we used the Cox & Munk (1954) glint model as described in Robinson et al. (2010, 2011). Figure 2 shows each of the spectral albedo inputs used over the wavelength range 0.2–2.5 μm.

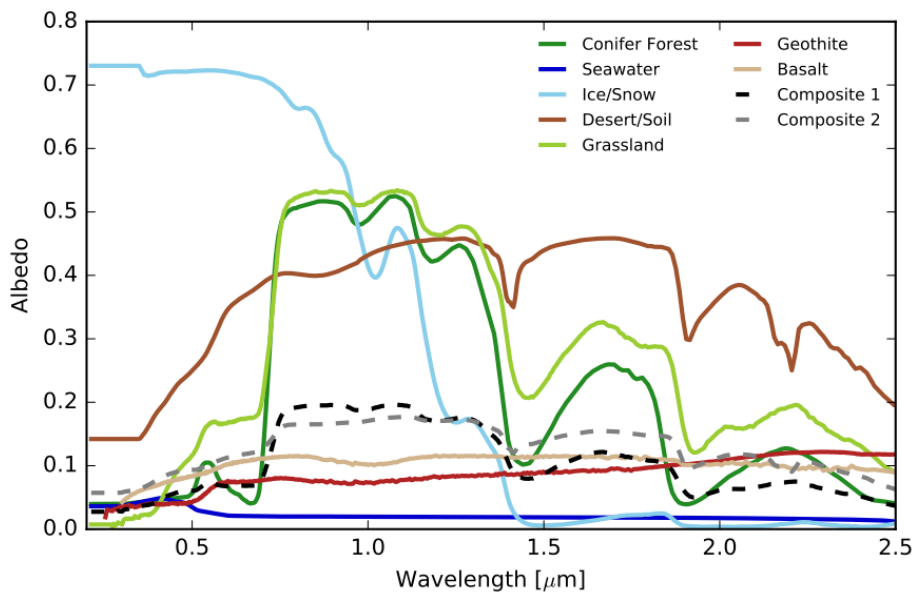

*Figure 2 - Input spectral surface albedos for modeled planetary scenarios. Composite 1 is a weighted average of 65.6% seawater, 13.6% grassland/brush, 4% conifer forest, 5.5% soil/desert (kaolinite), and 11.3% snow/ice. Composite 2 is 65.6% seawater, 23.1% soil/desert, and 11.3% snow/ice. All surface spectral albedos are sourced from the USGS spectral library (Clark et al., 2007), except for the conifer forest, which is from the ASTER spectral library (Baldridge et al., 2009)*

### 3.7.6 Aerosol Inputs.

Aerosols such as hazes and clouds are incorporated into SMART by providing input containing the altitude-dependent opacities as well as the phase function or particle asymmetry parameter ($g$) and the extinction, scattering, and absorption efficiencies ($Q_{ext}$, $Q_{scat}$, and $Q_{abs}$). We simulate aerosols in several of our planetary atmospheres: these include organic hazes, sulfuric acid clouds and haze, and water vapor clouds. Where we simulate organic hazes, we treat them as fractal aggregates using the fractal mean field approximation (Botet et al., 1997), and we use the hydrocarbon refractive indices measured by Khare et al. (1984). The particle sizes, number densities, and vertical distributions for organic hazes are derived from our *Atmos* simulations. We simulate sulfuric acid clouds and hazes in our Venus-like spectral models. Our nominal Venusian cloud profiles and particle size populations are based on Crisp (1986). Sulfuric acid refractive indices are derived from Palmer & Williams (1975). The cloud particle populations are modeled with log-normal size distributions, and the optical efficiencies are calculated using the code "Miescat" adapted from the code by Wiscombe (1980). Water vapor (stratocumulus) clouds are modeled using refractive indices from Hale & Querry (1973). Ice (cirrus) clouds are modeled as distributions of hexagonal crystals with scattering properties determined by a geometric optics approach (Muinonen et al., 1989). For spherical particles, a full Mie phase function is used; otherwise we use a Henyey-Greenstein phase function (Henyey & Greenstein, 1941). Patchy clouds on a disk-integrated spectrum are calculated in our 1D radiative transfer model using a weighted average of 50% clear-sky, 25% cirrus cloud, and 25% stratocumulus cloud.

## 4.0  Results.

To explore the possible evolutionary end states for Proxima Cen b, we present self-consistent atmospheres with constituent vertical mixing ratios and temperature-pressure profiles for the planetary states discussed in Section 2. Most of the simulations presented here are photochemically and climatically self-consistent unless otherwise noted. Following generation of these atmospheres, we simulate phase curves, and direct imaging and transmission spectra for these cases.  We then outline observational considerations and discuss detectability of spectral features that may be able to discriminate evolutionary states, habitability, and biosignatures.

*4.1 Comparison with Existing 3D GCM Simulations.*

Our computationally efficient, radiatively rigorous 1D climate models cannot self-consistently include 3D feedbacks and heat transport processes that can be studied in detail with more sophisticated 3D GCM models. Therefore, before presenting our climate results, we compare our model results with existing studies.  These inter-model comparisons can be valuable when ground truth is unobtainable, as agreement between dissimilar models can help to bolster each model's conclusions.

We compare our new VPL SMART Climate model with the recently submitted work by Turbet et al. (2016), who used a 3D GCM in a similar study examining the climate of Proxima Centauri b. Although our study's companion paper and the companion paper of Turbet et al. (2016) argue for synchronous rotation as a likely scenario for Proxima Centauri b (Barnes et al., 2016; Ribas et al., 2016), 1D models implicitly assume a rapidly-rotating planet to calculate a globally-averaged surface temperature. Consequently, the most relevant comparison is between our results and the spin-orbit resonance (asynchronous) results from Turbet et al. (2016). The atmospheres selected for comparison include $N_2$-dominated atmospheres with trace $CO_2$ and pure $CO_2$ atmospheres. The $N_2$ cases consist of 1 bar $N_2$ with 376 ppm $CO_2$, either with or without water. For the moist case, we use the water profile shown in Figure 7 of Turbet et al. (2016) to enable a close comparison. The pure $CO_2$ cases are for 1, 4, and 6 bar atmospheres. We use a wavelength independent surface albedo of 0.20, and a solar zenith angle of 60 degrees, which approximates the diurnal average in our 1D model.

Our VPL Climate model appears to be in reasonable agreement with the LMDz for the test cases chosen, with at most a discrepancy of 11K between the surface temperatures derived by the models. For the 1 bar $N_2$, 376 ppm $CO_2$ atmosphere we agree with Turbet et al., (2016) that this $CO_2$ amount is insufficient to maintain a global-averaged surface temperature above freezing. From Figure 3 of Turbet et al. (2016) we estimate their global average surface temperature to be around 235-245 K, which compares favorably with the 247K surface temperature obtained by our 1D VPL climate model. For the 1 bar $N_2$ case with $CO_2$ and $H_2O$, we calculate a global average surface temperature of 290 K compared to ~301 K from Fig. 7 of Turbet et al. Notably, the shape of the pressure-temperature profile is qualitatively similar, showing a double temperature inversion at the tropopause (Figure 3). Tropopause temperatures appear similar, although our temperature profile has a warmer mesosphere. For the pure $CO_2$ cases, our results also agree well with Turbet et al.'s 3:2 spin-orbit resonance cases. We find global average surface temperatures of 271, 306, and 320 K for 1, 4, and 6 bars of $CO_2$, respectively. From Figure 3 of Turbet et al., we estimate their global average surface temperatures are ~265 K (230 ~ 280 K from pole to equator), ~305K (270 ~ 320 K from pole to equator), and ~325 K (290 ~ 330 K from pole to equator) for 1, 4, and 6 bars respectively. Consequently our surface temperatures for pure $CO_2$ cases are all within ~6 K of Turbet et al.'s results, and we can reasonably reproduce the global average temperatures of the LMDz GCM for at least the non-synchronously rotating planets.

We have also previously compared our 1D *Atmos* climate model to the temperatures produced by the LMDz GCM and the CAM GCM for Archean Earth-like planets in Arney et al. (2016a). For a planet with 2% $CO_2$, 0.02% $CH_4$, and a surface albedo of 0.33, the LMDz model produces a global average temperature of 287 K (Charnay et al. 2013), which is 5 K warmer than our model's 282 K for this type of planet. For a planet with 6% $CO_2$, no $CH_4$, and an average albedo of 0.317, the CAM model produces a global average temperature of 287.9 K (Wolf and Toon 2013), and our model produces an average temperature of 285.3 K, a difference of 2.6 K. These comparable cases suggest that our 1D *Atmos* climate model can produce similar temperatures to GCM models, at least for non-synchronously rotating planets.

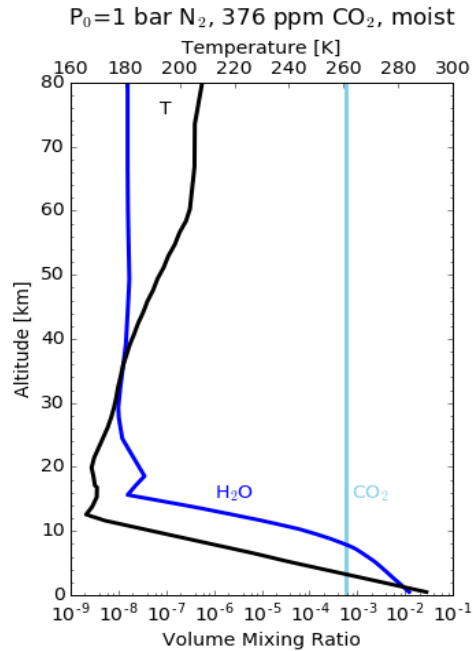

*Figure 3. Mixing ratios from Turbet et al. (2016; Fig. 7) for the $N_2$-dominated atmosphere with 376 ppm $CO_2$ and the water vapor profile (blue) as shown (retrieved from Turbet et al.), with the temperature profile (black) calculated by our SMART climate model. Our model obtains qualitatively similar results to Turbet et al., notably the double temperature inversion near the tropopause, and we obtain a surface temperature of 290 K versus their ~301 K.*

*4.2 Temperatures and Atmospheres.*

In this section, we show the simulated atmospheres and resulting temperature profiles for the end states of the evolutionary scenarios described in Barnes et al. (2016), including stages throughout the possibly evolutionary sequence from $O_2$-rich to $CO_2$-rich atmospheres, and possible habitable planetary environments.

*4.2.1 $O_2$-Rich Atmospheres.*

If Proxima Centauri b formed *in situ*, the extended pre-main sequence phase of its star may have driven substantial water loss and the buildup of potentially hundreds of bars of $O_2$ (Luger & Barnes, 2015; Barnes et al., 2016). The total $O_2$ generated depends on several parameters, including the initial volatile inventory of the planet and the rate of destruction or sequestration by geological processes, but plausible scenarios exist where substantial quantities of abiotic oxygen remain in the planet's atmosphere either alongside a remnant liquid water ocean or after complete desiccation of the planet (Barnes et al., 2016).

We used *Atmos* to self-consistently model the climate and trace gas abundance for two post-runaway, high-$O_2$ scenarios. In both cases the total pressure of the atmosphere is 10 bars, with 95% $O_2$ and 0.5% $CO_2$ by volume ($N_2$ is a filler gas, constituting the remaining atmospheric volume at each altitude level once all other major and trace gas species are accounted for). In the first high-$O_2$ case, we assume an ocean remains on the surface with a water vapor profile as described in Section 3.6 (i.e. it is "wet"). In the second case, we assume complete desiccation of the atmosphere and no surface ocean. The choice of 10 bars is motivated by a conservative compromise between the prediction of up to several hundreds of bars of $O_2$ (Barnes et al., 2016), uncertainties regarding the capacity of effects such as ozone shielding to arrest $O_2$ build up, and the extent to which a magma ocean may assimilate that $O_2$ (Schaefer et al., 2016). A 0.5% $CO_2$ mixing ratio is assumed for consistency in total $CO_2$ abundance (mixing ratio times surface pressure) with the modern and Early Earth scenarios described below, which have 5% $CO_2$ in a 1 bar atmosphere.

Figure 4 shows the temperature and gas mixing ratio profiles for these two cases. The primary differences between the "wet" and desiccated cases are the surface temperatures and $O_3$ profiles. The surface temperature for the "wet" case is 320 K versus 257 K for the desiccated scenario due to the significant greenhouse impact of water vapor, which is further enhanced relative to a hypothetical 1 bar case due to pressure broadening (e.g., Goldblatt et al., 2009). The presence of water vapor also significantly impacts the $O_3$ profile, because OH radicals from water photolysis are efficient at $O_3$ destruction. Thus, $O_3$ can build up to higher levels in the desiccated atmosphere, especially near the surface where the $H_2O$ mixing ratio would otherwise be high.

### 4.2.2 $CO_2$-Rich Atmospheres.

Although a remnant $O_2$ atmosphere from massive H escape is possible, if the planet is volcanically active then the atmosphere may evolve so that significant amounts of both $CO_2$ and $O_2$ exist simultaneously in the atmosphere (Section 4.2.2.1). The $O_2$ would originate from water loss during the pre-main sequence (Luger & Barnes 2015; Barnes et al., 2016) and the $CO_2$ from outgassing from the planetary interior.   In the absence of liquid water, the outgassed $CO_2$ would be unlikely to return to the mantle, as the carbonate-silicate cycle required to draw it down (Walker et al., 1981) will not operate without an active hydrological cycle. In our own Solar System, it has been suggested that Venus might have had an $O_2$- and $CO_2$-dominated atmosphere today if it had formed with a larger water inventory (Chassefière et al., 1996a).   However, if that water inventory is not initially present, or if sufficient time has elapsed that the planet has been able to sequester or lose its $O_2$, then the planet may be more Venus-like (Section 4.2.2.2) with a $CO_2$-dominated and largely desiccated atmosphere (< 40ppm of H).   Ultimately though, the atmosphere may become heavily H depleted, and in that case, a photochemical equilibrium will develop between $O_2$, CO, and $CO_2$ (Gao et al., 2015; Section 4.2.2.3).  We model these three cases below for Proxima Centauri b.

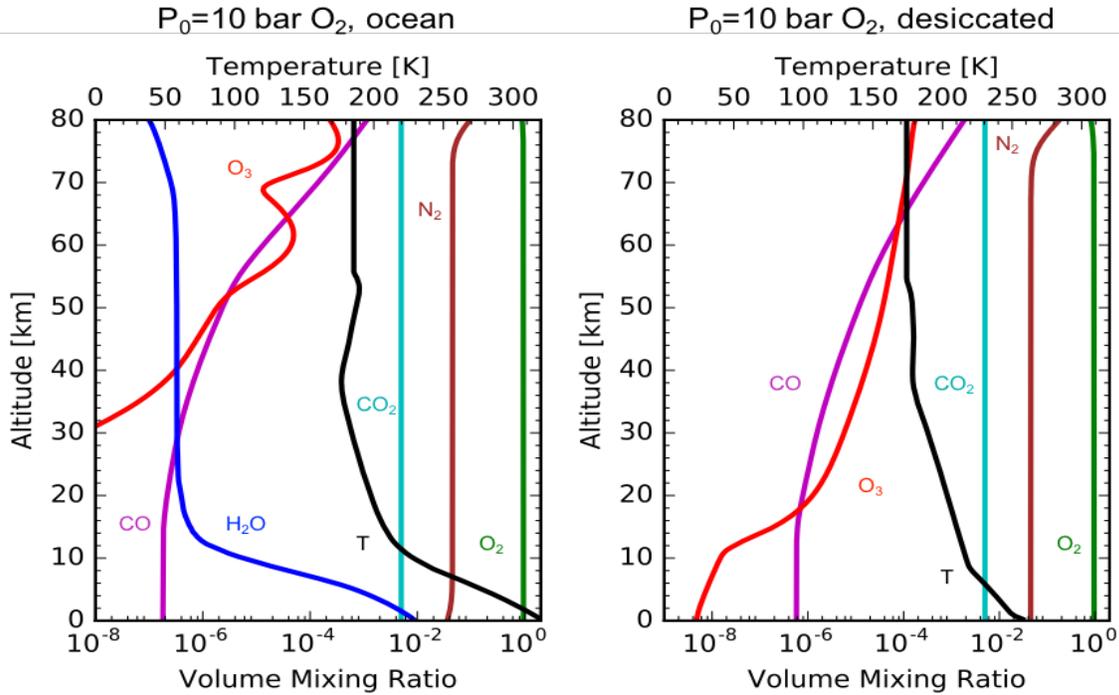

*Figure 4. Self-consistent high-O₂ (95%), post-runaway atmospheres with a surface ocean remaining (left) and completely desiccated (right). Differences in temperature and O₃ profiles are primarily driven by the presence or absence of water vapor.*

### 4.2.2.1 Evolved O₂/CO₂ Atmospheres.

We calculate the photochemistry and climate for a 10 and 90 bar mixed $CO_2$ and $O_2$ cases. These atmospheres consist of 45% $CO_2$, 45% $O_2$, with the rest of the atmospheric volume comprised of $N_2$ and photochemically-generated trace gases (primarily CO and $O_3$). We assume complete desication. We loosely coupled the photochemical component of *Atmos* to the SMART Climate model to calculate self-consistent gas mixing ratios and temperature profiles given in Figure 5. Here the hypothetical 10 bar $O_2$-$CO_2$ atmosphere for Proxima Centauri b would produce a surface temperature of 342 K, while the 90 bar case has a surface temperature of 393 K. Neither of these atmospheres would be habitable due to the lack of surface water.

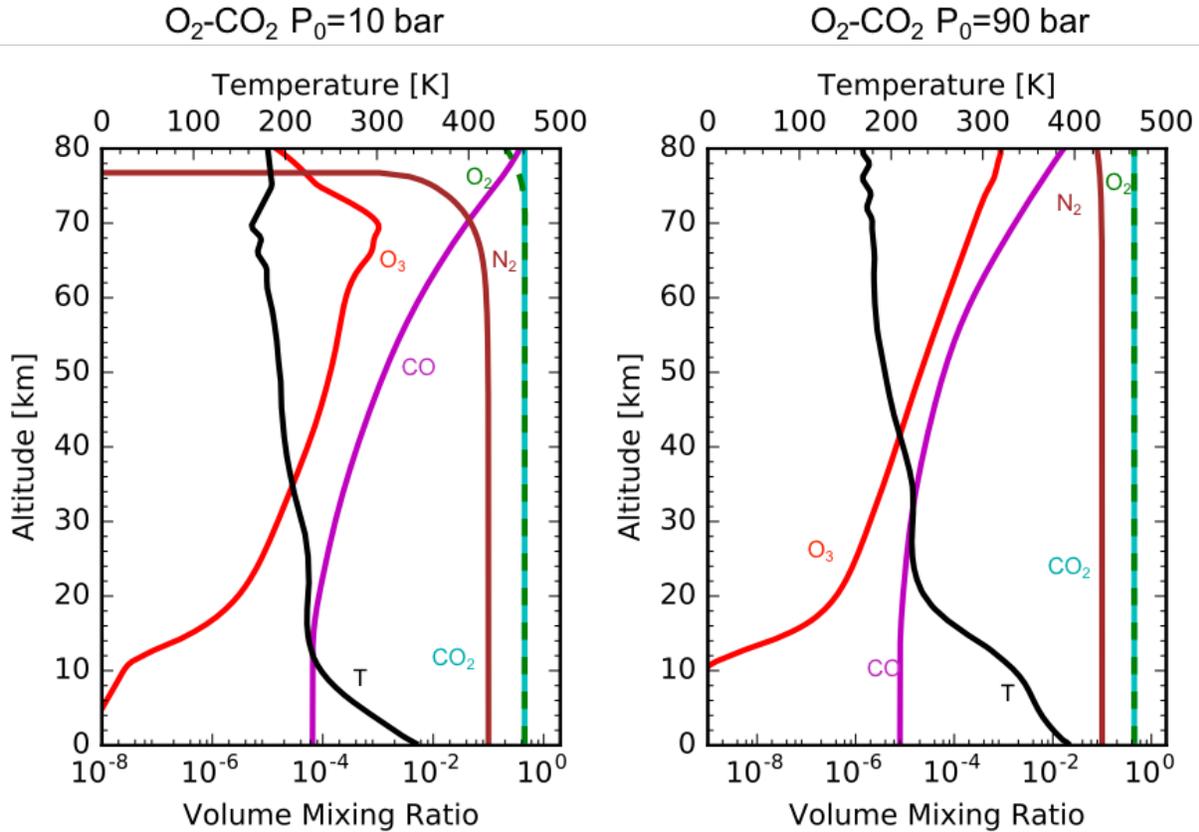

*Figure 5. Temperature and gas mixing ratio profiles for a 10 bar (left panel) and 90 bar (right panel) O₂-rich and CO₂-rich atmospheres with 45% CO₂, 45% O₂, and ~10% N₂.*

#### 4.2.2.2 Venus-like Atmospheres.

Venus is our nearest planetary neighbor and represents a class of worlds which may be common throughout the galaxy (Kane et al., 2014). Studies have suggested that Venus was more Earth-like in its past, possibly with liquid water as inferred from its high D/H ratio (McElroy et al., 1982; Donahue et al., 1992). However, the interpretation of Venus with early liquid water has been called into question by other studies (Grinspoon, 1993; Hamano et al., 2013). Still, in general, desiccation of exoplanets with liquid water by the runaway greenhouse process is likely to occur (Ingersoll, 1969; Kasting & Pollack, 1983; Kasting et al., 1984; Kasting, 1988; Goldblatt et al., 2012; Goldblatt et al., 2013). This may have occurred on Proxima Centauri b during its star's pre-main sequence phase (Luger & Barnes 2015) as discussed above. If Proxima Centauri b underwent a runaway greenhouse and lost its water due to water photolysis, subsequently losing both its hydrogen and oxygen, or if it formed without liquid water in the first place (Hamano et al., 2013), it may be Venus-like. Venus is unable to recycle its atmosphere back to its interior due to a lack of processes like plate tectonics, resulting in a 93 bar $CO_2$-dominated atmosphere.

Here, we simulate similar $CO_2$-dominated atmospheres using the VPL SMART Climate model, as the *Atmos* model is unable to simulate Venus-like atmospheres. SMART Climate was used to generate temperature pressure profiles for 10 and 90 bar $CO_2$ atmospheres containing 20 ppm $H_2O$, similar to the mixing ratio of $H_2O$ in the Venus lower atmosphere (e.g. Pollack et al., 1993; Chamberlain et al., 2013; Arney et al., 2014). These gases are assumed to be evenly mixed. Note that the simulations in this case are not photochemically-self consistent, nor do they include self-consistent cloud formation, but they are useful as a first-order approximation of a Venus-like atmosphere for Proxima Centauri b. Our simulations indicate that if Proxima Centauri b had a 10 bar $CO_2$ atmosphere with 20 ppm $H_2O$, its surface temperature would be 428 K; for a 90 bar $CO_2$ atmosphere also with 20 ppm $H_2O$, the surface temperature climbs to 568 K.

If Proxima Centauri b outgasses volcanic $SO_2$, it may be able to produce sulfuric acid ($H_2SO_4$) haze and cloud. In Venus' atmosphere, $H_2SO_4$ aerosols are generated photochemically through reactions such as (Yung & DeMore, 1984):

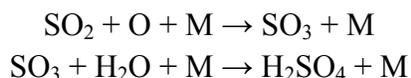

$$SO_2 + O + M \rightarrow SO_3 + M$$
$$SO_3 + H_2O + M \rightarrow H_2SO_4 + M$$

Note that the formation of $H_2SO_4$ requires the presence of $H_2O$ to react with $SO_3$. Sulfuric acid aerosols therefore cannot form if the atmosphere is completely water-free. Sulfuric acid is an efficient desiccant, and so water in the Venus cloud deck gets trapped in the $H_2SO_4$ droplets, producing a concentrated solution of $H_2SO_4$ and $H_2O$. To estimate where a cloud deck could form in our Venus-like atmospheres, we calculated a $H_2SO_4$ saturation vapor pressure (SVP) curve following the method described in Gao et al. (2014). If we assume an $H_2SO_4$ vapor mixing ratio of 4 ppm in the lower atmosphere—similar to measurements in Venus' atmosphere (Parkinson et al. 2014)—and a cloud droplet acid concentration of 85%, also similar to Venus' (e.g. Barstow et al. 2012)—we are able to correctly predict the altitude of the base of the Venus cloud deck at ~48 km. We assume these same parameters for the $H_2SO_4$ droplet water fraction and vapor pressure for Proxima Centauri b.

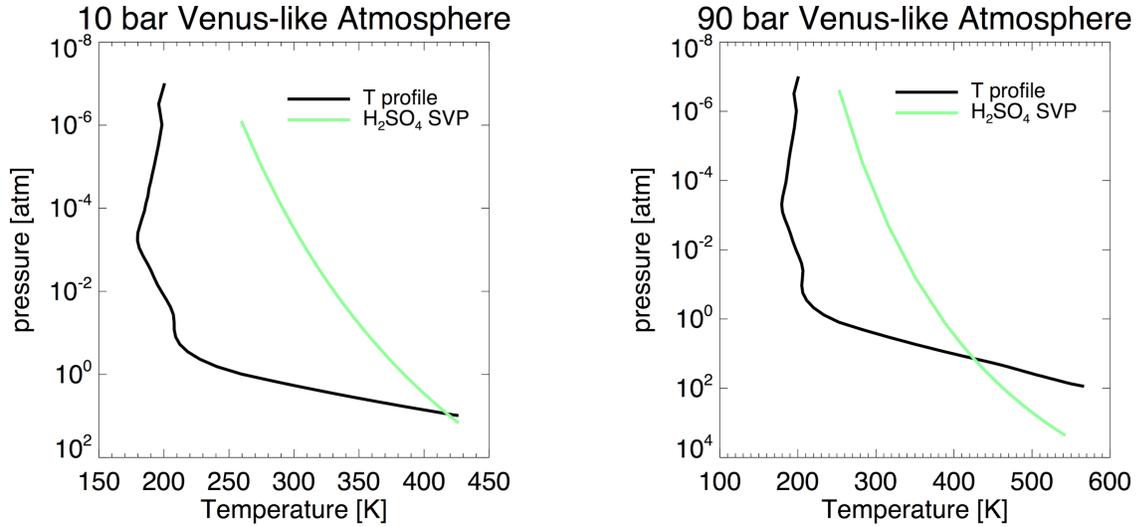

*Figure 6. The $H_2SO_4$ condensation curve and the temperature-pressure profile of a 10 bar $CO_2$ atmosphere (left) and 90 bar $CO_2$ atmosphere (right) for Proxima Centauri b. Clouds could condense where the green curve intersects the black temperature-pressure profile.*

The intersection of the SVP curves and the temperature-pressure profiles in Figure 6 show where $H_2SO_4$ condensation would occur; a similar technique has been used before to predict cloud condensation in the context of other exoplanets, such as GJ 1214b (Miller-Ricci et al., 2012). In the 10 bar $CO_2$ atmosphere, the $H_2SO_4$ condensation curve intersects the Proxima Centauri b temperature-pressure curve very close to the surface, at a pressure of 8.9 bars and an altitude of about 0.9 km. It may therefore be possible for $H_2SO_4$ condensation to occur close to the planet's surface. In a 90 bar $CO_2$ atmosphere, cloud condensation would occur at a pressure of about 14 bars, or 17 km in altitude.

To implement these sulfuric acid cloud decks in our model, we applied a scaling to the Venus cloud model of Crisp (1986) to move the base of the cloud deck to our predicted pressure, shifting the pressures of the entire cloud deck accordingly. However, this procedure does not include feedbacks on the atmospheric temperature structure from the formation of this $H_2SO_4$ deck. The clouds and haze of Venus reflect over 70% of the total incident solar radiation back to space, and the residual insolation passing through the cloud deck is strongly absorbed. The SMART model shows that only about 3% of the incident radiation at the top of the cloud deck ever reaches the surface of this atmosphere. The clouds themselves are effective absorbers of reradiated longwave radiation for wavelengths longer than 2.7 μm, contributing to the Venus greenhouse (Crisp 1986). These effects would combine to change the temperature structure of the atmosphere with a cloud deck in place, compared to the temperature profiles shown in Figure 6.

Future studies of the possibility of $H_2SO_4$ clouds on Proxima Centauri b should include self-consistent coupled climate-photochemical modeling of the processes that produce $H_2SO_4$ to determine the efficiency of $H_2SO_4$ production under the Proxima Centauri's UV spectral energy distribution. Our photochemical model cannot currently simulate Venus-like atmospheres, so we were unable to do this here. However, we ran a test of photochemical $H_2SO_4$ production in an Earthlike atmosphere with the photochemical model, and we find that 11 times as much $H_2SO_4$ is produced in the atmosphere of the planet orbiting Proxima Centauri compared to the planet orbiting the sun. It therefore does not seem that the spectral energy distribution of Proxima Centauri would be a barrier to the production of photochemical $H_2SO_4$.

*4.2.2.3 Desiccated $CO_2$ Atmosphere in Photochemical Equilibrium.*

If a $CO_2$-rich atmosphere becomes severely desiccated (< 1 ppm of H, significantly lower than on Venus today) over time, then it may generate significant amounts of $O_2$, to become $CO_2$- and $O_2$-rich again (Gao et al., 2015).  In the terrestrial planets of the solar system, including Venus and Mars, $O_2$ from $CO_2$ photolysis is efficiently returned to $O_2$ via catalytic reactions with $HO_x$ species, which originate primarily from the photolysis of $H_2O$. Gao et al. (2015) showed that oxygen can build up from the photolysis of $CO_2$ into CO and $O_2$ in desiccated M dwarf atmospheres on short timescales (~1 Myr) if the $H_2O$ abundance were sufficiently low, thus limiting the reactions that reform $CO_2$ from CO and $O_2$. Since Proxima Centauri b likely suffered from extreme water loss if it formed *in situ* (Barnes et al., 2016; Ribas et al., 2016), this scenario is a plausible end state for its atmosphere. This scenario is distinct from the case where abiotic $O_2$ is left remaining in the atmosphere from H escape, since the resulting equilibrium combinations of these gases ($CO_2$, CO, $O_2$, $O_3$) maintain abundances of C and O that are stoichiometrically consistent with pure $CO_2$.

However, our photochemical model cannot reproduce these types of atmospheres due to our model's treatment of $CO_2$ as an "inert" species, meaning its mixing ratio is fixed as an isoprofile. Therefore, to incorporate this scenario into the range of cases we explore, we use intermediate case 4 (0.0320 ppm hydrogen) mixing ratios from Gao et al. (2015), which have been converted from molar mass to volume mixing ratios and shown in Figure 7. Gao et al. (2015) assumed a surface temperature of 240 K, 1 bar of $CO_2$, an instellation of 1 $S_\odot$, and used the UV spectrum of GJ 436 (spectral type M2.5 V; Butler et al. 2004) from France et al. (2013). Since this is predominantly a $CO_2$ atmosphere, we use our SMART Climate model to simulate the climate of this planet. We use a surface albedo of goethite. We find the equilibrium surface temperature for this scenario is 298 K, which would be "habitable" if this planet did not already lack water (note this temperature is hotter than the 1 bar $CO_2$ atmospheric temperature we compute in Section 4.1 because the atmosphere here includes other gases and uses a darker surface albedo). It is important to note that for this particular case, the photochemistry is not self-consistent with the SED of Proxima Centauri, since we retrieved the atmospheric composition from literature. However, we self-consistently calculate the expected temperature profile based on the real stellar spectrum and instellation using the SMART Climate model.

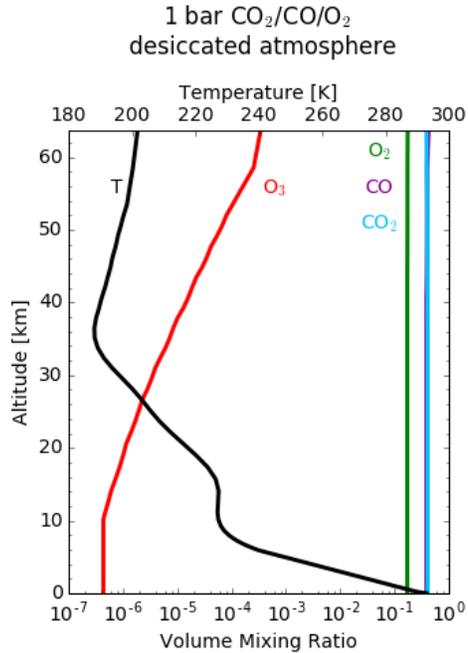

*Figure 7. Temperature and mass mixing ratios of desiccated $CO_2/O_2/CO$ atmosphere as a function of altitude. $CO_2$, $O_2$, and CO are the most abundant gases and result from photochemistry of outgassed $CO_2$. Only major and spectrally observable gases are shown. The temperature profile is the result of our SMART Climate model using these mixing ratios converted to volume mixing ratios from the molar mass mixing ratio values of Gao et al. (2015; case 4).*

### 4.2.3 Habitable Terrestrial Atmospheres.

If Proxima Centauri b is a "habitable evaporated core" that migrated from farther away from the star to its current position, or if it formed *in situ* with a 1% $H_2$ envelope, it could have escaped desiccation during its star's pre-main sequence super luminous phase, and it may be a planet hospitable to life. However, depending on where it formed in the planetary nebula, and its subsequent evolution, it could be either oxygen-rich or more reducing - as our own Earth's atmosphere has been over time. Here we model two terrestrial planet examples for a modern, oxidizing Earth-like atmosphere, and a more reducing Archean early Earth-like atmosphere.

*4.2.3.1 M Dwarf Modern Earth-like Atmosphere.*

If it survived the pre-main sequence phase and was endowed with and retained sufficient volatiles, Proxima Centauri b's atmosphere could be Earth-like (Barnes et al., 2016; Ribas et al., 2016), possessing a high-molecular weight, secondary atmosphere consisting of $N_2$, $CO_2$ and $H_2O$, and, if oxygenic photosynthesis evolved, $O_2$. However, even if Proxima Centauri b were Earth-like in all other respects, the different spectral energy distribution of its host star would significantly impact its photochemistry and therefore the mixing ratios of trace gas species such as $O_3$, $CH_4$, CO, and $N_2O$, given the same fluxes into the atmosphere and abundances of major species (e.g., Segura et al., 2003, 2005, 2007, 2010). Additionally, Proxima Centauri b's lower total instellation (0.65 $S_\odot$) would require higher larger greenhouse gas abundances to achieve the same globally averaged surface temperature as the Earth, though this is partially ameliorated by the shift of the stellar SED to red wavelengths and corresponding decrease in overall planetary albedo (Kopparapu et al., 2013).

We construct a hypothetical, "modern Earth" atmosphere for Proxima Centauri b that is consistent with the spectrum (including the UV portion) from the M5.5V host star, following the methodology of Segura et al. (2005). We determined the fluxes of trace gas species, like $CH_4$, required to produce their "modern" pre-industrial mixing ratios in our photochemical model (for an Earth-Sun case), and then applied those fluxes to the Proxima Centauri case to calculate the resultant mixing ratios in the Proxima Centauri b atmosphere, given the substantially different UV spectrum of the star. Unlike Segura et al. (2005), we used more conservative, preindustrial concentrations of these gases to determine the biogenic fluxes required for Earth's mixing ratios, rather than contemporary, anthropogenically influenced values. $CO_2$ and $H_2$ mixing ratios were fixed at 280 ppm and 53 ppb, and the pre-industrial mixing ratios from which we derived the surface fluxes were 500 ppb ($CH_4$), 50 ppb (CO), and 260 ppb ($N_2O$), all based on ice core data (Solomon et al., 2007; Haan & Raynaud, 1998; Flückiger et al., 2002). The modern day mixing ratios—which include anthropogenic sources—are considerably higher, being: ~1700 ppb, 100 ppb, and ~320 ppb, respectively. The fluxes required to reproduce these (lower) mixing ratios in our model for modern Earth orbiting the Sun are $5.6 \times 10^{10}$ molecules/cm$^2$/s ($CH_4$), $1.91 \times 10^{11}$ molecules/cm$^2$/s (CO), and $8.8 \times 10^{8}$ molecule/cm$^2$/s ($N_2O$), respectively. In both the modern, pre-industrial Earth and Proxima Centauri b cases, we assume a volcanic flux of $H_2S$ of $1.0 \times 10^{8}$ molecule/cm$^2$/s and a flux of $SO_2$ of $1.0 \times 10^{9}$ molecules/cm$^2$/s.

We applied these fluxes self-consistently to Proxima Centauri b with the coupled *Atmos* model described in Section 3.6, assuming that the biological and abiotic flux of these gases scales with surface area. We also prescribed a surface pressure of 1 bar, an $O_2$ mixing ratio of 21%, a $CO_2$ mixing ratio of 5% (to warm the planet and partially compensate for lower instellation), and an $N_2$ mixing ratio of 73%. This prescription produced surface mixing ratios of 830 ppm ($CH_4$), 58 ppm (CO), and 1.2 ppm ($N_2O$). Additionally, we used *Atmos* to calculate self-consistent profiles for $H_2O$, $O_3$, and temperature. A 5% $CO_2$ mixing ratio with the prior stated abundances of greenhouse gases $CH_4$ and $N_2O$ produces a converged surface temperature of 283 K. Figure 8 compares the gas and temperature mixing ratio profiles for preindustrial concentrations of biogenic trace gas species on Modern Earth (Figure 8, left panel) with those expected for Proxima Centauri b given similar bulk atmospheric constituents ($N_2$, $O_2$) and fluxes of trace gas species into the atmosphere, but with 5% $CO_2$ (Figure 8, right panel).

A notable difference compared to Earth is the elevated methane abundance on Proxima Centauri b, which is increased by a factor of ~1700 (830 ppm vs. 0.5 ppm). The cause of this effect was described in Segura et al. (2005) and in Section 2, and is related to the paucity of photons emitted by Proxima Centauri between 200 nm and 300 nm that drive the photochemical sequence of reactions that most efficiently destroy methane. This reaction sequence is (Segura et al., 2005):

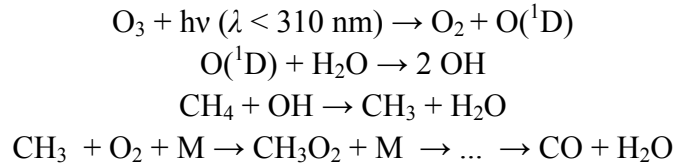

$$O_3 + h\nu \ (\lambda < 310 \ nm) \rightarrow O_2 + O(^1D)$$
$$O(^1D) + H_2O \rightarrow 2 \ OH$$
$$CH_4 + OH \rightarrow CH_3 + H_2O$$
$$CH_3 \ + O_2 + M \rightarrow CH_3O_2 + M \ \rightarrow ... \ \rightarrow CO + H_2O$$

Because this reaction primarily takes place in the lower stratosphere and troposphere, overlying $O_2$ shields photons with wavelengths less than 200 nm. $N_2O$ is enhanced, but only by a factor of ~5 (1.2 ppm vs. 0.26 ppm). This is also due to the relative lack of > 200 nm photons, where $N_2O$ has a significant photodissociative cross section. The water vapor abundance in the stratosphere of Proxima Centauri b would be elevated in this case, 40 ppm at 30 km vs. ~10 ppm for the Earth-Sun (see Figure 8 for profiles). This is due to the enhanced methane abundance in its upper atmosphere. The oxidation of methane by molecular oxygen produces additional water vapor through the reaction: $CH_4 + 2O_2 \rightarrow CO_2 + 2H_2O$ (Segura et al., 2005). The difference in surface water mixing ratio is due wholly to the 5 K difference in surface temperature. Additionally, the temperature profile in the upper atmosphere of Proxima Centauri b is lower primarily due to the lower near-UV fluxes and consequential lack of heating, despite a comparable ozone column. Overall, these results are consistent with those of Segura et al. (2005) and minor differences are due to the different UV spectrum of Proxima Centauri b, our choice of $CO_2$ mixing ratio, and the use of preindustrial fluxes for biogenic trace gases. It should be noted that our $O_3$ abundances here are calculated without considering ozone depletion from flares (e.g., Segura et al. 2010), which will be a key consideration for future work.

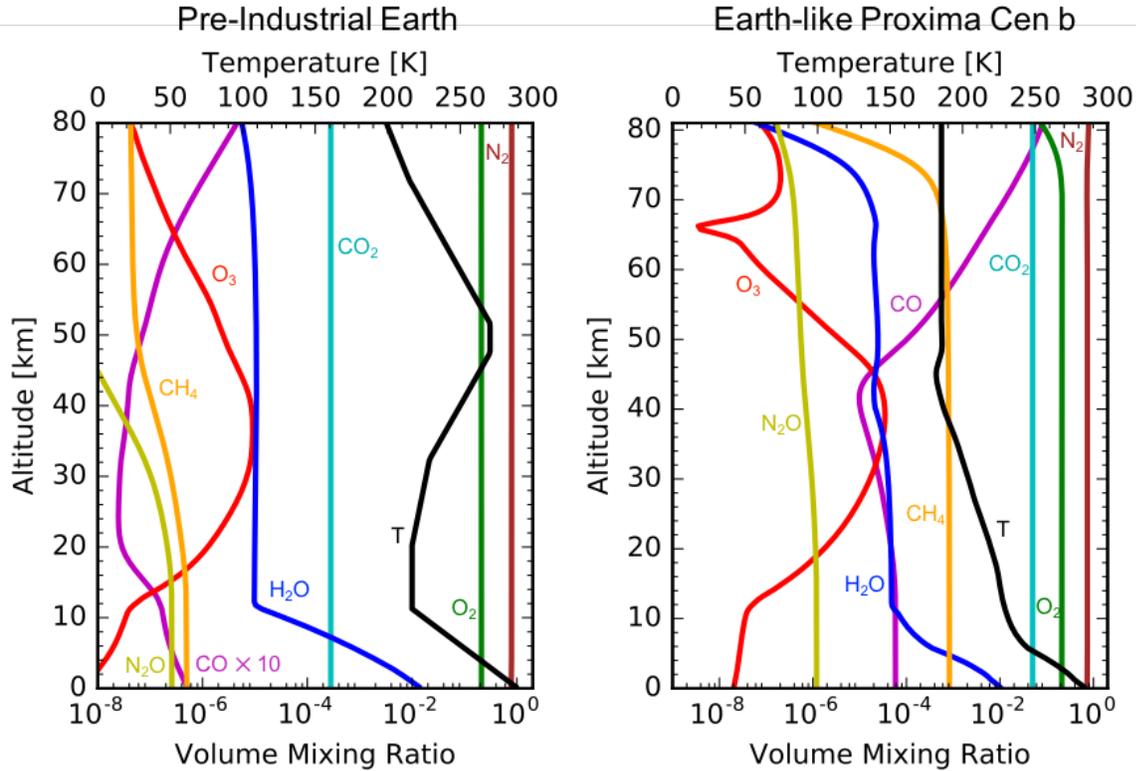

**Figure 8.** *Self-consistent temperature and gas mixing ratio profiles for preindustrial modern Earth orbiting the Sun (left) and a photochemically self-consistent Earth-like atmosphere for Proxima Centauri b (right). The mixing ratios of trace gas species CH₄, N₂O, and CO for Proxima Centauri b are determined from the flux required to produce their preindustrial concentration in Earth's atmosphere (see Section 4.2.3.1). Given the same fluxes as Earth, CH₄, CO, and N₂O would exist in Proxima Centauri b's atmosphere in much greater abundance.*

*4.2.3.2 M Dwarf Archean Earth-like Atmospheres.*

Early Earth represents another type of habitable, Earth-like planet whose spectral appearance and atmospheric composition are different from the modern planet. In particular, the Archean (roughly 4–2.5 billion years ago) atmosphere is generally believed to have been anoxic (e.g. Farquhar et al., 2000) and may have contained more $CO_2$ than the modern atmosphere (e.g. Driese et al., 2011), likely on the order of at least about 1% of the total atmosphere, and possibly more (Kanzaki & Murakami, 2015). Early Earth's atmosphere may also have contained a substantial amount of methane before the rise of atmospheric $O_2$. There is evidence that methane-producing metabolisms evolved as early as 3.5 billion years ago (Ueno et al., 2006), and methane can be produced by a variety of non-biological processes as well (Etiope & Sherwood-Lollar, 2013). Because methane can be produced by a number of biological and geological processes, methane-rich, anoxic Earth-like worlds may occur frequently elsewhere in the galaxy, especially given the propensity for methane to accumulate in Earth-like atmospheres around M dwarfs (Section 4.2.3.1). Earth-like methane-rich atmospheres may also form photochemically-induced organic hazes if the $CH_4/CO_2$ ratio exceeds 0.1 (e.g. Trainer et al., 2006; Domagal-Goldman et al., 2008; Haqq-Misra et al., 2008; Zerkle et al., 2012), cooling the planetary surface environment through the antigreenhouse effect and dramatically altering the spectral appearance of the planet.

We generated 1 bar Archean-like atmospheres with 5% $CO_2$ in the *Atmos* climate-photochemical model. This $CO_2$ amount is reasonable for an Earth-like planet based on constraints on the $CO_2$ abundance of the Archean atmosphere (Driese et al., 2011; Kanzaki & Murakami, 2015), and it is the same $CO_2$ amount used to model our modern Earth-analog planet. One of these atmospheres contained 1% $CH_4$ and the other contained 3% $CH_4$. Figure 9 shows the gas and temperature profiles of these two atmospheres. Note the large abundance of CO in these atmospheres from $CO_2$ photolysis that increases with increasing altitude. Figure 1 shows the UV spectrum of Proxima Centauri compared to the Sun, and it produces excess radiation compared to the sun for $\lambda$ < 170 nm, which overlaps with the peak of the $CO_2$ UV cross section. For an atmosphere with 5% $CO_2$ and 1% $CH_4$, the global average surface temperature of the planet is 289 K, which is similar to Earth's current average temperature. For an atmosphere with 5% $CO_2$ and 3% $CH_4$, a Titan-like organic haze forms, and the surface temperature decreases modestly to 285 K. Titan-like hazes can cool planetary surface temperatures by 20-25 K on Archean-like planets orbiting solar-type stars (Arney et al., 2016a), but these hazes produce less cooling on planets orbiting M dwarfs because M dwarf luminosity is mostly produced at wavelengths where these hazes are more transparent (Arney et al., 2016b).

Haze formation depends on the atmospheric $CH_4/CO_2$ ratio, and more reducing conditions enhance haze production. On an Archean-like planet orbiting a solar-type star, the $CH_4/CO_2$ ratio needed to initiate haze formation is ~0.2 (e.g. Trainer et al., 2006), but around Proxima Centauri, the $CH_4/CO_2$ ratio needed to initiate organic haze formation in an Archean-like atmosphere with 5% $CO_2$ is ~0.5, which is quite high, but easier to achieve given that the host star spectrum results in a longer atmospheric lifetime for $CH_4$ in a planetary atmosphere. Our previous work (Arney et al., 2016b) shows that haze-formation is more difficult on planets that generate larger quantities of photochemical oxygen radicals because these oxygen species consume haze-forming hydrocarbon gases. Therefore, planets with larger amounts of oxygen species in their atmospheres require higher $CH_4/CO_2$ ratios to form organic hazes. We find that the hazy Archean-like planet around Proxima Centauri produces 15, 1.9, and 1.5 times as much O, $O_2$, and OH, respectively, compared to a similar Archean-like planet around the sun. This is due to photolysis of $CO_2$ and water vapor. Thus, the higher $CH_4/CO_2$ ratio needed to generate haze around Proxima Centauri is consistent with our previous results.

.

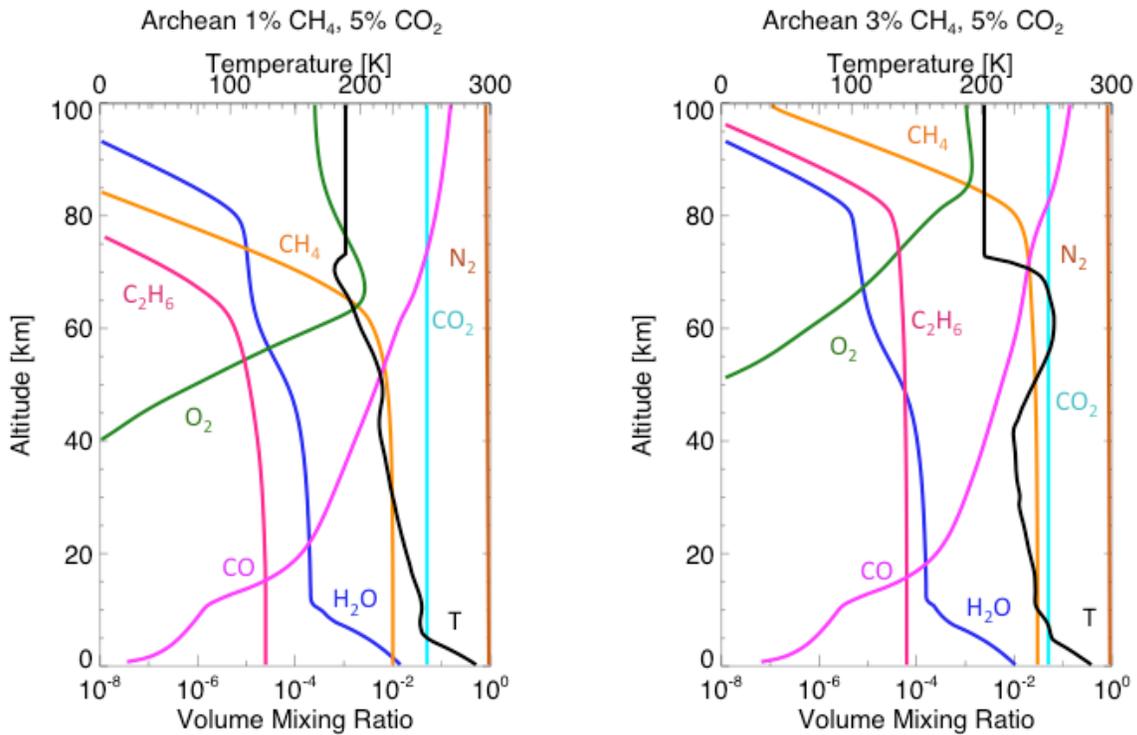

*Figure 9. Gas profiles and temperature structure of the Archean-analog atmospheres. The moderate temperature inversion seen in the right (3% CH₄) plot is due to UV absorption by the haze. UV-shielding by the haze in the 3% CH₄ plot (right) prevents photolysis of gases such as methane and ethane at higher altitudes than in the 1% CH₄ (left) plot. Note the temperature profile becomes an isoprofile above the top of the climate model grid when passed into the photochemical model.*

Figure 10 shows the haze particle number density and particle radii for the Archean-analog haze particles. Haze formation initiates high in the atmosphere, with the peak of the haze particle number density occurring at around 90 km in altitude. The photolysis of methane that initiates haze formation is apparent in the methane profiles in Figure 9 that decrease at higher altitudes. Haze shields methane (and other gases) from photolysis once it is produced, and so the hazy 3% CH₄ plot shows photolysis of gases occurring at higher altitudes than the haze-free 1% CH₄ plot. At these high altitudes, however, the particles are very small (~0.001 μm). Fractal particles begin to form at about 60 km in altitude when the particle radius reaches 0.05 μm. The coagulation of these small, simple spherical particles into fractals causes the decline in particle number density and the increase in particle radius at this altitude. The particles grow to a maximum radius of 0.57 μm through coagulation processes before they fall from the atmosphere.

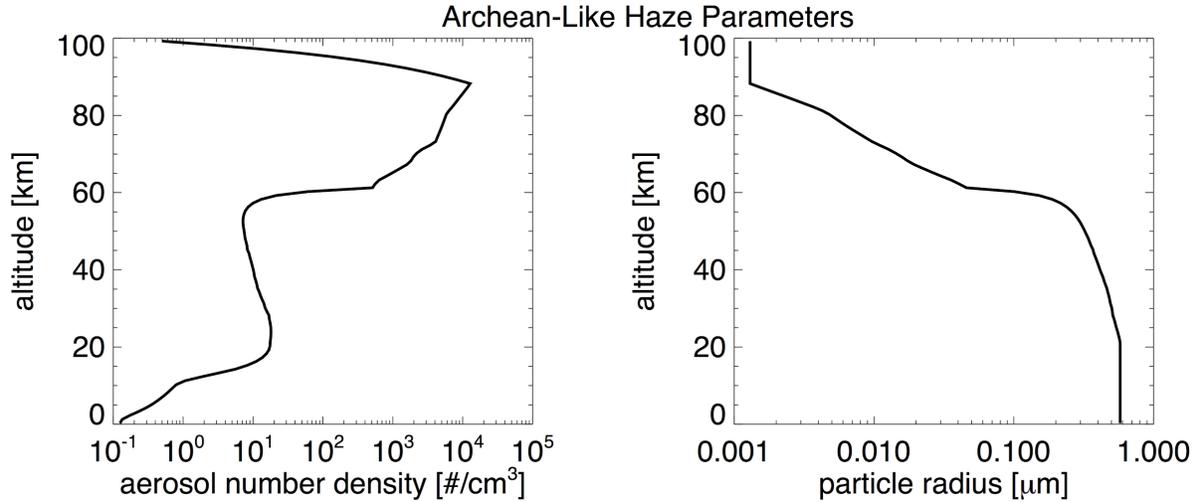

*Figure 10. Archean haze particle number density and particle radius. Note the sharp decrease in particle number density (and corresponding increase in particle size) at about 60 km in altitude where fractal particle formation begins to occur.*

*4.3 Simulated Planetary Phase Curves*

If Proxima Centauri b is not transiting, then observations of the planet's reflected and thermal phase variations may be our only option for detecting and characterizing its atmosphere before a next generation ground or space-based direct imaging coronagraph is built. Thermal phase curves have been used to study the atmospheric dynamics, thermal structure, and molecular composition of hot, thermally bright exoplanets (Cowan et al., 2007; Knutson et al, 2007; Crossfield et al., 2010; Cowan et al., 2012; Knutson et al., 2012; Zellem et al., 2014). Below, we estimate idealized thermal phase curve variations and planet-star contrast ratios through the mid-IR.

Simulated phase curves comparing the thermal emission from the planet to the stellar background flux are shown for each planet case in Figures 11–14. Each two-panel figure shows the planet-star contrast on the y-axis. The left panel shows the planetary phase evolution for different assumed day-night temperature contrasts in different integrated wavelength bins. We show two end-member cases, one with no day-night temperature contrast (solid color lines) and one with no flux from the nightside (dashed color lines). We also show an intermediate case where the nightside temperature is 20 K less than the dayside. Note that this intermediate case is not self-consistent. The true planet phase curve should fall somewhere between the two end-member cases. The right panel shows the planet-star contrast at full phase through the mid-IR wavelengths.

Figure 11 shows the self-consistent Earth-like planet, with the presence of a deep $CO_2$ feature at 15 microns. Figure 12 shows the hazy Archean-like planet. Figure 13 shows the desiccated $O_2$-rich planet. Figure 14 shows the 90 bar clear-sky Venus-like planet. The most evident feature across all the phase curve figures is the general rise in contrast with longer wavelengths, approaching $10^{-4}$ near 20 µm. Our planet-star contrast ratios are in agreement with those computed in the recently submitted papers, Turbet et al. (2016) and Kreidberg & Loeb (2016). Deviations from this steady rise in contrast are due to molecular absorption and emission in the planetary atmosphere, and afford the potential for detectability (Selsis et al., 2011). However, the actual detectability of these features ultimately depends on the amplitude of the phase curve, as the planet flux is not independently discriminated from the star, and the magnitude of the stellar variability at these wavelengths. In the mid-IR, the amplitude of the phase curves depends on the day-night temperature contrast, which is not self-consistently modeled in this work. These figures serve as a first order guide to what we might expect from multi-band phase-dependent observations in the mid-IR.

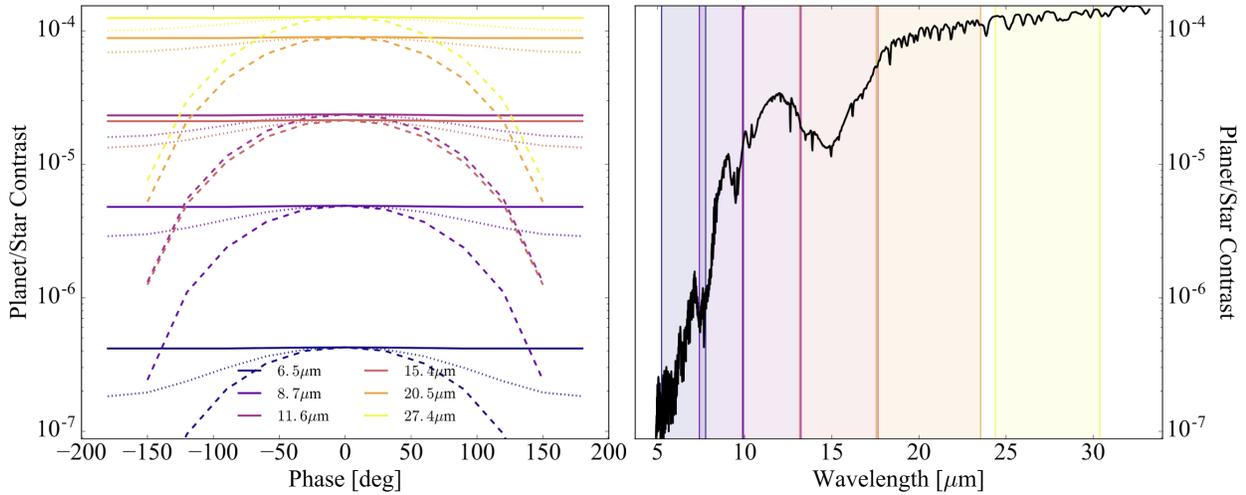

*Figure 11. Self-consistent Earth-like planet thermal phase contrast curves (left) and the corresponding planet-star contrast spectra at full phase (right). Colored lines and shaded regions indicate broadband wavelength binning consistent with R = 3. On the left, solid lines show the case with no day-night temperature contrast (no phase-dependent modulation), dashed lines show the case with no emergent flux from the nightside (maximal phase-dependent modulation), and dotted lines show an intermediate case with the nightside temperature 20 K less than the dayside temperature.*

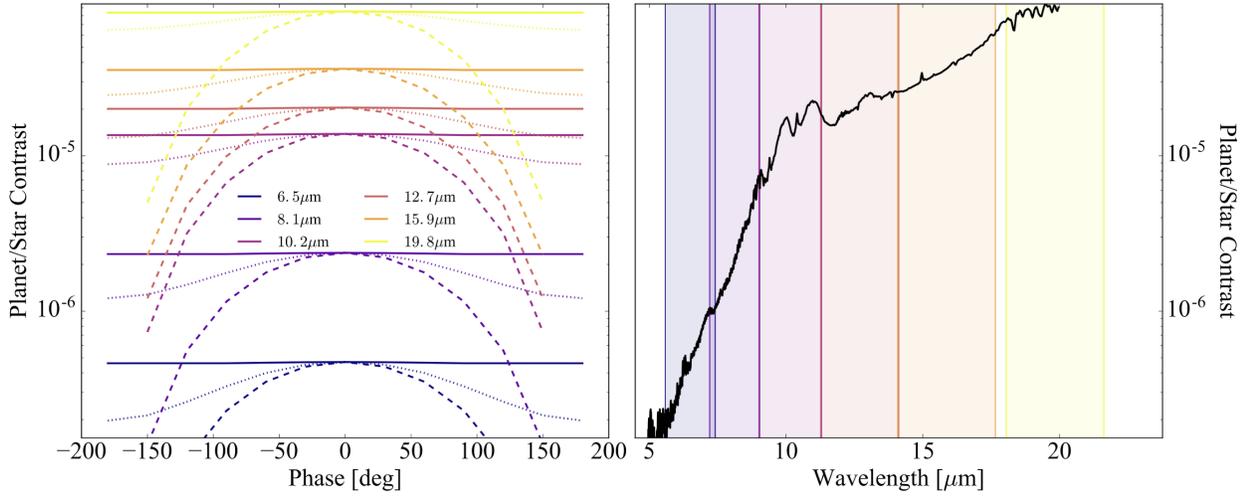

*Figure 12. Same as Fig. 11, but for the hazy Archean-like planet.*

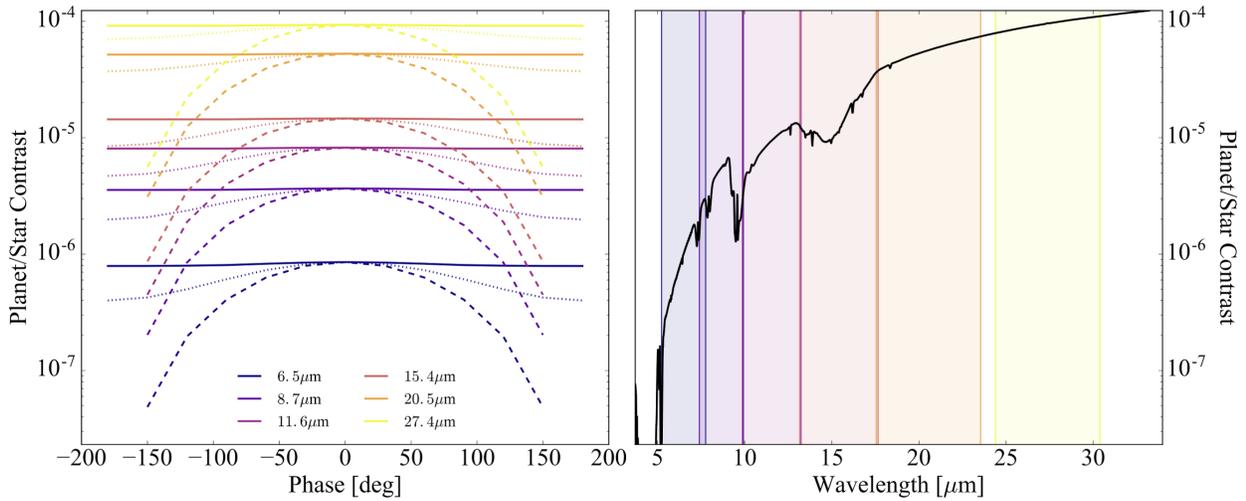

*Figure 13. Same as Fig. 11, but for the desiccated $O_2$-rich planet.*

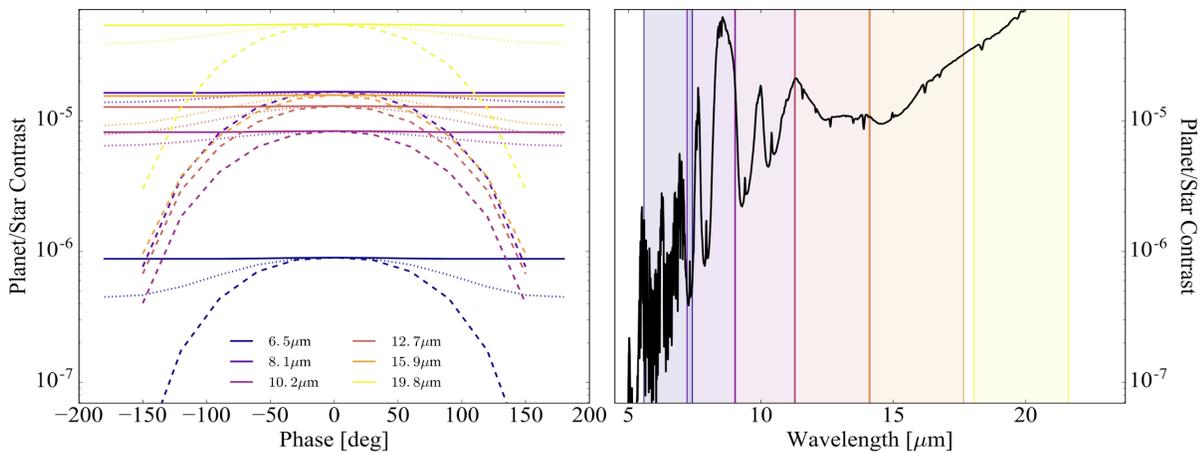

*Figure 14. Same as Fig. 11, but for the 90 bar clear-sky Venus-like planet.*

*4.3.1 Thermal Phase Curve Detection with JWST.*

We compute the detectability of thermal phase curves with the JWST Mid-Infrared Instrument (MIRI) that operates between 5-28 microns. In Section 4.3, we modeled thermal phase curves for four archetypal planet scenarios assuming three possible day-night temperature contrasts. Following the generalized JWST noise calculations from Greene et al. (2016) and the photon count rate formalism of Robinson et al. (2016), we estimate the integration times required to detect the quoted contrast ratios in each broad spectral band at a signal-to-noise ratio of 10 at each phase, assuming photon limited observations. Such a measurement is likely to be sufficient to resolve the shape of the phase curve, but ultimately the ability to resolve the shape will depend on the amplitude of the phase variations. In an idealized case, resolving the planet at both full and new phase can constrain the amplitude and a day-night brightness temperature contrast.

Using a wavelength-independent combined instrument plus JWST throughput of 0.27 (e.g. Greene et al., 2016), we find that an integration time of 21 hours at full phase can detect the planetary emission above the stellar flux with a signal-to-noise of 10 at 8.6 microns after binning to an R=3. These estimates are in agreement with the integration times adopted by Kreidberg & Loeb (2016). For our case with a nightside 20K cooler than the dayside, we find that an integration time of 62 hours is required to resolve the planetary signal at new phase.

Due to the rise of the planet-star contrast ratio towards longer wavelengths in the mid-IR and the presence of the strong 15 micron $CO_2$ band in many of our simulated planetary states, it would be advantageous to use the Medium Resolution Spectrograph (MRS) on JWST/MIRI, which is the only option for spectroscopy beyond 12 microns. It has been suggested that slit-losses with the integral field spectrometer will hinder exoplanet atmospheric characterization with MRS (e.g. Beichman et al., 2014; Kreidberg & Loeb, 2016). However, this may be less of a concern for thermal phase curves than transmission spectroscopy. The transmission spectroscopy signal scales with the stellar brightness, which decreases towards longer wavelengths and therefore requires bright targets. On the other hand, the thermal phase curve signal increases towards longer wavelengths as the planet-star contrast ratio increases. Although both the planet and star are faint in the Rayleigh-Jeans limit, less photons are needed to detect the planetary signal above the star.

Stellar variability and flaring in the mid-IR may severely obfuscate thermal phase curve studies, although little work has been done on M-dwarf variability at these wavelengths. In general, stellar variability is seen in WISE data out to 24 μm, but for M dwarfs it is fairly scant. Rotational variability is also likely to be seen at these wavelengths, but at the known rotation period of the star. For a starspot with an effective temperature about 500 K cooler than the photosphere and 10-20% the stellar radius, filling factors are large enough to get 3-8% optical flux modulations seen for Proxima Centauri (Davenport et al., 2016). At longer wavelengths, mid-IR modulation should be low, but non zero. For stars later than mid-M type, cloud-like modulation may become appreciable too (J. Davenport 2016, private communication). An in-depth study of the mid-IR variability of Proxima Centauri is needed to understand the impact of stellar variability on measuring planetary phase curves. It is interesting to note that pioneering work in terrestrial planet phase curves in the mid-IR and M-dwarf variability in the mid-IR will likely both occur simultaneously by observing Proxima Cen with JWST. In this case, disentangling stellar from planetary modulations will be highly important.

Further work is needed that combines photochemically and climatically self-consistent potential atmospheres for Proxima Centauri b with a full GCM treatment. Such modeling is necessary for making accurate phase curve predictions that are self-consistent between the altitude-dependent vertical structure and the longitude- and latitude-dependent horizontal structure.

Additionally, this study is limited to nearly edge-on inclinations. As the inclination shifts from edge-on to face-on the phase curve amplitude decays. More work is needed to explore the detectability of thermal phase curves over a complete range of possible inclinations.

*4.3.2 Detecting Ocean Glint in Phase Curves.*

One potential immediate application for JWST, in addition to the thermal phase curves discussed in Sections 4.3 and 4.5, is the detection of ocean glint, which would provide strong evidence for the presence of surface liquid water. However, as was shown in Robinson et al. (2010), strong glint signatures only occur for a subset of all possible orbital inclinations, as orbits near face-on do not allow for crescent-phase viewing geometries. Thus, only inclinations within ±30 degrees of edge-on have the potential to show strong signatures of glint. For this range of inclinations, the glint signature would be strongest at reflected-light wavelengths outside of Rayleigh scattering and gas absorption features. While the atmospheric composition of Proxima Cen b is currently unknown, key wavelength ranges of interest for glint detection would be between the 0.94, 1.1, 1.4, and 1.9 μm water vapor bands. At the phase angles of peak glint contributions for Earth (i.e., 120–165 degrees; Robinson et al., 2010), the planetary phase function will likely decrease the planet-to-star flux ratio by an order of magnitude from its full-phase value (Robinson et al., 2010) to near $10^{-8}$ (Turbet et al., 2016). Since glint could cause up to a 100% increase (i.e., doubling) in planet brightness at these phase angles and wavelengths, the precision of any phase curve measurements aimed at detecting a glint signature at reasonable signal-to-noise ratios would need to be several times smaller than $10^{-8}$. This estimate puts glint detection well outside the range of feasibility for JWST, whose systematic noise floor is usually predicted to be in the range of $10^{-5}$ (Greene et al., 2016). Direct imaging offers better prospects, as noise floors for these types of missions or instruments are typically an order of magnitude below the design contrast for the coronagraph or starshade, which would imply floors near $10^{-10}$ to $10^{-11}$ for next generation instruments. Here, the challenge for glint detection becomes angular separation—the planet-star separation at a phase angle near 150 degrees will be roughly 20 mas. Imaging the planet at this separation between the 0.94 and 1.1 μm water vapor bands would require an IWA of at least 1.5$\lambda/D$ for a 16 meter class telescope, or 3$\lambda/D$ for a 30 meter class telescope.

*4.4 Simulated Planetary Spectra.*

To predict the spectral observables of the planets simulated in our study, in this section, we present simulated direct imaging (Section 4.4.1) and transit transmission spectra (Section 4.4.2). Direct imaging observations will be possible with future starlight suppression technologies such as coronagraphs and starshades. Proxima Centauri b is not known to transit, but the transit spectra of our simulated worlds show several key diagnostic features that would help discriminate between habitable and uninhabitable scenarios, and are applicable not only to Proxima Cen b, but also to other terrestrial planets orbiting M dwarfs.

*4.4.1 Direct Imaging Spectra.*

To anticipate the spectral observables of Proxima Centauri b in reflected light observations that may become possible with future telescopes, we show simulated reflected light spectra generated with SMART in this section. In Section 4.6, we present these same spectra run through instrument noise models for several telescope configurations to predict which features will be observable with near-future technology.

*4.4.1.1 Direct Imaging Spectra for High-$O_2$ Atmospheres.*

Figure 15 shows clear sky reflectance spectra for 10 bar, $O_2$-dominated "wet" and desiccated cases. For the case with a surface ocean remaining, we assumed the second composite albedo described in Section 3.7.5 (ocean and continents with no vegetation), while we assumed a desert surface for the desiccated case. The spectra are dominated by $O_4$ bands in the UV-visible part of the spectrum at 0.345, 0.36, 0.38, 0.445, 0.475, 0.53, 0.57, and 0.63 μm, and $O_4$ absorption in the NIR at 1.06 and 1.27 μm (the latter overlaps with the $^1\Delta_g$ 1.27 μm $O_2$ band). The case with an ocean includes $H_2O$ bands that have been significantly pressure-broadened by the 10 bar atmosphere. Both scenarios contain significant UV absorption by $O_3$, $O_2$, and CO. Increased $O_3$ in the desiccated case produces stronger ozone absorption at 0.3 μm. At short wavelengths there is significant Rayleigh scattering from the large atmospheric mass seen in clear sky, which enhances the albedo even at relatively long wavelengths (~1 μm). This increases the contrast of the visible $O_4$ bands with their surrounding continua. The desiccated case is brighter overall at NIR wavelengths due to combination of a surface albedo that is brighter at these wavelengths and a lack of $H_2O$ absorption. Unlike the wet case, the desiccated case shows a strong CO absorption at 2.35 μm that is not obscured by deep $H_2O$ absorption.

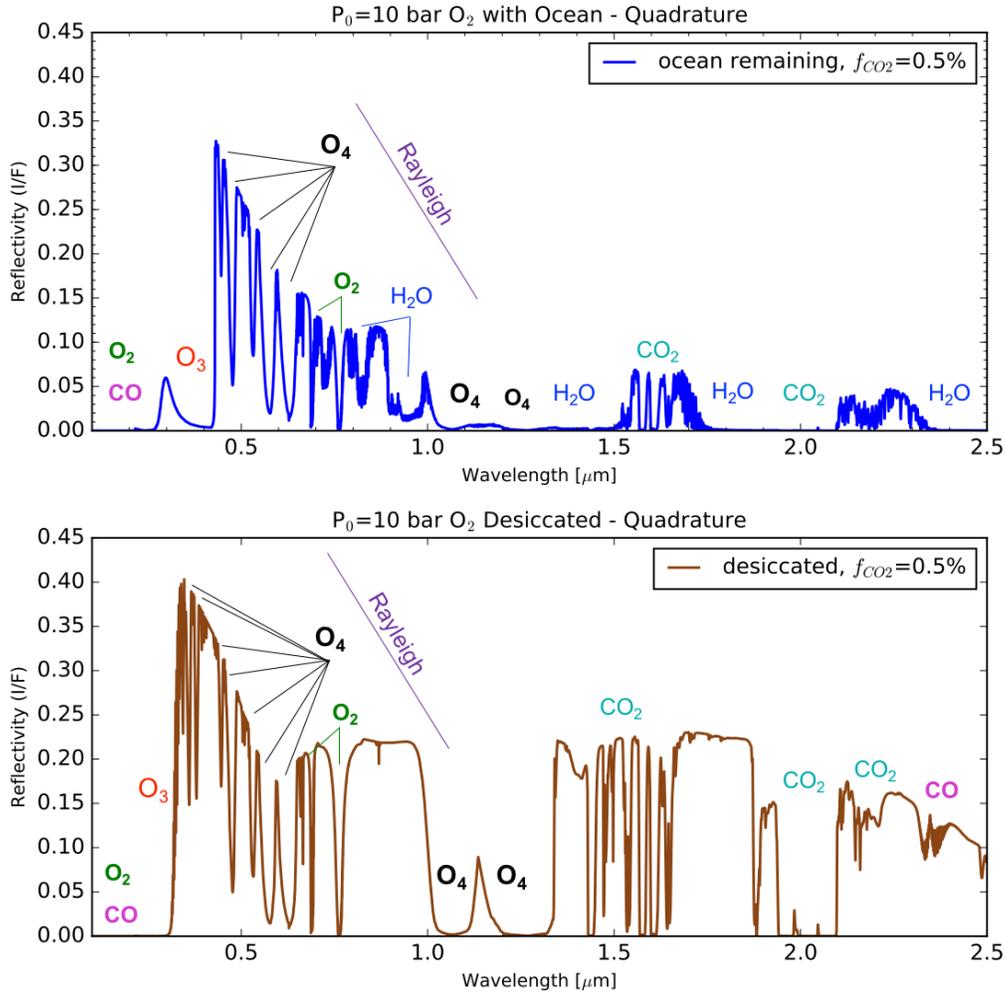

*Figure 15. Reflected light spectra of 10 bar, high-$O_2$ (95%) atmospheres with a surface ocean remaining (top) and completely desiccated (bottom). Note the strong $O_4$ bands present in the UV/VIS/NIR. Both atmospheres were stimulated with 0.5% $CO_2$.*

### 4.4.1.2 Direct Imaging Spectra for High-$CO_2$ Atmospheres

Figure 16 shows direct imaging reflectance spectra of desiccated, clear $O_2$-$CO_2$ atmospheres (45% $CO_2$, 45% $O_2$, 10% $N_2$) with $P_0 = 10$ and 90 bars at quadrature (half illuminated). To generate the spectra we assumed a desert surface albedo. We also assumed here that desiccation was complete, so there are no $H_2SO_4$ aerosols (which require water to form through photochemical processes). The resulting spectrum is rich in $O_4$ absorption features like Figure 15, but with additional strong $CO_2$ absorption throughout the NIR. Rayleigh scattering contributes to a high planetary albedo at wavelengths shorter than 1 μm, but this high albedo is only seen in the narrow continuum regions. The absence of high-altitude aerosols and consequent clear sky paths creates much stronger $CO_2$ absorption than in the simulated Venus spectrum.

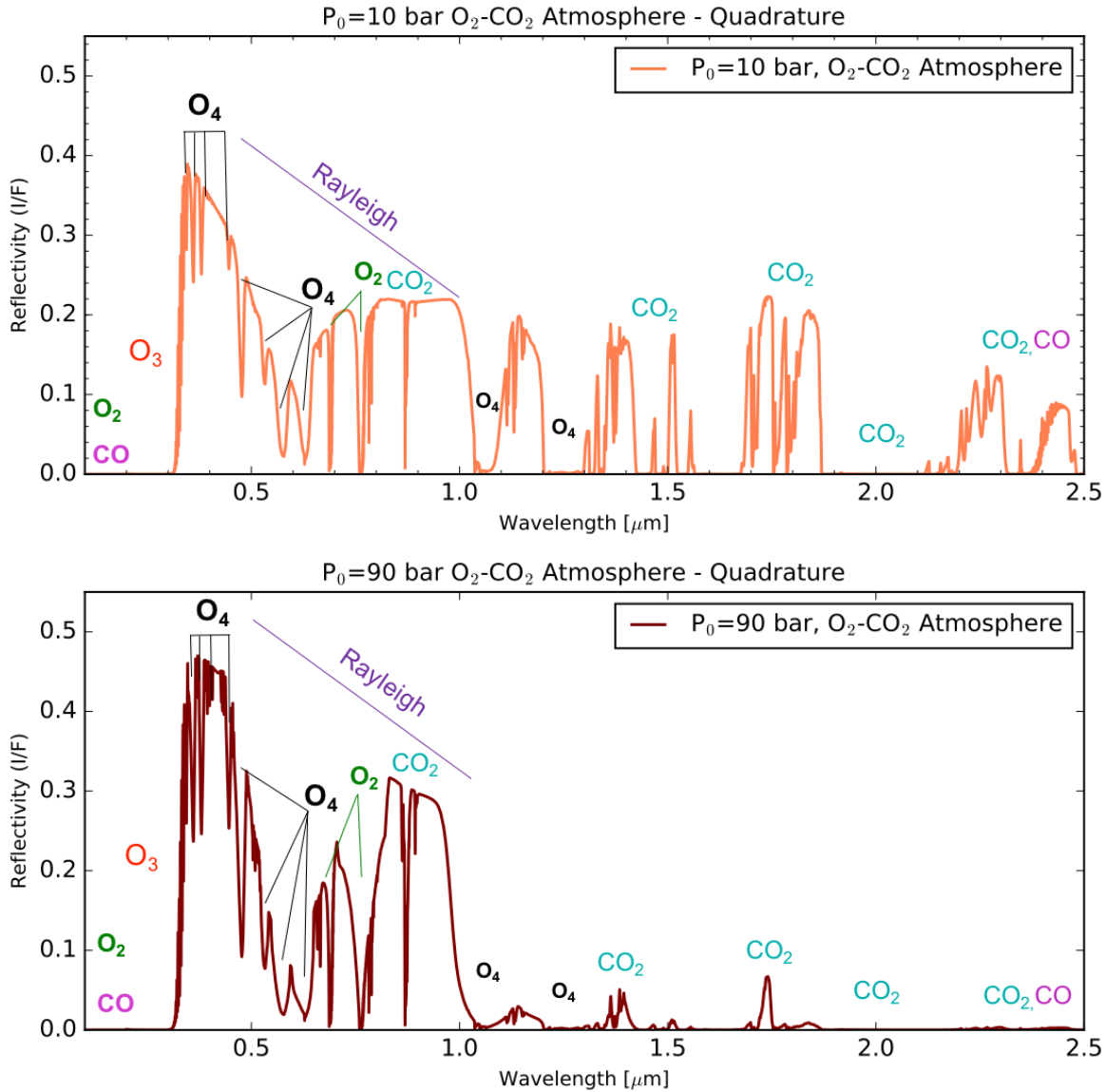

*Figure 16. Synthetic direct-imaging spectrum of a clear atmosphere $O_2$-$CO_2$ (45% $O_2$, 45% $CO_2$, 10% $N_2$) with surface pressures of 10 bar (top) and 90 bar (bottom) for comparison with $O_2$-only and Venus cases.*

Figure 17 shows reflected light spectra of Venus-like planets for our 10 bar and 90 bar simulations with an $H_2SO_4$ cloud deck. There are prominent $CO_2$ bands—particularly at 1.5 and 2 μm, with weaker $CO_2$ bands visible near 0.78, 0.87, 1.05, and 1.2 μm. Water vapor, present at 20 ppm, is observable at 1.35 and 1.85 μm. This uninhabitable planet shows that water in a spectrum is not an unambiguous sign of habitability. An absorption feature from the unknown UV absorber (Moroz et al., 1985), which is responsible for the absorption of most of the incident sunlight in the upper cloud deck in the solar system's Venus (Pollack et al. 1980), is included and is visible near 0.4 μm. Conspicuously absent are Earth-like absorption features from oxygen, ozone, and methane.

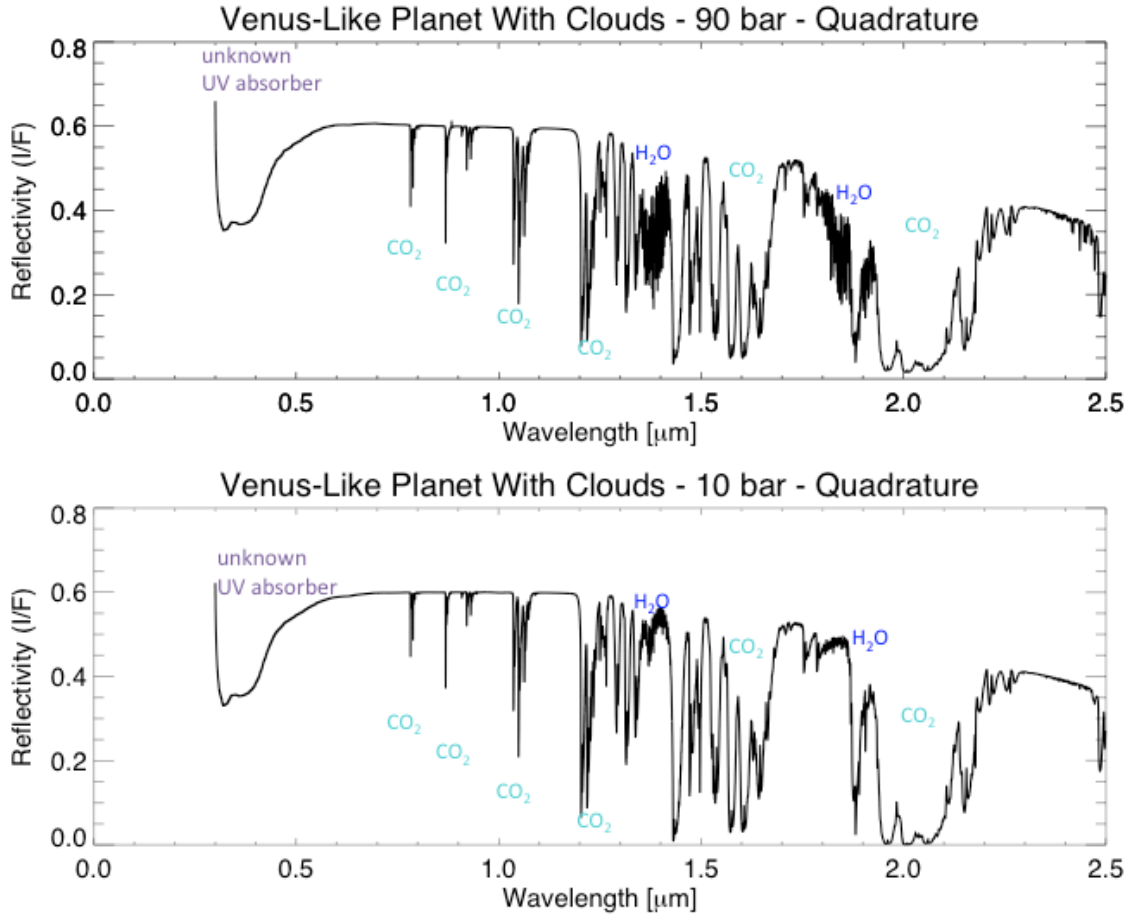

*Figure 17. Reflected light spectra of Venus-like worlds with 90 bar and 10 bar $CO_2$ atmospheres (top panel and bottom panel). Strong $CO_2$ absorption features are present at several wavelengths.*

Figure 18 shows the reflectance spectrum of the case of a desiccated $CO_2/O_2/CO$ atmosphere in photochemical equilibrium (Section 4.2.2.3; Gao et al., 2015). In calculating the reflectance spectrum, we assumed a Mars-like oxidized surface by using the spectral albedo of goethite (a form of iron oxide) from the USGS spectral database (Clark et al. 2007). No clouds are included because the desiccation of this atmosphere is almost complete (0.03 ppm $H_2O$). The spectrum contains strong absorption features from $CO_2$, $O_2$, $O_3$ and CO. The strength of the $CO_2$ and CO features are notably stronger than for an Earth-like case due to the higher abundance of these gases, and are also stronger than the Venus case due to clear views through the atmosphere that are not truncated by sulfuric acid clouds. As noted by Gao et al. (2015), the lack of $H_2O$ absorption bands would be an indicator that the $O_2$ is unlikely to be from a biological source.

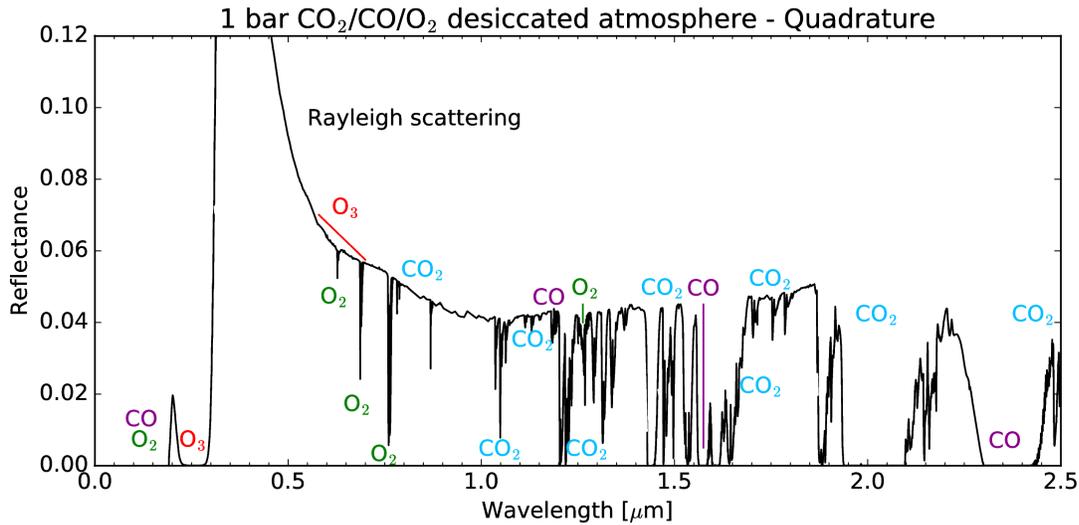

*Figure 18. Reflected light spectrum of a 1 bar desiccated $CO_2/O_2/CO$ atmosphere with an iron oxide surface. A desiccated planet with an outgassed $CO_2$ atmosphere can support a stable $CO_2/O_2/CO/O_3$ atmosphere without life (Gao et al., 2015). A lack of $H_2O$ is an indicator of the abiotic nature of the atmospheric $O_2$.*

### 4.4.1.3 Direct Imaging Spectra for Earth-like Atmospheres.

Figure 19 shows the reflectance spectrum of our photochemically self-consistent modern Earth case using the Earth composite surface spectrum in Section 3.7.5 as our surface spectral albedo. This reflectance spectrum is for a planet with a 50% cloud cover fraction. The composite spectrum is calculated by a weighted average of 50% clear-sky, 25% cirrus (ice) clouds placed at 8.5 km altitude (0.331 bar), and 25% stratocumulus (liquid) clouds placed at 1.5 km altitude (0.847 bar). The spectrum is rich with molecular features including $H_2O$, $O_2$, $O_3$, $CH_4$, and $CO_2$. These include: the $O_3$ Hartley bands at UV wavelengths < 0.38 μm and $O_3$ Chappuis bands from ~0.5-0.7 μm; Rayleigh scattering at blue wavelengths; $O_2$ absorption at 0.63, 0.69, 0.76 μm, and 1.27 μm; $H_2O$ absorption including at 0.65, 0.7, 0.73, 0.8, 0.95, 1.1, 1.4, and 1.8-2.0 μm; and $CH_4$ absorption at 0.9, 1.08, 1.35–1.41, 1.6–1.8, and 2.2–2.4 μm, and $CO_2$ at 1.6 and 2.0 μm. The spectral slope near 0.7 μm is partly due to the vegetation red edge (VRE) present in the composite albedo spectrum used as input to the model. Note that $N_2O$ lacks strong absorption bands at these short wavelengths and CO is not abundant enough to make a spectral impact.

The primary differences between the reflectance spectra of an Earth-analog planet orbiting Proxima Centauri b and the Earth-Sun case are substantially stronger $CH_4$ bands for the M dwarf planet. Figure 20 compares a simulated Earth-Sun case with that of Proxima Centauri b. Both spectra have 50% cloud cover. Other differences include stronger $CO_2$ bands at 1.6 and 2.0 μm due to higher $CO_2$ concentrations in the Proxima case and slightly stronger $O_3$ Chappuis bands, due primarily to the lower $H_2O$ abundance, which is a function of surface temperature.

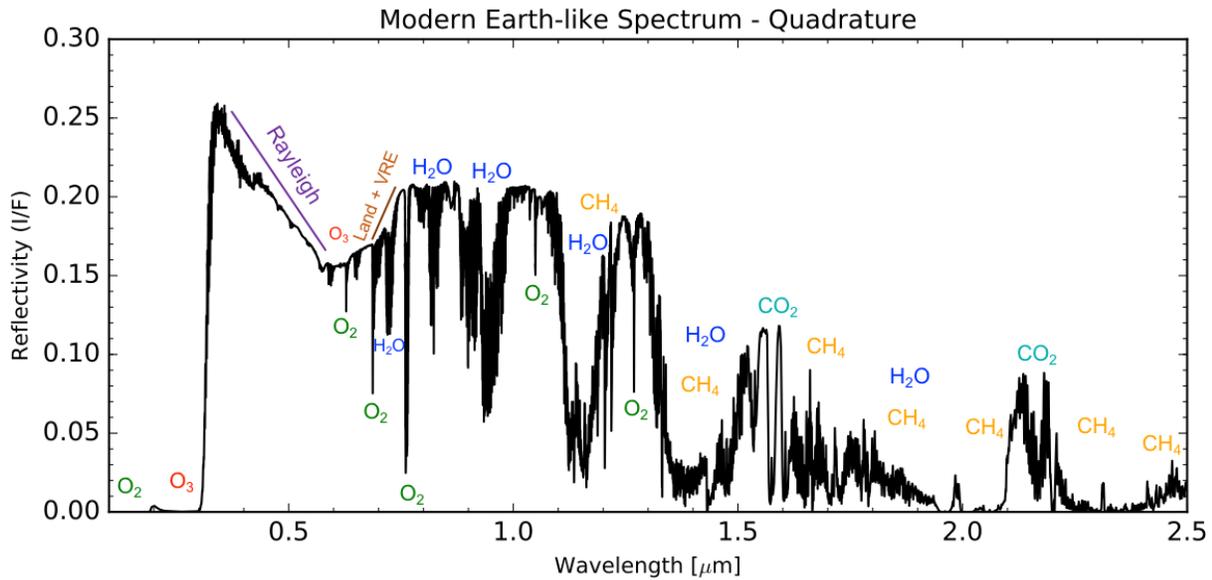

*Figure 19. Reflected light spectrum of Proxima Centauri b as an Earth-like planet with 21% $O_2$ and 5% $CO_2$. The ultraviolet to near-infrared spectrum contains features from Rayleigh scattering, $O_3$, $O_2$, $H_2O$, $CO_2$, and $CH_4$. A cloud cover fraction of 50% is simulated.*

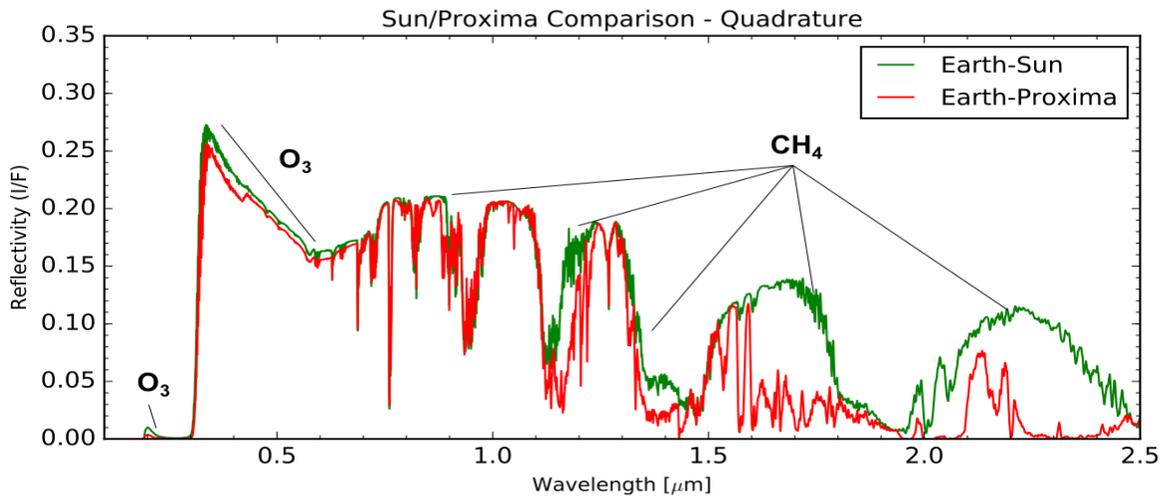

*Figure 20. A comparison of the reflection spectra of an Earth-Sun and Earth-Proxima case. The primary differences in the spectra are in the near-infrared and created by differences in total methane concentration, which is much higher for Proxima Centauri b. (Note that differences in the stellar SEDs have been divided out). A cloud cover fraction of 50% is simulated for both cases.*

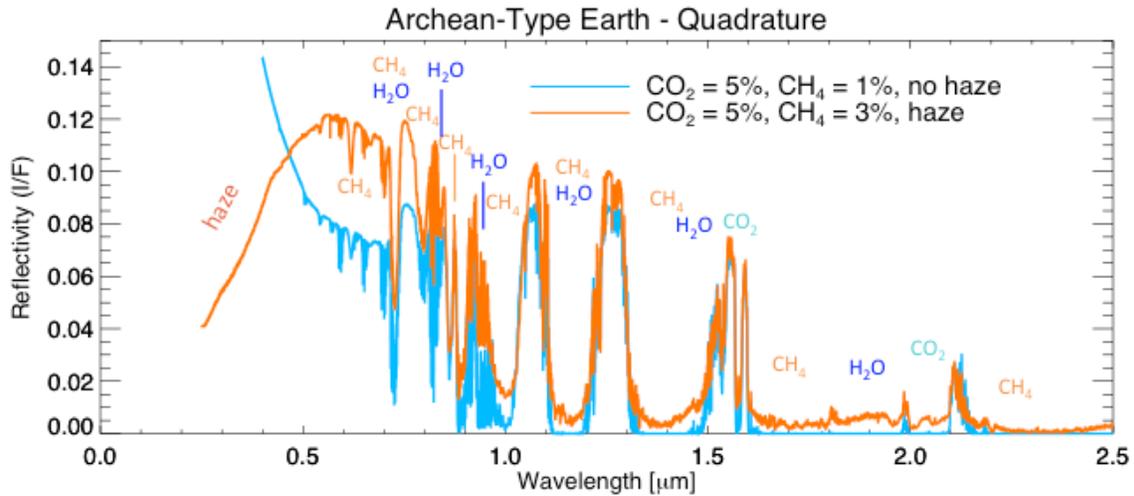

*Figure 21. Reflection spectra of Archean Earth-like planets with (orange) and without (blue) a photochemically self-consistent organic haze. Note the overlap of some $H_2O$ and $CH_4$ absorption bands and the strong haze absorption feature at short wavelengths. The surface albedo assumes the composite albedo described in Section 3.7.5 with no vegetation.*

Figure 21 shows reflected light spectra of Archean Earthlike planets with and without organic haze. The haze is a strong blue and UV wavelength absorber, causing the broad and deep decrease in reflectivity for wavelengths shorter than about 0.55 μm. However, despite its prominence, this feature would be more difficult to detect for a planet like Proxima Centauri b compared to a planet orbiting the sun (Section 4.5.2) because the star itself produces little flux at these wavelengths. The haze also produces back-scattering at longer wavelengths, which is the cause of the increase in brightness of the hazy spectrum relative to the haze-free spectrum at wavelengths longer than 0.55 μm.

Water, carbon dioxide, and methane all produce absorption bands for wavelengths less than 2.5 μm. Several of the prominent $CH_4$ and $H_2O$ bands overlap, including the ones near 1.15 and 1.4 μm. $H_2O$ absorption is present at several wavelengths including 0.65, 0.7, 0.73, 0.8, 0.95, 1.1, 1.4, and 1.8–2.0 μm; $CH_4$ absorption is present at 0.62, 0.73, 0.8, 0.9, 1.08, 1.35–1.41, 1.6–1.8, and 2.2–2.4 μm; $CO_2$ is present 1.6 and 2.0 μm, and weakly near 1.25 μm.

*4.4.2 Transmission Spectra.*

Although Proxima Centauri b is not known to transit, this possibility has not yet been ruled out. We therefore present transit transmission spectra of our simulated atmospheres that are also potentially relevant to other worlds orbiting M dwarfs.

*4.4.2.1 Transmission Spectra for High-$O_2$ Atmospheres.*

Figure 22 shows the transmission spectra of the modeled 10 bar, $O_2$-dominated post-runaway atmospheres for the cases of water remaining and complete desiccation. The spectra are in general very similar with strong $O_3$ and $CO_2$ features, and broad $O_4$ features at 1.06 and 1.27 μm. The NIR $O_4$ bands at 1.06 and 1.27 μm may be observable with the JWST NIRISS[10] instrument, which may confirm the existence of an $O_2$-dominated atmosphere (Schwieterman, et al. 2016). For the "wet" 10 bar $O_2$ case, the 6.3 μm water band is prominent, although water is not seen at shorter wavelengths, and there is significantly less $O_3$ absorption at 0.5–0.7 and 9.7 μm. This is due to the presence of $H_2O$ water vapor in the upper atmosphere. Hydroxyl (OH) radicals from $H_2O$ photolysis can efficiently remove $O_3$ from the atmosphere (Segura et al., 2005). $O_3$ can build up to higher concentrations at most altitudes in the desiccated case. Weak $O_4$ features are apparent at 0.57 and 0.63 μm for the "wet" case, which are otherwise overwhelmed by the Chappuis feature centered near 0.6 μm in the desiccated case.

*4.4.2.2 Transmission Spectra for High-$CO_2$ Atmospheres.*

Figure 23 shows the synthetic transit transmission spectrum for 10 and 90 bar mixed $O_2$-$CO_2$ atmospheres. These spectra are similar to Figure 22 (bottom panel), though additional weaker $CO_2$ bands appear that partially overlap with the 1.06 and 1.27 μm $O_4$ bands. There are only slight differences in the spectra between the 10 and 90 bar cases because the portion of the atmosphere probed by transmission is almost entirely above 10 bar pressures, though the atmosphere is optically thick at higher altitudes for the 90 bar case.

Transit transmission spectra of our 10 bar and 90 bar $CO_2$-rich atmospheres with sulfuric acid cloud decks are shown in Figure 24. The transit transmission spectra of these Venus-like worlds exhibit flat, featureless spectra at visible wavelengths reminiscent of similar spectra observed on exoplanets that likely have cloud or haze layers of various compositions (e.g. Knutson et al., 2014; Kreidberg et al., 2014).

---

[10] http://www.stsci.edu/jwst/instruments/niriss

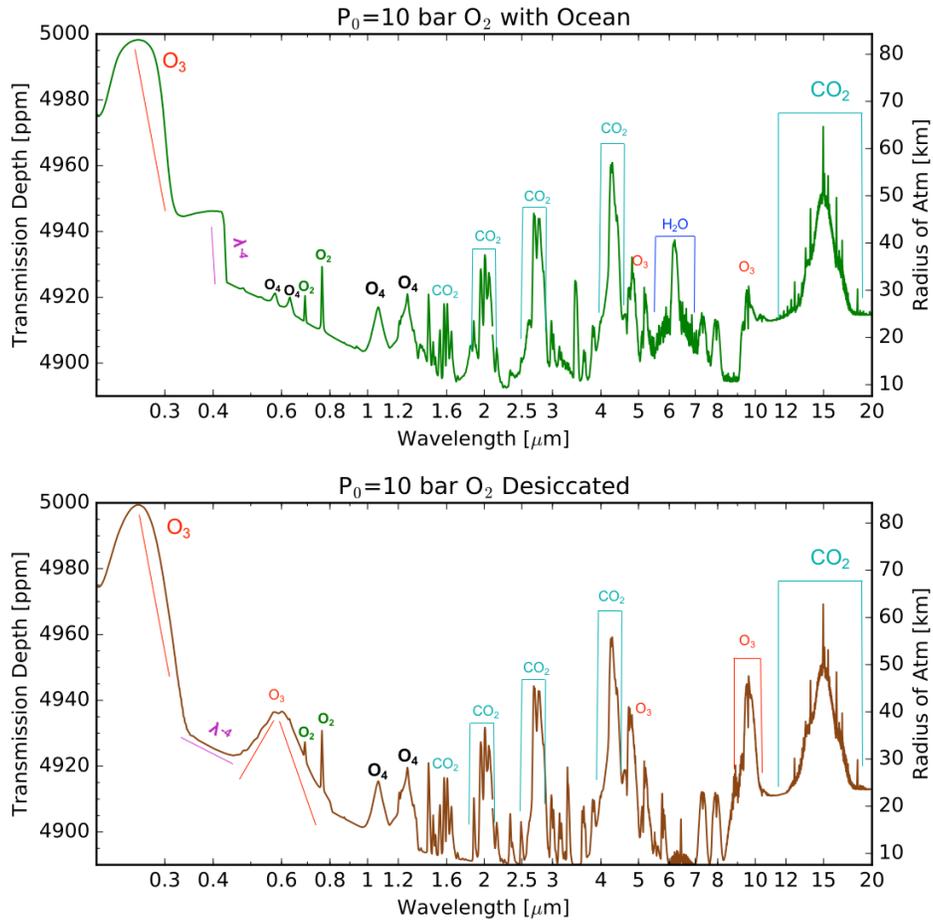

*Figure 22. Transmission spectra of $P_0$=10 bar, 95% $O_2$ atmospheres with a surface water ocean (top panel) and with complete desiccation (bottom panel). The contrasts in $O_3$ absorption are ultimately due to the presence and absence of water vapor in the model atmospheres.*

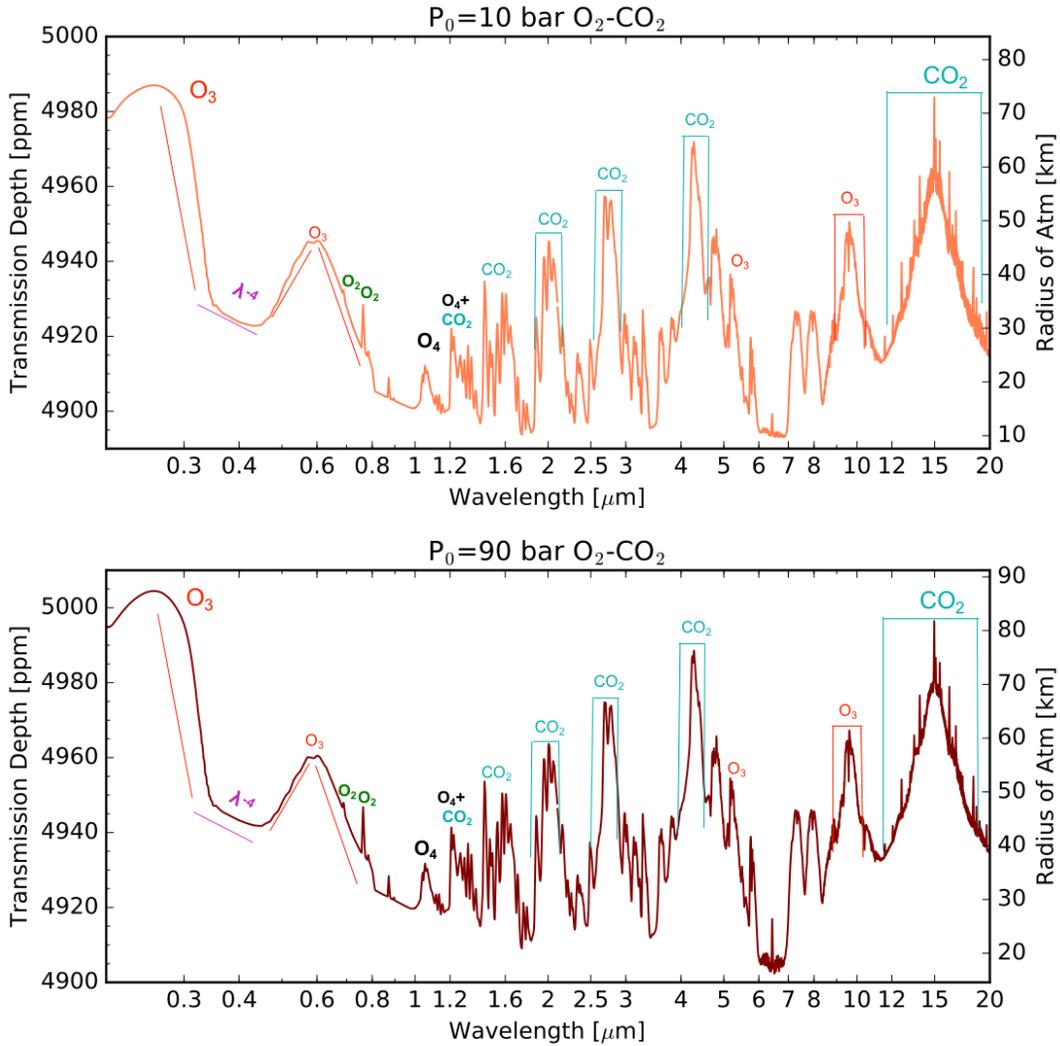

Figure 23. *Same as Figure 22 but for O₂-CO₂ atmospheres with P₀=10 bars (top) and P₀ = 90 bars (bottom). The two spectra are almost identical except that the atmosphere is optically thick at higher altitudes in the 90 bar case.*

As discussed in Ehrenreich et al. (2012), a Venus-like transmission spectrum is dominated by $H_2SO_4$ Mie scattering at $\lambda < 2.7$ μm and by $H_2SO_4$ absorption at $\lambda > 2.7$ μm. Because of this, a broad "V" shape can be seen in the Venus spectrum centered near 2.7 μm, especially in the 90 bar atmosphere with a thicker cloud deck. An almost stepwise increase in the $H_2SO_4$ imaginary refractive index by about two orders of magnitude occurs near 2.7 μm, allowing its absorption to dominate at longer wavelengths. This behavior allows $H_2SO_4$ absorption features to be apparent in Venus' transit transmission spectrum at wavelengths longer than 2.7 μm, providing a means to directly identify sulfuric acid aerosols in the spectrum of a transiting exoplanet. Spectral features from the sulfuric acid clouds are present near 3 μm, 5.8 μm, 8.6 μm, 9.7 μm, and 11.2 μm that correspond to peaks in the $H_2SO_4$ extinction coefficients. These features may be detectable at a level of 3–15 ppm. Despite the opacity of Venus-like $H_2SO_4$ clouds, there are also $CO_2$ absorption features visible near 2, 2.7, 4.5, and 15 μm. The strongest of these features may be observable at about 20–30 ppm. The most detectable is likely the 4.5 μm $CO_2$ feature; although the 15 μm feature is stronger, it occurs at a wavelength where the host star is much dimmer.

Figure 25 shows the transit transmission spectrum of a 1 bar $CO_2/CO/O_2$ atmosphere assuming very low hydrogen abundances (~0.03 ppm H). In contrast to other high-$CO_2$ cases, this spectrum contains significant CO bands at 2.35 and 4.6 μm, which, together with strong $CO_2$ bands, suggests active $CO_2$ photolysis (Harman et al., 2015, Schwieterman et al., 2016).

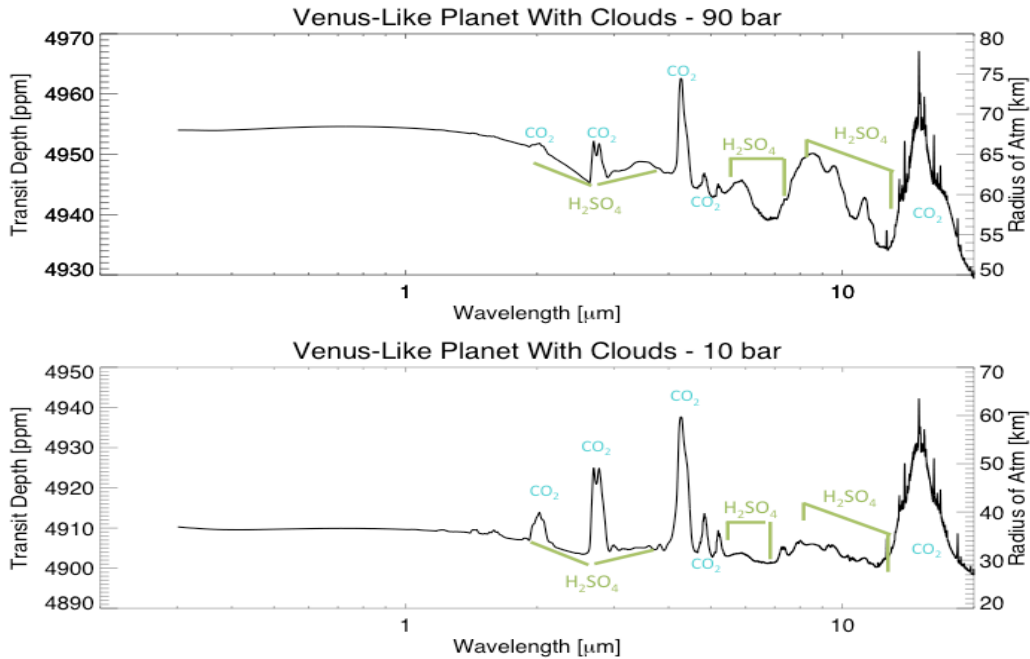

*Figure 24. Transit transmission spectra of Venus-like worlds with 90 bar and 10 bar $CO_2$ atmospheres (top panel and bottom panel, respectively). $H_2SO_4$ absorption features can be seen at IR wavelengths, and the spectrum is flat at wavelengths shorter than 1 μm.*

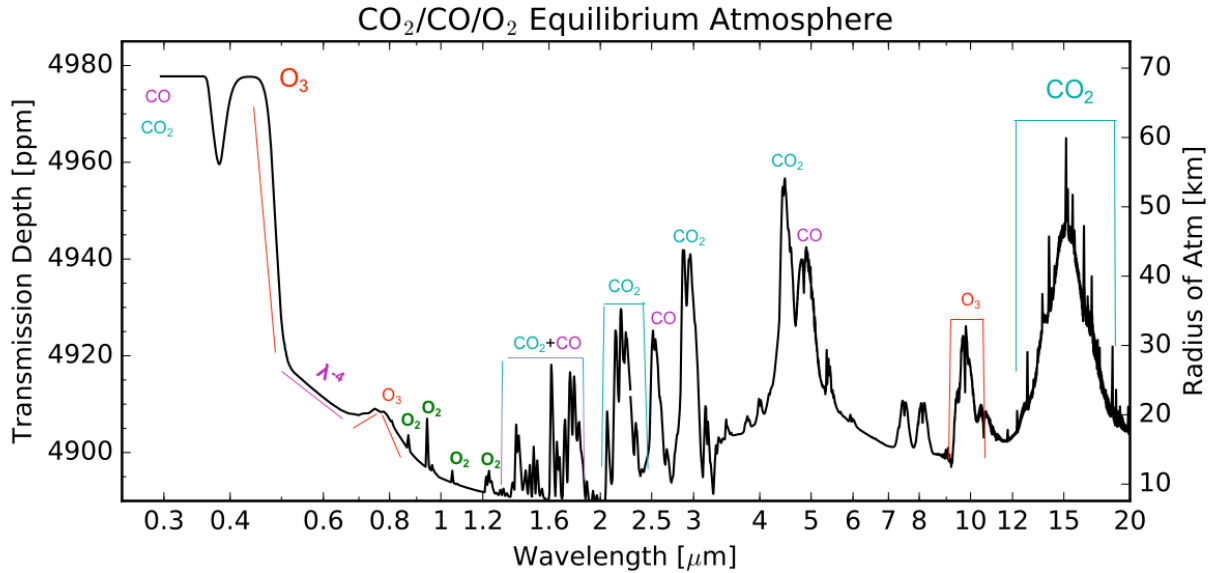

*Figure 25. Transit transmission spectrum of a heavily H-depleted CO₂/O₂/CO atmosphere in photochemical equilibrium from Section 4.2.2.3 (also see Gao et al., 2015). Note the strong CO bands at 2.35 and 4.6 μm compared to other cases.*

*4.4.2.3 Earth-like Transmission Spectra.*

Figure 26 shows a simulated transmission spectrum of the photochemically self-consistent modern Earth orbiting Proxima Centauri, as described in Section 4.2.3.1. The simulated spectrum does not contain clouds and aerosols, although refraction limits altitudes probed to those above ~12km, and the majority of clouds are below this altitude and so would not affect the spectrum even if they were present. Strong, broad features from ozone (0.2–0.3, 0.5–0.7, 4.8, and 9.7 μm), $CH_4$ (0.9, 1.1, 1.4, 2.4, 3.5, 7.5 μm), and $CO_2$ (1.6, 2.0, 2.7, 4.3, and 15 μm) are present. The water bands are weak because stratospheric water vapor abundance is low (even though it is significantly higher than for the Earth-Sun; see Figure 8). Additionally, $H_2O$ bands overlap significantly with $CH_4$, but $CH_4$ abundances are higher at stratospheric altitudes. Figure 27 illustrates the impact of $CH_4$ and $H_2O$ on the transmission spectrum by individually removing their contributions, and provides additional evidence that the 1.1 and 1.4 μm features are due to $CH_4$, and not $H_2O$ in these transmission spectra.

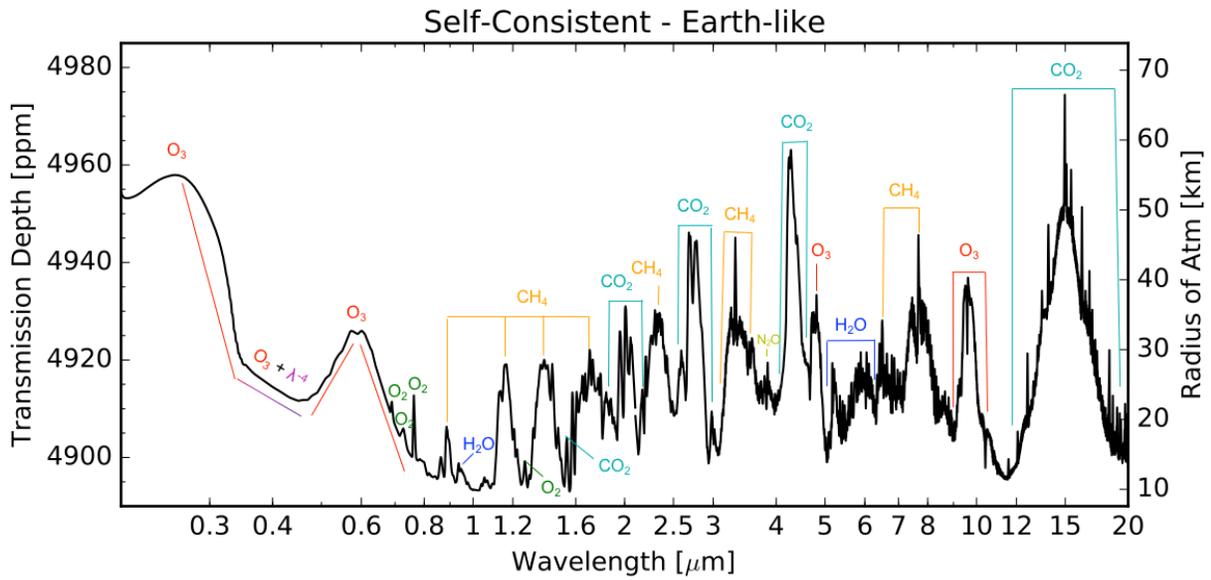

*Figure 26. Transmission spectrum (0.2–20 µm) of a photochemically self-consistent modern Earth with pre-industrial biological fluxes of CH₄, CO and N₂O orbiting Proxima Centauri b. The spectrum is given in terms of expected transmission depth (left y-axis) and tangent height of the atmosphere (right y-axis). Major absorption features are labeled.*

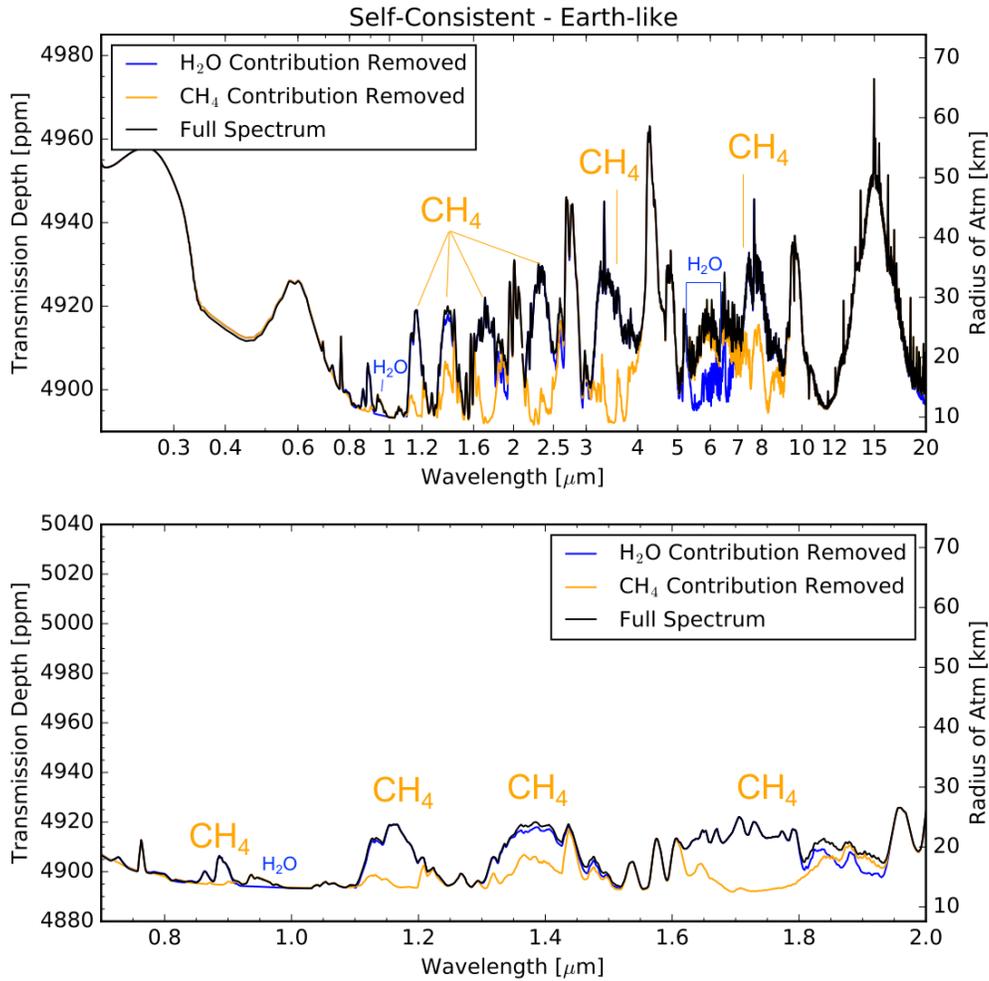

*Figure 27. Sensitivity test showing the relative contributions of $CH_4$ and $H_2O$ to the Earth-Proxima transmission spectrum for the full spectral range (0.2–20 µm; top panel) and a restricted spectral range from 0.7 to 2.0 µm (bottom panel). The weak $H_2O$ features result from water being mostly concentrated in the troposphere level, below ~12 km, although water features would be modestly visible in the absence of $CH_4$ features that swamp them and raise the effective tangent height to drier altitudes.*

Figure 28 shows the transit transmission spectra for Archean-like planets with and without organic haze. Gaseous spectral features become weaker in the presence of a haze, and the depth probed into the atmosphere is reduced. The absorption features near 1 μm are present at about 15–42 ppm in the haze-free spectrum, and at 5–23 ppm in the hazy spectrum. The haze becomes more transparent at longer wavelengths, so the difference between the hazy and haze-free spectra diminishes as wavelength increases. A spectral slope caused by the Rayleigh scattering is apparent at wavelengths shorter than 0.5 μm in the clear-sky spectrum. The hazy spectrum exhibits a spectral slope that continues into the NIR, which is caused by wavelength-dependent haze extinction. Features from $CH_4$ (0.62, 0.73, 0.8, 0.9, 1.2, 1.4, 2.4, 3.5, 7.5 μm) and $CO_2$ (2.0, 2.7, 4.3, and 9.5, 10.5, and 15 μm) may be detectable with a mission like JWST (Arney et al. 2016b). A haze absorption feature that occurs near 6 μm, which overlaps with $C_2H_6$ (6.5, 12 μm), may also be detectable with the JWST MIRI instrument (Arney et al. 2016b).

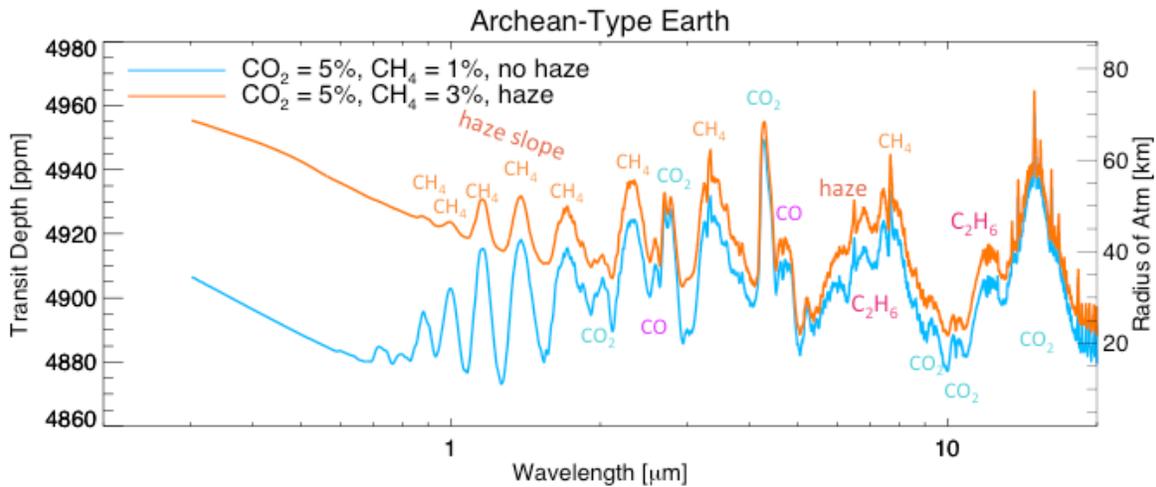

*Figure 28. Archean Earthlike planets with and without organic haze orbiting Proxima Centauri. A haze absorption feature is present near 6 μm, and haze produces the spectral slope continuing into the NIR for the hazy spectrum (orange).*

### 4.5 Observational Considerations for Direct Imaging.

In this section, we describe the types of observations that may be possible to make with direct imaging telescopes in the coming decades. We discuss the challenges of making these observations and show simulated spectra of Proxima Centauri b using our direct imaging coronagraph noise model in Section 4.6.2.

### 4.5.1 Inner Working Angle Constraints.

Observations of Proxima Centauri b will be constrained by the inner working angles of the observatories that may come online in the next decades. The inner working angle (IWA), which is relevant to both coronagraphic and starshade starlight suppression technologies, defines the angular distance from the center of the field-of-view within which starlight suppression rapidly worsens. The IWA is typically expressed as $n\lambda/D$ where $n$ is a small-valued constant (here, we assume $n$ = 1.22–3). The diffraction limit, which is the smallest angular separation a telescope mirror can resolve, is represented by $n$ = 1.22. An effective IWA better than the primary mirror's diffraction limit could be achieved for a starshade-telescope system because most of the starlight never reaches the primary mirror, and so the star's central point spread function is strongly suppressed below the diffraction limit. The angular separation between Proxima Centauri and planet b is 37 milliarcseconds (mas), which is the minimum IWA required to see this planet at quadrature. Note that to observe the planet at any other phase, a smaller IWA will be required.

Note also that the IWA is strongly wavelength-dependent, and so for the same mirror size, the IWA is larger for a longer wavelength. For a given star-planet separation with a given telescope diameter, this means that there is a maximum wavelength beyond which the wavelength-dependent IWA is larger than the planet-star separation, meaning that the target is effectively unobservable. For spectral observations of the target, this means that there is an effective long-wavelength cutoff, beyond which the spectrum is not available. Table 3 presents long-wavelength cutoffs for IWAs for seven different telescope sizes ranging from a 2.4 m telescope such as WFIRST (Spergel et al. 2015) to a 40 m ground-based observatory. These IWAs make no assumptions about starlight suppression technologies; they instead only show the wavelength cutoffs for three potential IWAs.

Table 3.  Inner Working Angles for future telescopes

| Case | $\lambda$ (IWA = $3\lambda/D$) | $\lambda$ (IWA = $2\lambda/D$) | $\lambda$ (IWA = 1.22 $\lambda/D$) |
|---|---|---|---|
| **Ground (40 m)** | 2.41 μm | 3.61 μm | 5.92 μm |
| **Ground (30 m)** | 1.80 μm | 2.71 μm | 4.44 μm |
| **LUVOIR (16 m)** | 0.96 μm | 1.44 μm | 2.37 μm |
| **LUVOIR (10 m)** | 0.60 μm | 0.90 μm | 1.48 μm |
| **LUVOIR (8 m)** | 0.48 μm | 0.72 μm | 1.18 μm |
| **HabEx (6.5 m)** | 0.39 μm | 0.59 μm | 0.96 μm |
| **HabEx (4 m)** | 0.24 μm | 0.36 μm | 0.59 μm |
| **WFIRST (2.4 m)** | 0.14 μm | 0.21 μm | 0.36 μm |

The ability to characterize Proxima Centauri b varies widely with telescope diameter, and Proxima Centauri b is unlikely to be directly imaged with the current generation of 8–10 m telescopes. Among the mirror diameters simulated here, the best case scenario is the 40 m ground-based observatory if an IWA = $1.22\lambda/D$ is achievable: this configuration can reach a wavelength of 5.92 μm. However, observing out to this wavelength is difficult in practice due to the thermal background of the sky and telescope (see Section 4.5.2). Additionally, an IWA that reaches the diffraction limit for a segmented telescope may be too optimistic. Nevertheless, large ground-based telescopes have the potential to characterize Proxima Centauri b at shorter wavelengths where the atmospheric background and transmission are more favorable. Even without optimistic assumptions about the IWA; a 30 m ground-based telescope can still reach 1.8 μm for IWA = $3\lambda/D$, and a 40 m ground-based telescope can reach 2.4 μm for the same IWA. However, these Earth-bound observatories must contend with atmospheric turbulence, and the limitations of adaptive optics to counteract this. These adaptive optics systems generally perform better at longer wavelengths (e.g. Bouchez et al., 2014), where the IWA is not favorable for Proxima Cen b. Nonetheless, the European Extremely Large Telescope (E-ELT)'s first generation adaptive optics (AO) instrumentation is planned to function for wavelengths 0.8–2.4 μm[11], and the GMT's planned adaptive optics wavelength range is 1–25 μm (Lloyd-Hart et al., 2006). Both of these wavelength ranges do not encompass the oxygen A band near 0.76 μm, so it would be advantageous to stretch their AO capabilities to shorter wavelengths.

The most optimistic scenario from space simulated here is a 16 m class LUVOIR (Dalcanton et al., 2015) telescope with IWA = $1.22\lambda/D$, which would provide access out to 2.37 μm. This would allow measurement of the 2 μm $CO_2$ feature, which is the strongest $CO_2$ feature at wavelengths shorter than 4 μm and may provide the best chance of constraining $CO_2$ abundance in Earth-like atmospheres that lack strong $CO_2$ bands elsewhere. For smaller telescopes, a starshade could in principle extend their wavelength coverage to wavelengths longer than the primary mirror's diffraction limit, and possibly out to similar wavelengths observable with a larger mirror depending on the starshade size and the starshade-telescope separation. A large starshade-telescope separation distance would probably be required to observe Proxima Centauri b in the near-infrared. For instance, for a given wavelength and IWA, the telescope-starshade separation distance, $z$, is given by $z = F \times \lambda \,/\, \mathrm{IWA}^2$, where $F$ is the Fresnel number. To observe out to 1 μm, Fresnel numbers greater than a factor of a few require starshade-telescope separations that are a large fraction of the distance between the Earth and the moon. An additional difficulty is that observations with a starshade and a small telescope would have a large PSF (on the order of the size of the star's habitable zone) due to the small size of the mirror. This is especially problematic given that Proxima Centauri is near the plane of the Milky Way galaxy, so the problem of background contamination will likely be significant. A large PSF in a crowded field may make disentangling the photons from planet b and background sources difficult. Due to the small angular separation between Proxima Centauri b and its parent star, it will be important to push technology towards as small an IWA as possible for the best chance of characterizing this interesting system.

---

[11] http://www.eso.org/public/teles-instr/e-elt/e-elt-instr/maory/

### 4.5.2 Coronagraph Model Simulations.

In this section, we show simulated spectra from the previous sections convolved with our coronagraph noise model for different telescope mirror sizes to illustrate the challenges and opportunities for observing Proxima Centauri b with possible future observatories. We show simulated observations for the 10 bar $O_2$ planet with an ocean (Figure 29), the desiccated $O_2$ planet (Figure 30), the Venus-like planet (Figure 31), the desiccated $CO_2/O_2/CO$ planet (Figure 32), the self-consistent modern Earth-like planet (Figure 33), and the hazy Archean Earth-like planet (Figure 34). All of the spectra in this section are simulated with spectral resolution ($\lambda/\Delta\lambda$) of $R = 70$. In the figures below, we choose an IWA of $1.22\lambda/D$ for our nominal simulations to show what may be theoretically possible at the diffraction limit. However, in our coronagraph noise model figures, we also indicate where the $2\lambda/D$ and $3\lambda/D$ cutoffs occur. Note that coronagraph performance tends to decrease with proximity to the IWA, so larger error bars than shown here would be expected at wavelengths near the IWA cutoff. Note also, as mentioned in Section 4.5.1, that a starshade could theoretically observe a smaller IWA than the diffraction limit of the primary mirror and could allow observations to longer wavelengths than shown here.

In the figures below, we show simulated observations for telescope sizes corresponding to HabEx (6.5 m), LUVOIR (16 m) and a large ground-based observatory (30 m). Integration times assume a 5% throughput for a coronagraph and telescope instruments. This is based on WFIRST (Robinson et al., 2016), but it could be improved with off-axis telescope architectures or if the coronagraph has fewer reflections.

Direct imaging of terrestrial exoplanets may become possible from the ground (e.g. Males et al., 2014, Quanz et al., 2015). For the 30 m ground-based telescope simulations, although we include the atmospheric transmittance (Figure 35) and thermal emission from the telescope mirror and sky, our simulations are optimistic because they do not include atmospheric airglow emission, or the wavelength-dependent performance of adaptive optics, and they assume that absorption features from the planet can be perfectly separated from absorption features in Earth's atmosphere. This latter issue will make it difficult to definitively detect gases in any exoplanet's atmosphere if they overlap with strong telluric features. This latter issue could be resolved by going to very high resolution (e.g. $R = 100,000$) to disentangle individual planetary spectral lines from Earth's atmosphere's absorption using the Doppler shift of the planet's motion relative to Earth (Snellen et al., 2015) although this technique is necessarily limited to the very brightest targets.

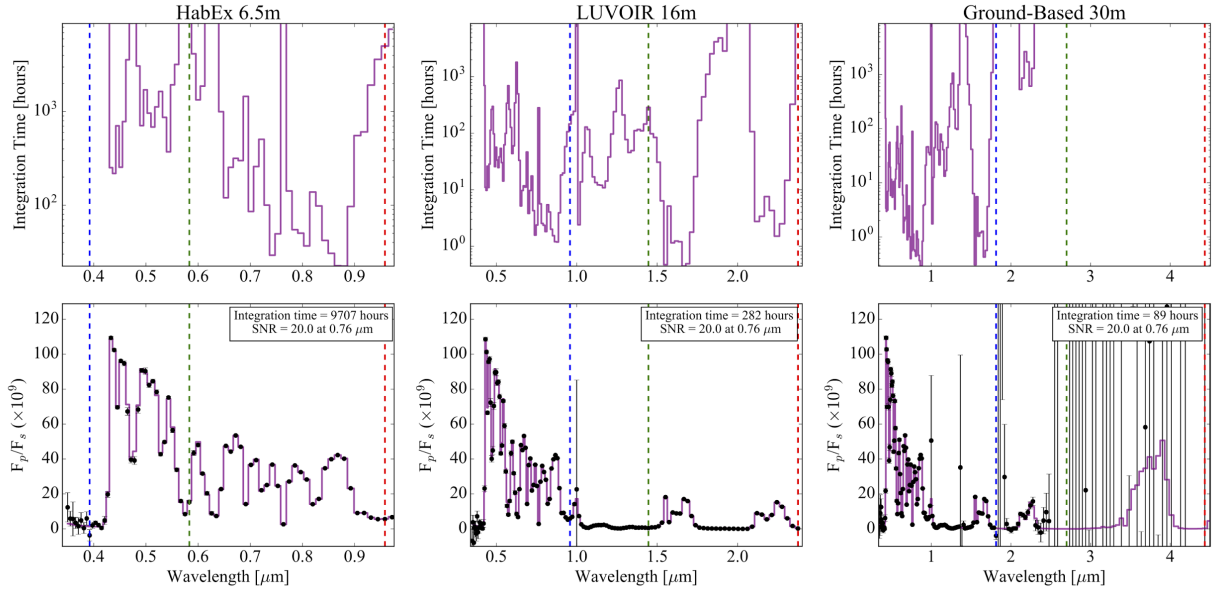

*Figure 29. Simulated coronagraph integration times (top) and reflectance spectra (bottom) for the "wet" 10 bar O₂-rich planet using three different future telescope concepts: HabEx (left), LUVOIR (middle), and a 30 m ground-based telescope (right). Dashed vertical lines are placed at IWA = 1λ/D (red), 2λ/D (green), and 3λ/D (blue), to show the long-wavelength limit for the given telescope diameter and planet-star angular separation. All spectra are shown for a spectral resolution of R = 70. The integration times given are required to achieve S/N = 20 in each resolution element, while the errors on each simulated observation (black) represent the estimated 1σ error bars for the integration time needed, per nulling bandwidth, to achieve S/N = 20 in the resolution element at the 0.76 µm oxygen A-band.*

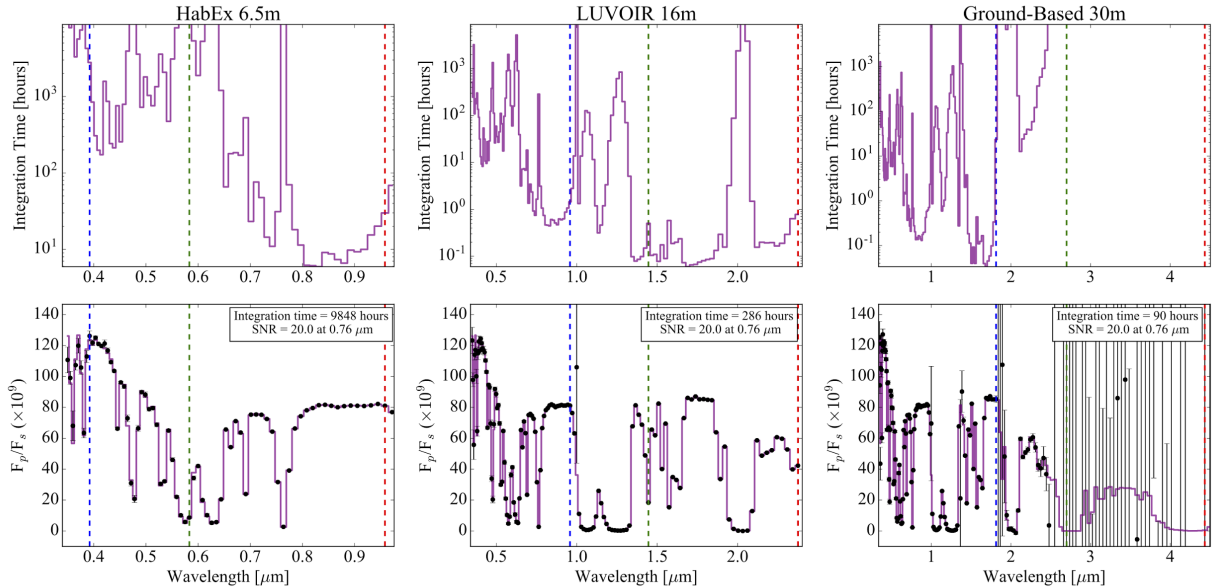

*Figure 30. Same as Fig. 29 but for the dry 10 bar O₂ planet.*

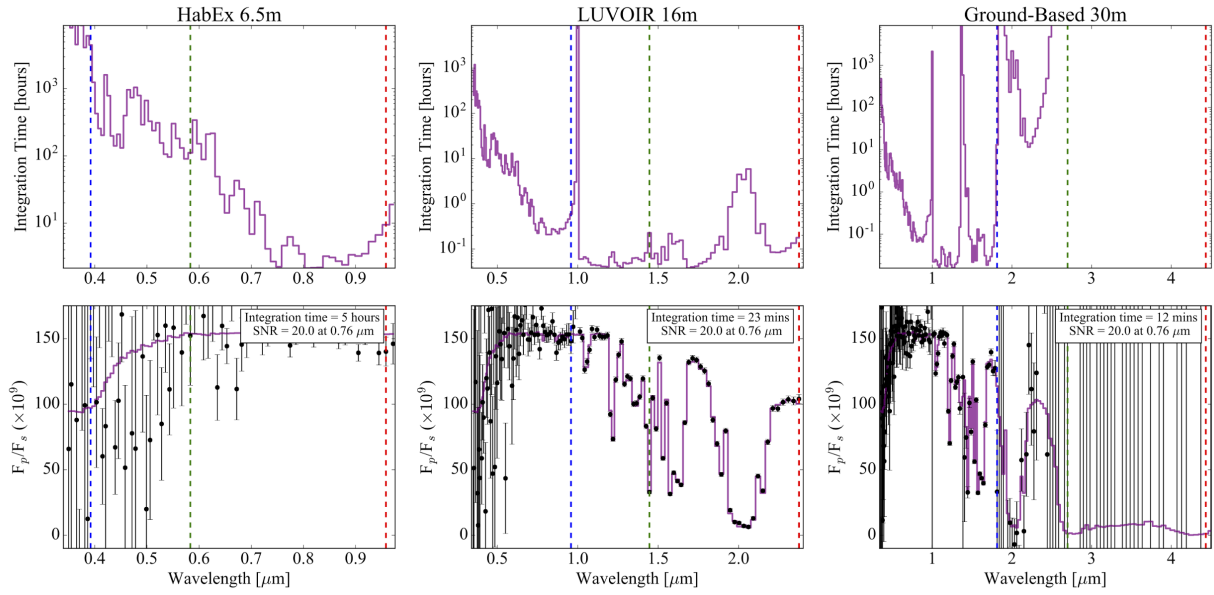

*Figure 31. Same as Fig. 29 but for the Venus planet.*

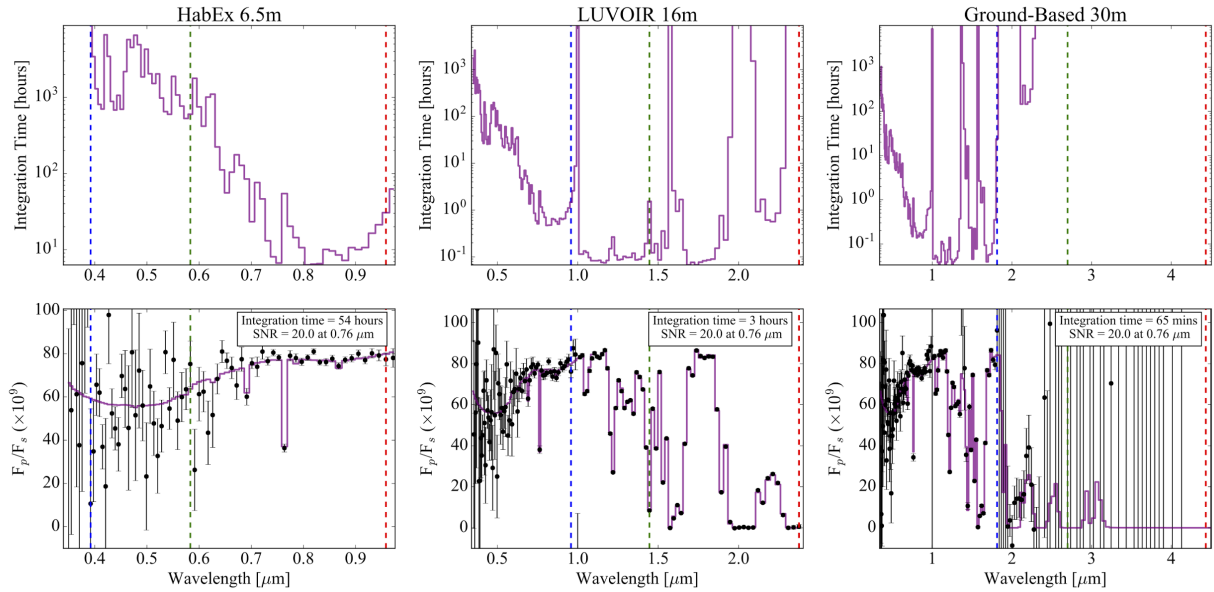

*Figure 32. Same as Fig. 29 but for the desiccated $CO_2/O_2/CO$ planet*

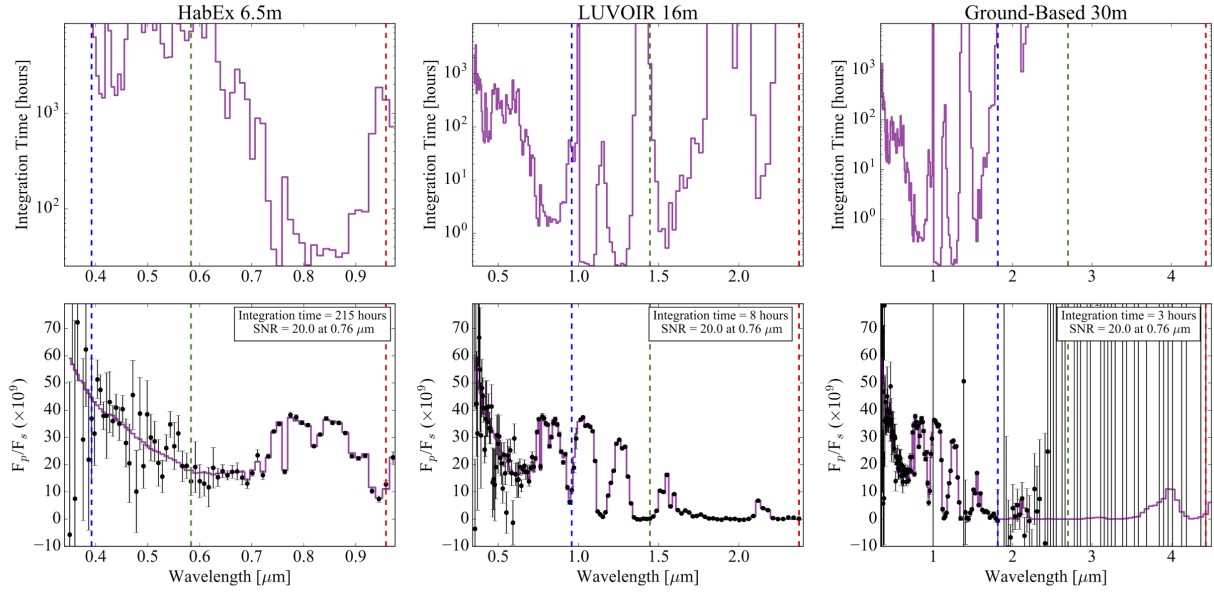

*Figure 33. Same as Fig. 29 but for the modern Earth.*

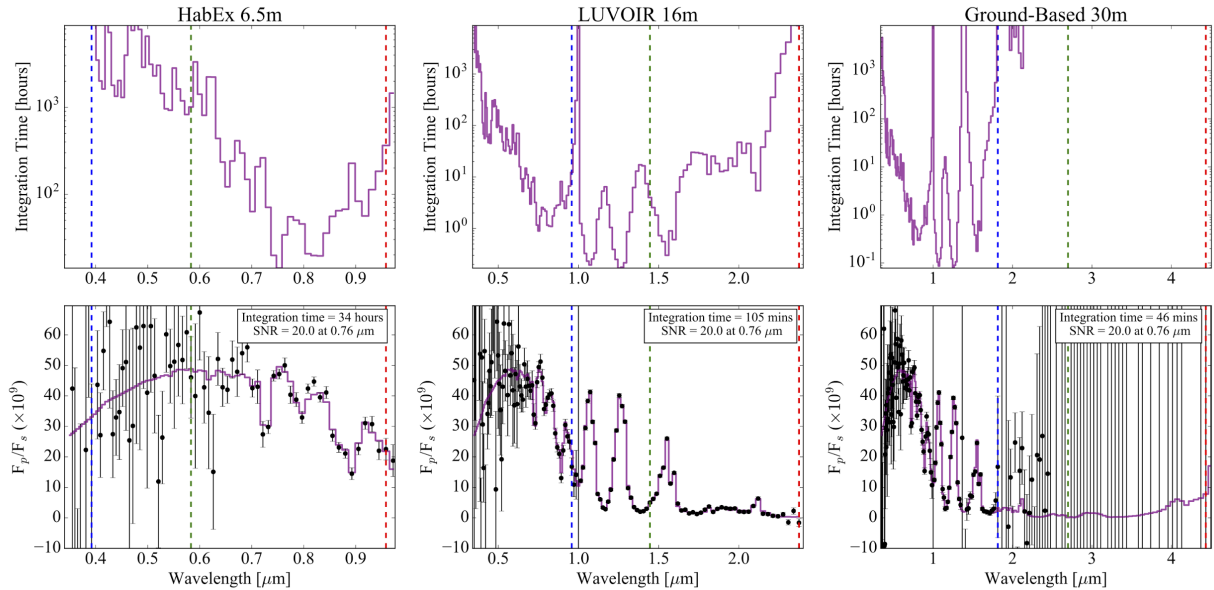

*Figure 34. Same as Fig. 29 but for the Archean Earth.*

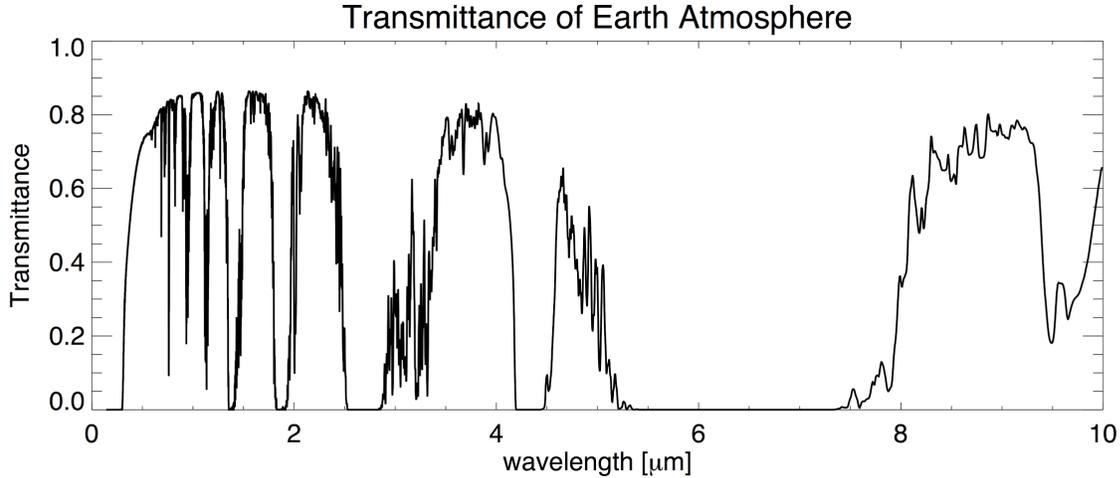

*Figure 35. Earth's assumed atmospheric transmittance for simulations of ground-based observations.*

The top panel of each figure shows the integration times in each resolution element required to achieve a signal-to-noise ratio (S/N) of 20. The bottom panel of each figure shows the simulated spectrum assuming that the planet was observed for the required time to achieve S/N = 20 at the bottom of the oxygen A band—a measurement that is sufficient for retrieving the oxygen abundance (Robinson et al., 2016)—although we could also assess concentration from the width of the feature or from combinations of features that appear at different concentrations. As a result, planets with strong oxygen absorption (e.g. Figures 29 and 30), and consequently lower planet-star flux contrasts in the oxygen A band, require longer integration times to achieve a signal-to-noise ratio of 20 in that spectral element than a planet with little to no oxygen absorption (e.g. Figure 31). However, since the simulated observation is representative of the oxygen A band integration time *for all wavelengths*, planets with large oxygen absorption are depicted in these figures with higher signal-to-noise for spectral elements *outside* of the oxygen A band. In practice, the full visible to near-IR spectrum will be constructed from independent exposures for each coronagraphic nulling bandwidth, and different integration times will be selected for each null based on the expected contrast at that wavelength. In all simulations, there is poorer signal-to-noise at visible wavelengths compared to NIR wavelengths, because the star is dim at visible wavelengths compared to the NIR. The vertical bars illustrate the IWA cutoffs corresponding to $1.22\lambda/D$ (red), $2\lambda/D$ (green), and $3\lambda/D$ (blue).

For the 30 m ground-based simulations, the thermal background from Earth's surface and atmosphere makes observations longward of about 2.5 μm difficult, so IWA = $1.22\lambda/D$ may not actually meaningfully increase our ability to characterize Proxima Centauri b compared to a more modest IWA of $2\lambda/D$ for this type of observatory. In the optimistic scenario where atmosphere can be cleanly corrected for, the base of the oxygen A band for the modern Earth analog planet could be detected with S/N = 20 in 3 hours, allowing this spectrum to be observed with 3 hours per coronagraph bandwidth. Even if the true observing time is longer in reality, this short integration time for a very high-confidence detection suggests that such an observation may be possible for more conservative scenarios.

For a 16 m LUVOIR-class space-based telescope, good signal-to-noise is achievable for the entire wavelength range captured by wavelengths shorter than the $1.22\lambda/D$ cutoff except for the shorter visible wavelengths where the host star is dim. The base of the oxygen A band on the modern Earth analog could be observed at S/N = 20 in 8 hours with such an observatory. Note that although IWA = $1.22\lambda/D$ could theoretically allow observations out to 2.37 μm, actually making observations at these wavelengths can only be achieved if the telescope mirror is cooled; our simulations assume a mirror temperature of 200 K, but the thermal emission from warm "room temperature" mirrors make observations longward of about 1.6 μm extremely difficult. If an optimistic IWA of $1.22\lambda/D$ is achievable, cooling the telescope would be highly beneficial to allow access to these longer wavelengths that would allow measurement of, e.g., the strong $CO_2$ band near 2 μm to help constrain the efficiency of the Proxima Centauri b's greenhouse and its atmospheric state. On the other hand, if an IWA of only $2\lambda/D$ is possible, cooling the mirror would not be necessary for observations of Proxima Centauri b since this IWA would only allow access to wavelengths < 1.44 μm, and these wavelengths are relatively unaffected by the thermal emission of the mirror. If an IWA of only $3\lambda/D$ is possible, a 16 m LUVOIR telescope could observe out to 0.96 μm. This would still allow access of the oxygen A band, but a large fraction of this wavelength range would occur at wavelengths where the star is dim and the spectrum lacks strong absorption features.

Observing Proxima Centauri b is naturally more difficult, but still possible, with a smaller telescope like the HabEx concept. A 6.5 m HabEx telescope could observe the deepest part of the oxygen A band at S/N = 20 in over 200 hours. However, the oxygen A band is not accessible with a 6.5 m telescope if the IWA achievable is not better than $2\lambda/D$, which cuts off the spectrum longward of 0.59 μm. On the other hand, if HabEx were to fly with a starshade, this could make it possible to observe a smaller IWA and achieve a wider coverage of wavelength space. However, the ability to discriminate the light of the planet from background sources will still be hampered by the PSF of the smaller mirror, and integration times will be long compared to a larger mirror.

*4.6 What might Proxima b look like to the human eye?*

In addition to all out spectral simulations, we used the Python package ColorPy[12] to visualize the apparent visual color of the range of possible atmospheric states for Proxima Centauri b examined here. ColorPy uses the 1931 *Commission Internationale de l'Eclairage (CIE)* color matching functions (Smith & Guild, 1931) to calculate the apparent, human-perceived color from an intensity spectrum. We modified ColorPy to take our top-of-atmosphere planetary spectra (Section 4.4) as inputs for these color calculations. Figure 36 shows the visual spectra (380–720 nm) and calculated colors for the star, Proxima Centauri, and our modern Earth-like case with 50% cloud cover, Archean (early) Earth, $O_2$-dominated, and Venus-like scenarios. We find that worlds with clear sky paths, i.e. without global cloud or haze cover, appear blue despite the redness of Proxima Centauri. This somewhat counterintuitive result is primarily due to three effects. Firstly, the semi-empirical visual spectrum of Proxima Centauri does contain a small intensity enhancement at blue wavelengths (Figure 1). Secondly, while Proxima Centauri's SED peaks at long wavelengths (~1 µm), the Wien tail is very steep and this limits the intensity pickup at the red wavelengths where the human color perceptive range ends. Finally, and most importantly, the blue photons that do interact with the planetary atmosphere are Rayleigh scattered just as for the Earth's atmosphere. For the $O_2$-dominated cases with substantial (10 bar) atmospheres, the blue color enhancement overwhelms the redness of the star that is encapsulated in the human visual range. The modern Earth case appears lavender because Rayleigh scattering and ocean reflection enhances the blue, but reflection by continents and clouds enhances reflection of red light. The haze-covered Archean Earth and the Venus-like cases are redder than modern Earth and are surprisingly lilac-colored since they do not exhibit a significant spectral impact from Rayleigh scattering. Therefore, for plausible sets of both habitable and uninhabitable atmosphere states, Proxima Centauri b might appear as a pale blue, or pale purple, dot. However, this calculation is preliminary pending a fully empirical panchromatic spectrum of Proxima Centauri.

---

[12] ©Mark Kness: http://markkness.net/colorpy/ColorPy.html

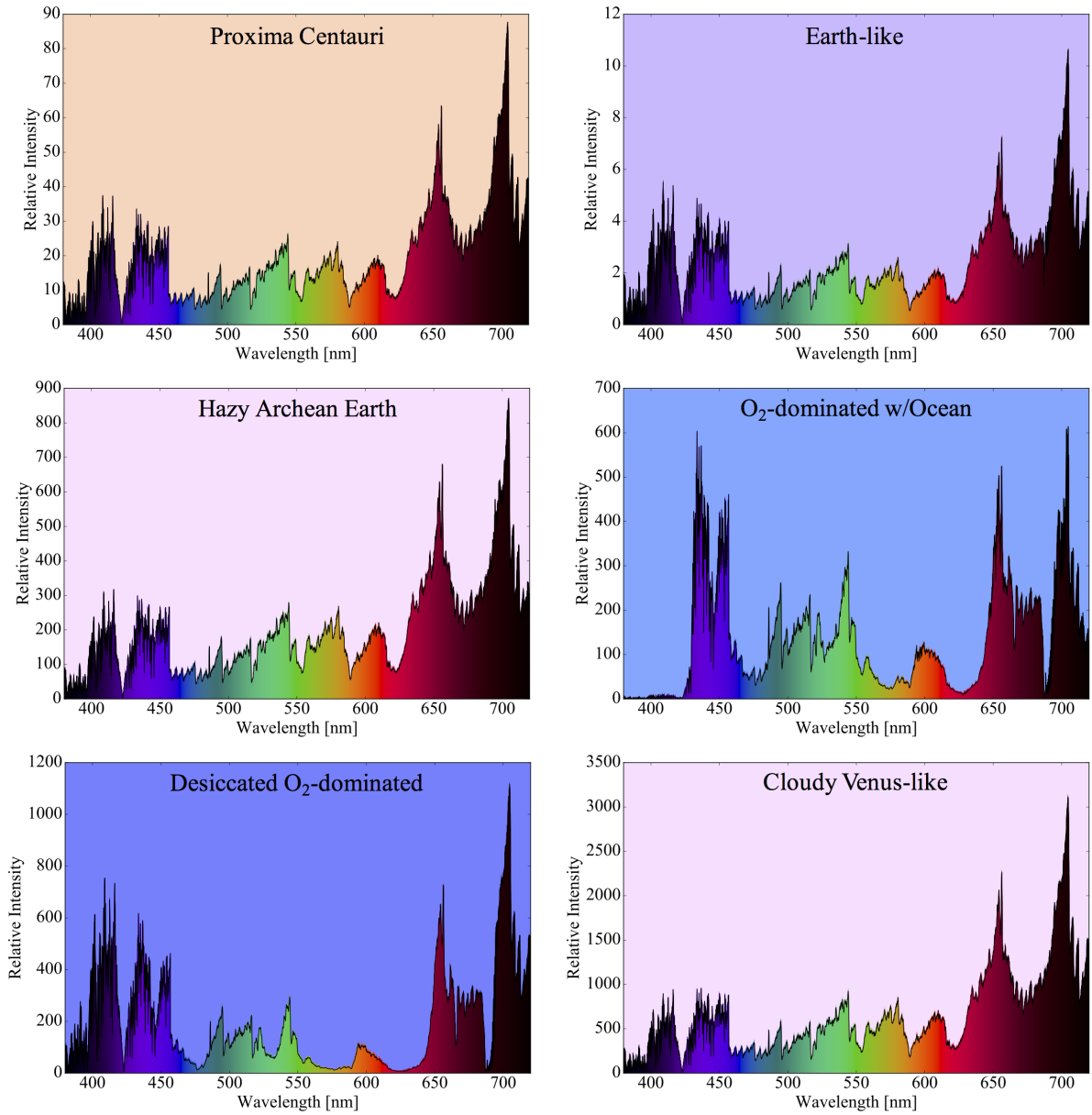

*Figure 36. Optical spectra and the corresponding colors that the human eye would perceive for Proxima Centauri and five of our simulated planetary states. From upper left to lower right: the M5.5V red dwarf star Proxima Centauri, Earth-like with 50% cloud cover, hazy Archean, O₂-dominated with ocean, desiccated O₂-dominated, and cloudy Venus-like. The background of each plot is colored as it would appear to the human eye. The shaded area under the curve represents the individual color of each wavelength at the simulated spectral resolution. Each planetary spectrum represents a convolution of the stellar spectrum with the planetary albedo, which is dictated by the planet's surface and atmospheric composition. Furthermore, each perceived color represents a convolution of the planetary spectrum with the human eye.*

**5.0 Discussion.**

The discovery of Proxima Cen b provides a truly remarkable opportunity to learn about the evolution of terrestrial planets orbiting M dwarfs, as well informing our understanding of the range of habitable environments in the Universe. However, although it sits in the habitable zone, which confers a higher probability of being able to support liquid water on its surface, the evolution of the star and planet can work to maintain or destroy habitability for Proxima Cen b (Barnes et al., 2016; Ribas et al., 2016). Consequently, even though Proxima Cen b has the advantages of size and current position in the habitable zone, its habitability is by no means certain, and will depend strongly on the evolutionary path taken by the planet, and its current interaction with its parent star. Our study has simulated several possible planetary states for Proxima Centauri b that show both habitable and uninhabitable environments.

*5.1 The Habitability of Proxima Centauri b.*

We find that Proxima Cen b inspires three fundamental scientific questions: "What are the evolutionary paths for Proxima Cen b, and for terrestrial planets around other late type M dwarfs?", "Is Proxima Cen b currently habitable?" and "How can we best discriminate between possible environmental states for Proxima Cen b, and for other M dwarf planets?". The first question was explored in detail in our companion paper, Barnes et al., (2016) which showed that despite the planet's location in its star's habitable zone, this world may not be water-rich, or habitable. The star's luminous pre-main sequence phase represents the biggest barrier to its habitability, and if the planet formed *in situ*, it may have been desiccated early in its history. However, if the planet retained some of its initial water inventory or migrated to its current orbit from the outer solar system after the pre-main sequence phase, it may yet be hospitable to life. As formulated here, we identified planetary environmental states that spanned the post-runaway, oxygen-dominated worlds of Luger & Barnes (2015) through terrestrial planet evolution that would lose or sequester the $O_2$, increase atmospheric $CO_2$ from geological processes, and progressively desiccate the planet. Habitable scenarios are also possible, especially if $H_2$ envelopes protect terrestrial planets that formed *in situ*, or volatile-rich bodies migrate in from beyond the snowline (Luger et al., 2015). We note however that although these scenarios are representative, they are still likely a small sample of the possible configurations for M dwarf planet environments.

But is Proxima Centauri b currently habitable? To start to address that question, our climate-photochemical modeling of the possible end-state atmospheres for the evolutionary paths shows environmental states ranging from uninhabitable due to water loss and/or surface temperatures that are too high or low, to worlds that could support liquid water. The uninhabitable planet environments for Proxima Cen b include the desiccated, 10 bar high-$O_2$ planet with 0.5% $CO_2$, which has a globally-averaged surface temperature of only 257 K, and so would be cold and dry.

Other uninhabitable planets fell into the hot and dry category, with the evolved, desiccated $O_2/CO_2$ atmospheres exhibiting surface temperatures of 342–383 K.   The Venus-like planets, with a larger atmospheric $CO_2$ fraction and pressures up to 90 bars, showed extremely high surface temperatures of 428–568 K at Proxima Centauri b's position within its habitable zone.  These results highlight the considerable impact of planetary evolution and atmospheric composition on habitability, even for planets in the habitable zone.   For the desiccated (< 1ppm H) atmospheres that initially have 1 bar of $CO_2$, but then photochemically produce $O_2$ and CO, which remain stable in their desiccated atmosphere, habitable surface temperatures of 298 K were obtained.  However, this extremely desiccated planet would not be habitable due to its lack of water, despite the apparently clement climatic conditions.  In this case it is interesting to note that the photochemical destruction of $CO_2$ had a significant impact on the planetary climate.   Because of the importance of greenhouse gas abundance for the final climate state, future work should concentrate on understanding the likely outgassing history and plausible atmospheric compositions for terrestrial exoplanets.

The habitable planetary states were also quite varied.  If the runaway process halts before all the planetary water inventory is lost, then even a 10 bar $O_2$ planet with 0.5% $CO_2$ and an ocean exhibits a globally-averaged surface temperature of 320 K, primarily due to the greenhouse effect of the atmospheric water vapor.  This is significantly warmer than the Earth's current 288K globally averaged surface temperature, but still potentially habitable.   However, it is debatable whether life could start on this planet, as the strongly oxidizing atmosphere would be challenging for the stability of the organic molecules required for the origin of life.   We also considered more Earth-like cases, including a modern, but photochemically self-consistent Earth-like planet whose atmospheric composition is consistent with the UV radiation from the parent star.  The star's UV spectrum resulted in longer lifetimes for, and a buildup of, greenhouse gases such as $CH_4$ and $N_2O$, as was previously seen by Segura et al., (2005).  This emphasizes the need for both photochemical models and knowledge of the UV spectrum of the host star when trying to model and interpret information on the environments of planets orbiting M dwarfs.     For this simulated environment we can obtain cold, but habitable surface conditions even with current Earth-like $CO_2$ abundances, but $CO_2$ abundances near 5% are required to warm the planet to 283 K, which is close to Earth's current surface temperature of 288 K.    For the Archean Earth analog, 5% $CO_2$ was also specified, but the higher methane fraction resulted in a warmer surface temperature of 289 K in the haze-free case, and when $CH_4$ was increased to 3% a haze formed, but only modestly cooled the planet because hydrocarbon hazes are relatively transparent at the longer emitted by late type M dwarfs (Arney et al., 2016b).

So, how should we best use observations to discriminate between these possible environmental states, not only for Proxima Cen b, but for other M dwarf planets as well?   The post-runaway $O_2$-dominated worlds are best discriminated by the presence of strong $O_4$ bands in the visible and NIR for direct imaging, and in the NIR for transmission.   Since the atmosphere is strongly oxidizing, and if desiccation has occurred, the presence of aerosols other than dust may be unlikely, making the planet quite bright in the Rayleigh scattering tail.    The desiccated case also shows strong ozone absorption because the $HO_x$ products that would destroy it are not present.    Water may also be detected in direct imaging, in the case where some of the remnant ocean is retained, but the strong $O_4$ would sound a warning that the $O_2$ was most likely formed post-runaway, and not from a photosynthetic biosphere (Schwieterman et al., 2016).    As the planet evolves to also include abundant $CO_2$ in addition to the $O_2$, strong $CO_2$ bands are seen in both the direct imaging and transmission spectra.  Venus-like worlds are dominated by $CO_2$ features, especially near 1.5 and 2.0 microns, although weaker bands can also be seen even into the visible at 0.78, 0.87, 1.05, and 1.2 microns.  In our simulations, weak water vapor was also seen in the spectrum, because the specified water vapor abundance of 20 ppm is evenly mixed throughout the atmospheric column and exceeds the water vapor abundance above the Venus cloud tops, which is closer to 6 ppm on the nightside (Schofield et al., 1982) where spectra are typically observed/modeled.   These planets were also able to form sulfuric acid hazes that increase the planetary albedo and can be spectrally identified by NIR absorption features that would be visible in transit transmission spectra (Figures 25 and 28).  The presence of an $H_2SO_4$ haze may also imply active volcanism on an exoplanet, and may hint at internal or tidal energy driving volcanic activity. The amount of $SO_2$ in the Venusian atmosphere required to maintain the sulfuric acid cloud deck has been estimated to be in excess of equilibrated conditions by a factor of 100, implying a source within the past 20 million years (Bullock et al., 2001) or possibly within two million years (Fegley et al., 1989). Detection of $H_2SO_4$ would therefore likely be a sign of some level of geological activity.  For the highly desiccated, photochemically stable $CO_2/CO/O_2$ atmospheres, no $H_2O$ would be detectable, but $CO_2$ and CO would be prominent at 2.35 μm, and $O_4$ would be absent.   The strong $CO_2$ and CO would also be a useful indication that the $O_2$ in the atmosphere was likely not biologically produced (Schwieterman et al., 2016).

For the habitable planets water dominates the simulated direct imaging spectra, with strong bands starting at 0.95 μm and continuing through the near-infrared.  $O_2$ is also seen, but $O_4$ is not significant.  The other strong discriminant in these cases is the presence of $CH_4$, which is not seen in the strongly oxidizing, post-runaway terrestrial atmospheres.  $CH_4$ has its strongest bands in the near-infrared, and interestingly, water absorption features frequently overlap with stronger methane bands. Figures 19 and 21 show how the strong NIR water features at 1.15 and 1.4 μm overlap with methane features. This is particularly problematic given that the UV spectrum of the parent M5.5V drives the significant photochemical accumulation of $CH_4$ in the atmosphere of even the modern Earth.  In transmission water is not readily seen for these habitable planets.   This is largely due to the fact that habitable planets have cold traps that keep water in the troposphere and desiccate the stratosphere.  Due to the effects of refraction (e.g. Misra et al., 2013) transmission observations predominantly probe the stratosphere of habitable terrestrial planets where water is scarce.  Strong water vapor bands in a transit transmission spectrum would indicate a planet with a wet stratosphere, which may indicate a planet undergoing a moist or runaway greenhouse. Additionally, the photochemically-favored build up of $CH_4$ works to "cover" the weak stratospheric water signal spectrally, and also provides additional opacity to make it more difficult to see into the troposphere.   In direct imaging, water vapor can still be seen due to the more direct line of sight, potentially to the planetary surface.   Quantifying water and methane abundances in these types of atmospheres without degeneracy will require access to "clean" methane and water bands that do not overlap.   The water band near 0.95 μm, is relatively clean, as is the methane band near 1.7 μm.

*5.2 Potential for Future Observations.*

Because Proxima Cen b is so close to Earth, it offers us the unprecedented opportunity to characterize a terrestrial exoplanet that orbits within its star's habitable zone. The proximity of this target offers obvious major advantages (e.g. shorter integration times and a larger IWA relative to more distant planets orbiting M dwarfs), but there are still difficulties.

Because Proxima Cen b orbits so close to its dim host star, the angular separation between it and its star is only 37 mas at maximum elongation. This narrow separation means that to be able to directly image this target out to longer wavelengths, larger telescope apertures are favored to overcome IWA limitations, and they can shorten the required integration times to obtain a good signal relative to smaller telescopes. Large ground-based telescopes may be able to measure the planet's spectrum with the aid of coronagraphy and adaptive optics, but the thermal signature from the telescope and atmosphere will make ground-based direct imaging observations very difficult at wavelengths longer than about 2.5 μm.  However, the wavelengths between 1 and 2.5 μm are still extremely useful for atmospheric characterization due to absorption signatures from potentially diagnostic gases, including $CO_2$ and $CH_4$.

JWST may be able to measure thermal phase curves of Proxima Centauri b, and may even be able to acquire transit spectra if the planet transits its star.  From an observational standpoint, both phase curve observations and spectra could be diagnostic.   Thermal phase curves can provide evidence for the existence of an atmosphere and constraints on the heat transport, thermal structure, and molecular composition (Stevenson et al., 2014; Cowan et al., 2015).   Detecting ocean glint would also be extremely diagnostic for surface water, but is likely to be beyond the capabilities of JWST for Proxima Cen b.  Ocean detection may need to wait for a larger aperture space-based coronagraph or starshade mission, or possibly a large aperture ground-based telescope with coronagraphic capabilities.

To understand Proxima Cen b's atmosphere, it will also be important to thoroughly characterize its host star. Through photochemical reactions, the star's UV spectral energy distribution can change the atmospheric state of Proxima Cen b sufficiently to distinguish it from an analog planet orbiting the Sun (Section 4.2.3.1). As discussed above, methane builds up to considerably higher amounts detectable in the spectrum of a modern Earth analog planet around Proxima Centauri, but this does not occur around a solar-type star. To study anticipated effects such as this, it is critical to examine potential atmospheres of this planet using photochemical models with the correct stellar parameters for Proxima Centauri. Additionally, Proxima Centauri is an active star and our study here has not examined the impact of stellar flares on planet b's atmosphere and photochemistry. Future work will examine the impact of flares on the photochemistry, habitability, surface environment, and detectability of spectroscopic signatures.

## 6.0 Conclusions.

Proxima Cen b provides an exciting opportunity to learn about the evolution of terrestrial planets orbiting M dwarfs, and the nature and extent of habitable environments in the Universe. However, we find that its evolutionary history and current volatile inventory will strongly impact its habitability, even though it resides in the habitable zone. Our prior paper (Barnes et al., 2016) predicted different evolutionary scenarios as forced by the likely evolution of its star. In this paper, we used coupled climate-photochemistry models to simulate several plausible states for the current environment of Proxima Cen b, for those various evolutionary scenarios. We find several post-runaway states that are uninhabitable either due to extreme water loss or inclement surface temperatures. In particular, a dense Venus-like $CO_2$ atmosphere will result in extremely high surface temperatures at Proxima Cen's current semi-major axis. However, several evolutionary scenarios may lead to possibly habitable planetary environments, including $O_2$-rich atmospheres that retain a remnant ocean after extreme water loss, and other, more Earth-like scenarios. These later scenarios include a possible terrestrial planet or volatile rich core having been protected against the bright early star by a hydrogen envelope, or the late migration via orbital instabilities of a terrestrial world from exterior to the habitable zone. In these more Earth-like cases, only modest amounts of carbon dioxide (5%) or methane (1-3%) are required to warm the planetary surface. These different scenarios can be discriminated between using observations of thermal phase curves, ocean glint, and spectroscopic diagnostics including $O_2$, $O_3$, $O_4$, CO, $CO_2$, $H_2O$, and $CH_4$, all of which absorb in the 0.3–2.5 µm region. Further information on its planetary characteristics will provide a crucial opportunity for comparative planetology - by broadening our understanding of how terrestrial planetary evolution and processes that operate in our own Solar System may be impacted by a very different host star. Upcoming observations with large ground and space-based telescopes may help to illuminate the intriguing environment of our nearest exoplanetary neighbor.

## Acknowledgments

We thank Guillem Anglada-Escude and his team for sharing their results with us. This work was supported by the NASA Astrobiology Institute's Virtual Planetary Laboratory Lead Team, funded through the NASA Astrobiology Institute under solicitation NNH12ZDA002C and Cooperative Agreement Number NNA13AA93A. This work was facilitated through the use of advanced computational, storage, and networking infrastructure provided by the Hyak supercomputer system at the University of Washington. D. P. Fleming is supported by an NSF IGERT DGE-1258485 fellowship.

## References

Anglada-Escudé, G., Amado, P. J., Barnes, J., Berdiñas, Z. M., Butler, R.P.,. Coleman, G.A.L, de la Cueva, I., Dreizler, S., Endl, M., Giesers, B., Jeffers, S. V., Jenkins, J. S., Jones, H.R.A., Kiraga, M., Kürster, M., López-González, M., Marvin, C. J., Morals, N., Morin, J., Nelson, R.P., Ortiz, J. L., Ofer, A., S.-J., Paardekooper, Reiners, A., Sijme-Jan Paardekooper, Reiners, A., Rodríguez, E., Rodríguez-López, C., Sarmiento, L.F., Strachan, J.P., Tsapras, Y., Tuomi, M., Zechmeister, M. (2016),  A terrestrial planet candidate in a temperate orbit around Proxima Centauri, *Nature*, vol. 536 (7617) , pp. 437-440.

Arney, G., Meadows, V., Crisp, D., Schmidt, S. J., Bailey, J., & Robinson, T. (2014). Spatially resolved measurements of H2O, HCl, CO, OCS, SO2, cloud opacity, and acid concentration in the Venus near‐infrared spectral windows. *Journal of Geophysical Research: Planets*, 119(8), 1860-1891.

Arney, G. N., Domagal-Goldman, S. D., Meadows, V. S., Wolf, E. T., Schwieterman, E., Charnay, B., Claire, M., Hébrard, E., and Trainer. M., (2016a), The Pale Orange Dot: The Spectrum and Habitability of Hazy Archean Earth**,** *Astrobiology*, in press.

Arney, G. N., Meadows, V. S., Domagal-Goldman, S. D., Deming, D., Robinson, T. D. Tovar, G., Wolf, E. T., and Schwieterman, E., (2016b), Pale Orange Dots: The Impact Of Organic Haze On The Habitability And Detectability Of Earthlike Exoplanets, *The Astrophysical Journal*, submitted.

Arnold, L., Gillet, S., Lardière, O., Riaud, P., & Schneider, J. (2002). A test for the search for life on extrasolar planets-Looking for the terrestrial vegetation signature in the Earthshine spectrum. *Astronomy & Astrophysics*, *392*(1), 231-237.

Armstrong, J. C., Barnes, R., Domagal-Goldman, S., Breiner, J., Quinn, T. R., & Meadows, V. S. (2014). Effects of extreme obliquity variations on the habitability of exoplanets. *Astrobiology*, *14*(4), 277-291.

Baldridge, A.M., Hook, S.J., Grove, C.I. & Rivera, G. (2009) The ASTER spectral library version 2.0. Remote Sensing of Environment 113:711–715.

Baraffe, I., Chabrier, G., Allard, F., and Hauschildt, P. H. (1998) Evolutionary models for solar metallicity low-mass stars: mass-magnitude relationships and color-magnitude diagrams, *Astr. & Astroph.*, v.337, p.403-412.

Baraffe, I., Homeier, D., Allard, F., & Chabrier, G. (2015). New evolutionary models for pre-main sequence and main sequence low-mass stars down to the hydrogen-burning limit. Astronomy & Astrophysics, 577, A42.

Baranov, Y. I., Lafferty, W. J., & Fraser, G. T. (2004). Infrared spectrum of the continuum and dimer absorption in the vicinity of the O2 vibrational fundamental in O 2/CO 2 mixtures. *Journal of molecular spectroscopy*, *228*(2), 432-440.

Barnes, R., Raymond, S. N., Jackson, B., & Greenberg, R. (2008). Tides and the evolution of planetary habitability. Astrobiology, 8(3), 557-568.

Barnes, R., Jackson, B., Greenberg, R., & Raymond, S. N. (2009). Tidal limits to planetary habitability. *The Astrophysical Journal Letters*, *700*(1), L30.

Barnes, R., Mullins, K., Goldblatt, C., Meadows, V. S., Kasting, J. F., & Heller, R. (2013). Tidal Venuses: triggering a climate catastrophe via tidal heating. *Astrobiology*, *13*(3), 225-250.

Barnes, R., Deitrick, R., Luger, R., Driscoll, P., Quinn, T., Fleming, D., Guyer, B., McDonald, D. V., Meadows, V. S., Arney, G., Crisp, D., Domagal-Goldman, S. D., Lincowski, A., Lustig-Yaeger, J., Schwieterman, E. (2016). The Habitability of Proxima Centauri b 1: Evolutionary Scenarios. *Astrobiology*, submitted.

Barstow, J.K., Tsang, C.C.C., Wilson, C.F., Irwin, P.G.J., Taylor, F.W., McGouldrick, K., Drossart, P., Piccioni, G. and Tellmann, S., (2012). Models of the global cloud structure on Venus derived from Venus Express observations. *Icarus*, *217*(2), pp.542-560.

Bazot, M., Bouchy, F., Kjeldsen, H., Charpinet, S., Laymand, M., & Vauclair, S. (2007). Asteroseismology of α Centauri A-Evidence of rotational splitting. *Astronomy & Astrophysics*, *470*(1), 295-302.

Beichman, C., Benneke, B., Knutson, H., Smith, R., Lagage, P. O., Dressing, C., Latham, D., Lunine, J., Birkmann, S., Ferruit, P., & Giardino, G. (2014). Observations of transiting exoplanets with the James Webb Space Telescope (JWST). *Publications of the Astronomical Society of the Pacific*, *126*(946), 1134.

Benedict, G.F., McArthur, B., Nelan, E., Story, D., Whipple, A.L., Shelus, P.J., Jefferys, W.H., Hemenway, P.D., Franz, O.G., Wasserman, L.H., Duncombe, R.L., van Altena, W. & Fredrick,

L.W. (1998). Photometry of Proxima Centauri and Barnard's Star Using Hubble Space Telescope Fine Guidance Sensor 3: A Search for Periodic Variations, *ApJ,* 116, pp. 429.

Benneke, B., & Seager, S. (2012). Atmospheric retrieval for super-earths: uniquely constraining the atmospheric composition with transmission spectroscopy. *The Astrophysical Journal*, *753*(2), 100.

Bétrémieux, Y., & Kaltenegger, L. (2014). Impact of Atmospheric Refraction: How Deeply Can We Probe Exo-Earth's Atmospheres During Primary Eclipse Observations?, *Ap. J.*, *791*(1), 7.

Boyajian, T.S., von Braun, K., van Belle, G., McAlister, H.A., ten Brummelaar, T.A., Kane, S.R., Muirhead, P.S., Jones, J., White, R., Schaefer, G., Ciardi, D., Henry, T., López-Morales, M., Ridgway, S., Gies, D., Jao, W.-C., Rojas-Ayala, B., Parks, J.R., Sturmann, L., et al. (2012) Stellar Diameters and Temperatures. II. Main-Sequence K- and M-Stars. *The Astrophysical Journal* 757:112.

Botet, R., Rannou, P. & Cabane, M. (1997) Mean-field approximation of Mie scattering by fractal aggregates of identical spheres. *Applied Optics* 36:8791–8797.

Bouchez, A.H., Acton, D.S., Biasi, R., Conan, R., Espeland, B., Esposito, S., Filgueira, J.M., Gallieni, D., McLeod, B.A., Pinna, E. and Santoro, F., 2014, July. The Giant Magellan telescope adaptive optics program. In *SPIE Astronomical Telescopes+ Instrumentation* (pp. 91480W-91480W).

Brown, R. A. (2005). Single-visit photometric and obscurational completeness. The Astrophysical Journal, 624(2), 1010.

Bullock, M. A., & Grinspoon, D. H. (2001). The recent evolution of climate on Venus. *Icarus*, *150*(1), 19-37.

Butler, R.P., Vogt, S.S., Marcy, G.W., Fischer, D.A., Wright, J.T., Henry, G.W., Laughlin, G. and Lissauer, J.J., 2004. A Neptune-Mass Planet Orbiting the Nearby M Dwarf GJ 436. *The Astrophysical Journal*, *617*(1), p.580.

Chamberlain, S., Bailey, J., Crisp, D. & Meadows, V. (2013) Ground-based near-infrared observations of water vapour in the Venus troposphere. *Icarus* 222:364–378.

Charnay, B., Forget, F., Wordsworth, R., Leconte, J., Millour, E., Codron, F. & Spiga, A. (2013) Exploring the faint young Sun problem and the possible climates of the Archean Earth with a 3-D GCM. *Journal of Geophysical Research: Atmospheres* 118:10,414–10,431.

Chassefière, E. (1996a) Hydrodynamic escape of oxygen from primitive atmospheres: applications to the cases of Venus and Mars. *Icarus* 552:537–552.

Chassefière, E. (1996b) Hydrodynamic escape of hydrogen from a hot water-rich atmosphere: The case of Venus. *Journal of Geophysical Research: Planets* 101:26039–26056.

Claret, A. (2000) A new non-linear limb-darkening law for LTE stellar atmosphere models. *Astronomy and Astrophysics* 363:1001–1005.

Clark, R.N., Swayze, G.A., Wise, R., Livo, E., Hoefen, T., Kokaly, R. & Sutley, S.J. (2007) USGS digital spectral library splib06a: *U.S. Geological Survey, Digital Data Series 231, http://speclab.cr.usgs.gov/spectral.lib06.*

Collinson, G. A., et al. (2016), The electric wind of Venus: A global and persistent "polar wind"-like ambipolar electric field sufficient for the direct escape of heavy ionospheric ions, *Geophys. Res. Lett.*, 43, 5926–5934.

Cowan, N. B., Agol, E., & Charbonneau, D. (2007). Hot nights on extrasolar planets: mid-infrared phase variations of hot Jupiters. *Monthly Notices of the Royal Astronomical Society*, *379*(2), 641-646.

Cowan, N. B., Machalek, P., Croll, B., Shekhtman, L. M., Burrows, A., Deming, D., ... & Hora, J. L. (2012). Thermal phase variations of WASP-12b: Defying predictions. *The Astrophysical Journal*, *747*(1), 82.

Cowan, N. B., Greene, T., Angerhausen, D., Batalha, N. E., Clampin, M., Colon, K., Crossfield, I. J. M., Fortney, J. J., Gaudi, B. S., Harrington. J., Iro, N., Lillie, C. F., Linsky, J. L., Lopez-Morales, M., Mandel, A. M., and Stevenson, K. B., on behalf of Exo PAG SAG-10 (2015). Characterizing Transiting Planet Atmospheres through 2025. *Publications of the Astronomical Society of the Pacific*, *127*(949), 24.

Cox, C. & Munk, W. (1954) Measurement of the Roughness of the Sea Surface from Photographs of the Sun's Glitter. Journal of the Optical Society of America 44:838.

Crisp, D. (1986). Radiative forcing of the Venus mesosphere: I. solar fluxes and heating rates. *Icarus*, *67*(3), 484-514.

Crisp, D. (1997). Absorption of sunlight by water vapor in cloudy conditions: A partial explanation for the cloud absorption anomaly. *Geophysical Research Letters*, *24*(5), 571-574.

Crossfield, I. J., Hansen, B. M., Harrington, J., Cho, J. Y. K., Deming, D., Menou, K., & Seager, S. (2010). A new 24 μm phase curve for υ Andromedae b. *The Astrophysical Journal*, *723*(2), 1436.

Dalcanton, J., Seager, S., Aigrain, S., Battel, S., Brandt, N., Conroy, C., Feinberg, L., Gezari, S., Guyon, O., Harris, W., Hirata, C., Mather, J., Postman, M., Redding, D., Schiminovich, D., Stahl, P. H., Tumlinson, J. (2015) From Cosmic Birth to Living Earths: The Future of UVOIR Space Astronomy, arXiv:1507.04779

Davenport, J. R. (2016). The Kepler Catalog of Stellar Flares. *The Astrophysical Journal*, accepted.

De Bergh, C., Bezard, B., Owen, T., Crisp, D., Maillard, J. P., & Lutz, B. L. (1991). Deuterium on Venus: observations from Earth. *Science*, *251*(4993), 547-549.

Domagal-Goldman, S. D., Segura, A., Claire, M. W., Robinson, T. D., & Meadows, V. S. (2014). Abiotic ozone and oxygen in atmospheres similar to prebiotic Earth. *The Astrophysical Journal*, *792*(2), 90.

Domagal-Goldman, S.D., Kasting, J.F., Johnston, D.T. & Farquhar, J. (2008) Organic haze, glaciations and multiple sulfur isotopes in the Mid-Archean Era. *Earth and Planetary Science Letters* 269:29–40.

Domagal-Goldman, S. D., & Meadows, V. S. (2010, October). Abiotic buildup of ozone. In *Pathways Towards Habitable Planets* (Vol. 430, p. 152)

Domagal-Goldman, S. D., Meadows, V. S., Claire, M. W., & Kasting, J. F. (2011). Using biogenic sulfur gases as remotely detectable biosignatures on anoxic planets. *Astrobiology*, *11*(5), 419–441.

Donahue, T. M., & Hodges, R. R. (1992). Past and present water budget of Venus. *Journal of Geophysical Research: Planets*, *97*(E4), 6083-6091.

Dressing, C. D., Spiegel, D. S., Scharf, C. A., Menou, K., & Raymond, S. N. (2010). Habitable climates: the influence of eccentricity. *The Astrophysical Journal*, *721*(2), 1295.

Driese, S. G., Jirsa, M. A., Ren, M., Brantley, S. L., Sheldon, N. D., Parker, D., & Schmitz, M. (2011). Neoarchean paleoweathering of tonalite and metabasalt: Implications for reconstructions of 2.69 Ga early terrestrial ecosystems and paleoatmospheric chemistry. *Precambrian Research*, *189*(1), 1-17.

Driscoll, P. E., & Barnes, R. (2015). Tidal heating of Earth-like exoplanets around M stars: thermal, magnetic, and orbital evolutions. *Astrobiology*, *15*(9), 739-760.

Ehrenreich, D., Vidal‑Madjar, A., Widemann, T., Gronoff, G., Tanga, P., Barthélemy, M., Lilensten, J., Lecavelier des Etangs, A. & Arnold, L. (2012) Transmission spectrum of Venus as a transiting exoplanet. *Astronomy & Astrophysics* 537:L2.

Etiope, G., & Sherwood Lollar, B. (2013). Abiotic methane on Earth. *Reviews of Geophysics*, *51*(2), 276-299.

Farquhar, J., Bao, H., & Thiemens, M. (2000). Atmospheric influence of Earth's earliest sulfur cycle. *Science*, *289*(5480), 756-758.

Fegley, B., & Prinn, R. G. (1989). Estimation of the rate of volcanism on Venus from reaction rate measurements. *Nature*, *337*(6202), 55-58.

Flückiger, J., Monnin, E., Stauffer, B., Schwander, J., Stocker, T.F., Chappellaz, J., Raynaud, D. & Barnola, J.-M. (2002) High-resolution Holocene N2O ice core record and its relationship with CH4 and CO2. *Global Biogeochemical Cycles* 16:10–1–10–8.

France, K., Froning, C.S., Linsky, J.L., Roberge, A., Stocke, J.T., Tian, F., Bushinsky, R., Désert, J.M., Mauas, P., Vieytes, M. and Walkowicz, L.M. (2013). The Ultraviolet Radiation Environment Around M Dwarf Exoplanet Host Stars. *The Astrophysical Journal*, *763*(2), p.149.

García-Muñoz, A., Zapatero Osorio, M. R., Barrena, R., Montañes-Rodríguez, P., Martín, E. L., Pallé, E., (2012) Glancing Views of the Earth: From a Lunar Eclipse to an Exoplanetary Transit, *The Astrophysical Journal*, 755(2), 103.

Gao, P., Zhang, X., Crisp, D., Bardeen, C. G., & Yung, Y. L. (2014). Bimodal distribution of sulfuric acid aerosols in the upper haze of Venus. *Icarus*, *231*, 83-98.

Gao, P., Hu, R., Robinson, T.D., Li, C., and Yung, Y.L. (2015). Stabilization of $CO_2$ atmospheres on exoplanets around M dwarf stars." *Ap. J.,* 806, 249-261.

Gates, D. M., Keegan, H. J., Schleter, J. C., Weidner, V. R. (1965) Spectral Properties of Plants, *Appl. Optics*, 4(1), 11-20.

Gierasch, P. & Goody, R. (1968). A study of the thermal and dynamical structure of the martian lower atmosphere. *Planetary and Space Science,* 16(5), 615-646.

Goldblatt, C., Claire, M. W., Lenton, T. M., Matthews, A. J., Watson, A. J., & Zahnle, K. J. (2009). Nitrogen-enhanced greenhouse warming on early Earth. *Nature Geoscience*, 2(12), 891-896.

Goldblatt, C., & Watson, A. J. (2012). The runaway greenhouse: implications for future climate change, geoengineering and planetary atmospheres. *Philosophical Transactions of the Royal Society of London A: Mathematical, Physical and Engineering Sciences*, *370*(1974), 4197-4216.

Goldblatt, C., Robinson, T. D., Zahnle, K. J., & Crisp, D. (2013). Low simulated radiation limit for runaway greenhouse climates. *Nature Geoscience*, *6*(8), 661-667.

Goldblatt, C. (2016). Tutorial models of the climate and habitability of Proxima Centauri b: a thin atmosphere is sufficient to distribute heat given low stellar flux. arXiv preprint arXiv:1608.07263.

Greenblatt, G. D., Orlando, J. J., Burkholder, J. B., & Ravishankara, A. R. (1990). Absorption measurements of oxygen between 330 and 1140 nm. *Journal of Geophysical Research: Atmospheres*, 95(D11), 18577-18582.

Greene, T. P., Line, M. R., Montero, C., Fortney, J. J., Lustig-Yaeger, J., & Luther, K. (2016). Characterizing transiting exoplanet atmospheres with JWST. *The Astrophysical Journal*, *817*(1), 17.

Grinspoon, D. (1993) Implications of the high D/H ratio for the sources of water in Venus' atmosphere. *Nature* 363:428–431.

Gruszka, M., & Borysow, A. (1997). Roto-translational collision-induced absorption of $CO_2$ for the atmosphere of Venus at frequencies from 0 to 250 cm−1, at temperatures from 200 to 800 K. *Icarus*, *129*(1), 172-177.

Guzmán-Marmolejo, A., Segura, A., & Escobar-Briones, E. (2013). Abiotic production of methane in terrestrial planets. *Astrobiology*, 13(6), 550-559

Haan, D. & Raynaud, D. (1998) Ice core record of CO variations during the last two millennia: atmospheric implications and chemical interactions within the Greenland ice. Tellus, Series B: Chemical and Physical Meteorology 50B:253–262.

Hale, G. M., & Querry, M. R. (1973). Optical constants of water in the 200-nm to 200-μm wavelength region. *Applied Optics*, 12(3), 555-563.

Hamano, K., Abe, Y., & Genda, H. (2013). Emergence of two types of terrestrial planets on solidification of magma ocean. *Nature*, *497*(7451), 607-610.

Harman, C. E., Schwieterman, E.W., Schottelkotte, J.C., Kasting, J.F. (2015) Abiotic $O_2$ Levels on Planets around F, G, K, and M Stars: Possible False Positives for Life?" *Ap. J.*, 812(2), 137.

Haqq-Misra, J.D., Domagal-Goldman, S.D., Kasting, P.J. & Kasting, J.F. (2008) A revised, hazy methane greenhouse for the Archean Earth. *Astrobiology* 8:1127–37.

Hawley, S. L., & Pettersen, B. R. (1991). The great flare of 1985 April 12 on AD Leonis. The Astrophysical Journal, 378, 725-741.

Hawley, S.L., Allred, J.C., Johns‑Krull, C.M., Fisher, G.H., Abbett, W.P., Alexseev, I., Avgoloupis, S.I., Deustua, S.E., Gunn, A., Seiradakis, J.H., Sirk, M.M, and Valenti, J. A. (2003) Multiwavelength Observations of Flares on AD Leonis. *Ap. J.* 597(1), 535–54.

Heller, R., Leconte, J. and Barnes, R., (2011). Tidal obliquity evolution of potentially habitable planets. *Astronomy & Astrophysics*, *528*, p.A27.

Henyey, L.C. & Greenstein, J.L. (1941) Diffuse radiation in the Galaxy. *The Astrophysical Journal* 93:70.

Hermans, C., Vandaele, A.C., Carleer, M., Fally, S., Colin, R., Jenouvrier, A., Coquart, B. and Mérienne, M.F., (1999). Absorption cross-sections of atmospheric constituents: NO2, O2, and H2O. *Environmental Science and Pollution Research*, *6*(3), pp.151-158.

Hitchcock D.R., and Lovelock, J.E. (1967) Life detection by atmospheric analysis. *Icarus*, 7, 149-159.

Husser, T., von Berg, S.W., Dreizler, S., Homeier, D., Reiners, A., Barman, T. & Hauschildt, P.H. (2013). A new extensive library of PHOENIX stellar atmospheres and synthetic spectra. *A&A*, 553, A6.

Ingersoll, A. P. (1969). The runaway greenhouse: A history of water on Venus. *Journal of the atmospheric sciences*, *26*(6), 1191-1198.

Jackson, B., Barnes, R., & Greenberg, R. (2008). Tidal heating of terrestrial extrasolar planets and implications for their habitability. Monthly Notices of the Royal Astronomical Society, 391(1), 237-245.

Joshi, M. M., Haberle, R. M., & Reynolds, R. T. (1997). Simulations of the atmospheres of synchronously rotating terrestrial planets orbiting M dwarfs: conditions for atmospheric collapse and the implications for habitability. *Icarus*, *129*(2), 450-465.

Kane, S. R., Kopparapu, R. K., & Domagal-Goldman, S. D. (2014). On the Frequency of Potential Venus Analogs from Kepler Data. *The Astrophysical Journal Letters*, *794*(1), L5.

Kanzaki, Y., & Murakami, T. (2015). Estimates of atmospheric CO 2 in the Neoarchean–Paleoproterozoic from paleosols. *Geochimica et Cosmochimica Acta*, *159*, 190-219.

Kasper, M.E., Beuzit, J.-L., Verinaud, C., Yaitskova, N., Baudoz, P., Boccaletti, A., Gratton, R.G., Hubin, N., Kerber, F., Roelfsema, R., Schmid, H.M., Thatte, N.A., Dohlen, K., Feldt, M., Venema, L. & Wolf, S. (2008) EPICS: the exoplanet imager for the E-ELT. In N. Hubin, C. E. Max, & P. L. Wizinowich, eds. *Adaptive Optics Systems. Edited by Hubin, Norbert; Max, Claire E.; Wizinowich, Peter L. Proceedings of the SPIE, Volume 7015, article id. 70151S, 12 pp. (2008).* p. 70151S.

Kasting, J. F., Liu, S. C., & Donahue, T. M. (1979). Oxygen levels in the prebiological atmosphere. *Journal of Geophysical Research, 84*(C6), 3097.

Kasting, J. F., & Pollack, J. B. (1983). Loss of water from Venus. I. Hydrodynamic escape of hydrogen. *Icarus*, *53*(3), 479-508.

Kasting, J. F., Pollack, J. B., & Crisp, D. (1984). Effects of high CO2 levels on surface temperature and atmospheric oxidation state of the early Earth. *Journal of Atmospheric Chemistry*, *1*(4), 403-428.

Kasting, J., Pollack, J. & Ackerman, T. (1984) Response of Earth's atmosphere to increases in solar flux and implications for loss of water from Venus. *Icarus* 57:335–355.

Kasting, J. F. (1988). Runaway and moist greenhouse atmospheres and the evolution of Earth and Venus. *Icarus*, *74*(3), 472-494.

Kasting, J. F., Whitmire, D. P., & Reynolds, R. T. (1993). Habitable zones around main sequence stars. *Icarus*, *101*(1), 108-128.

Kasting, J. F. (1997). Habitable zones around low mass stars and the search for extraterrestrial life. In *Planetary and Interstellar Processes Relevant to the Origins of Life* (pp. 291-307). Springer Netherlands.

Kouveliotou, C., Agol, E., Batalha, N., Bean, J., Bentz, M., Cornish, N., Dressler, A., Figueroa-Feliciano, E., Gaudi, S., Guyon, O., Hartmann, D., Kalirai, J., Niemack, M., Ozel, F., Reynolds, C., Roberge, A., Straughn, K.S.A., Weinberg, D. & Zmuidzinas, J. (2014) Enduring Quests-Daring Visions (NASA Astrophysics in the Next Three Decades). *arXiv:* 1401:3741.

Kelley, D.S., Karson, J.A., Früh-Green, G.L., Yoerger, D.R., Shank, T.M., Butterfield, D.A., Hayes, J.M., Schrenk, M.O., Olson, E.J., Proskurowski, G. and Jakuba, M., (2005). A serpentinite-hosted ecosystem: the Lost City hydrothermal field. *Science*, *307*(5714), pp.1428-1434.

Kiang, N.Y., Segura, A., Tinetti, G., Govindjee, Blankenship, R.E., Cohen, M., Siefert, J., Crisp, D., and Meadows, V.S. (2007) Spectral signatures of photosynthesis. II. Coevolution with other stars and the atmosphere on extrasolar worlds. *Astrobiology* 7:252–274.

Khare, B.N., Sagan, C., Thompson, W.R., Arakawa, E.T., Suits, F., Callcott, T.A., Williams, M.W., Shrader, S., Ogino, H., Willingham, T.O. & Nagy, B. (1984) The organic aerosols of Titan. *Advances in Space Research* 4:59–68.

Kharecha, P., Kasting, J. F., Siefert, J. (2005) A coupled atmosphere-ecosystem model of the early Archean Earth, *Geobiology*, 3, 53-76.

Kitzmann, D., Patzer, A. B. C., & Rauer, H. (2013). Clouds in the atmospheres of extrasolar planets-IV. On the scattering greenhouse effect of CO2 ice particles: Numerical radiative transfer studies. *Astronomy & Astrophysics*, *557*, A6.

Knutson, H. A., Charbonneau, D., Allen, L. E., Fortney, J. J., Agol, E., Cowan, N. B., ... & Megeath, S. T. (2007). A map of the day–night contrast of the extrasolar planet HD 189733b. *Nature*, *447*(7141), 183-186.

Knutson, H. A., Lewis, N., Fortney, J. J., Burrows, A., Showman, A. P., Cowan, N. B., ... & Désert, J. M. (2012). 3.6 and 4.5 μm phase curves and evidence for non-equilibrium chemistry in the atmosphere of extrasolar planet HD 189733b. *The Astrophysical Journal*, *754*(1), 22.

Knutson, H. A., Benneke, B., Deming, D., & Homeier, D. (2014). A featureless transmission spectrum for the Neptune-mass exoplanet GJ 436b. *Nature*, 505(7481), 66-6

Kopparapu, R.K., Ramirez, R., Kasting, J.F., Eymet, V., Robinson, T.D., Mahadevan, S., Terrien, R.C., Domagal-Goldman, S., Meadows, V. and Deshpande, R., 2013. Habitable zones around main-sequence stars: new estimates. *The Astrophysical Journal*, *765*(2), p.131.

Kopparapu, R., Wolf, E. T., Haqq-Misra, J., Yang, J., Kasting, J. F., Meadows, V., Terrien, R., and Mahadevan, S. (2016). The inner edge of the habitable zone for synchronously rotating planets around low-mass stars using general circulation models. *The Astrophysical Journal*, *819*(1), 84.

Kreidberg, L., Bean, J. L., Désert, J. M., Benneke, B., Deming, D., Stevenson, K. B., Seager, S., Berta-Thompson, Z., Seifahrt, A., & Homeier, D. (2014). Clouds in the atmosphere of the super-Earth exoplanet GJ 1214b. *Nature*, 505(7481), 69-72.

Kreidberg, L., Loeb, A. (2016). Prospects for Characterizing the Atmosphere of Proxima Centauri b. *The Astrophysical Journal, Letters*, submitted. http://arxiv.org/abs/1608.07345

Lafferty, W.J., Solodov, A.M., Weber, A., Olson, W.B. & Hartmann, J.M. (1996) Infrared collision-induced absorption by N(2) near 4.3 µm for atmospheric applications: measurements and empirical modeling. Applied optics 35:5911–5917.

Lammer, H., Kasting, J. F., Chassefière, E., Johnson, R. E., Kulikov, Y. N., & Tian, F. (2008). Atmospheric escape and evolution of terrestrial planets and satellites. *Space Science Reviews*, *139*(1-4), 399-436.

Leconte, J., Wu, H., Menou, K., & Murray, N. (2015). Asynchronous rotation of Earth-mass planets in the habitable zone of lower-mass stars. *Science*, *347*(6222), 632-635.

Livengood, T.A., Deming, L.D., A'Hearn, M.F., Charbonneau, D., Hewagama, T., Lisse, C.M., McFadden, L.A., Meadows, V.S., Robinson, T.D., Seager, S. & Wellnitz, D.D. (2011) Properties of an Earth-like planet orbiting a Sun-like star: Earth observed by the EPOXI mission. *Astrobiology* 11:907–30.

Lissauer, J. J. (2007). Planets formed in habitable zones of M dwarf stars probably are deficient in volatiles. *Ap. J. Lett.*, *660*(2), L149.

Lloyd-Hart, M., Angel, R., Milton, N., Rademacher, M., Codona, J., (2006) Design of the adaptive optics systems for GMT. *Proc. SPIE 6272, Advances in Adaptive Optics II*, 62720E; doi:10.1117/12.672444.

Luger R., and Barnes R. (2015) Extreme water loss and abiotic $O_2$ buildup on planets throughout the habitable zones of M dwarfs. *Astrobiology* 15, 119-43.

Luger, R., Barnes, R., Lopez, E., Fortney, J., Jackson, B., & Meadows, V. (2015). Habitable evaporated cores: transforming mini-Neptunes into super-Earths in the habitable zones of M dwarfs. *Astrobiology*, *15*(1), 57-88.

Males, J. R., Close, L. M., Guyon, O., Morzinski, K., Puglisi, A., Hinz, P., Follette, K.B., Monnier, J.D., Tolls, V., Rodigas, T.J.,Weinberger, A., Boss, A., Kopon, D., Wu, Y.-L., Esposito, S., Riccardi, A., Xompero, M., Briguglio, R., Pinna, E. (2014). Direct imaging of exoplanets in the habitable zone with adaptive optics. In *SPIE Astronomical Telescopes and Instrumentation* (pp. 914820-914820). International Society for Optics and Photonics.

Manabe, S. & Strickler, R.F. (1964). Thermal Equilibrium of the Atmosphere with a Convective Adjustment, *Journal of the Atmospheric Sciences,* 21(4), pp. 361-385.

Manabe, S. & Wetherald, R.T. (1967) Thermal Equilibrium of the Atmosphere with a Given Distribution of Relative Humidity. *Journal of the Atmospheric Sciences* 24:241–259.

Maté, B., Lugez, C., Fraser, G. T., & Lafferty, W. J. (1999). Absolute intensities for the $O_2$ 1.27 µm continuum absorption. *Journal of Geophysical Research: Atmospheres*, 104(D23), 30585-30590.

Matvienko, A.S. & Orlov, V.V. (2014). Dynamic status of the wide multiple system α Centauri + Proxima, *Astrophysical Bulletin,* 69(2), pp. 205-210.

Maurin, A. S., Selsis, F., Hersant, F., & Belu, A. (2012). Thermal phase curves of nontransiting terrestrial exoplanets-II. Characterizing airless planets. *Astronomy & Astrophysics*, *538*, A95.

McElroy, M. B., Prather, M. J., & Rodriguez, J. M. (1982). Escape of hydrogen from Venus. *Science*, *216*(4540).

Meadows, V. S., & Crisp, D. (1996). Ground‐based near‐infrared observations of the Venus nightside: The thermal structure and water abundance near the surface. *Journal of Geophysical Research: Planets*, *101*(E2), 4595-4622.

Meadows, V.S. (2006) Modelling the Diversity of Extrasolar Terrestrial Planets. *Proceedings of the International Astronomical Union* 1:25–34.

Meadows, V.S. (2016) Reflections on $O_2$ as a biosignature in exoplanetary atmospheres, *Astrobiology*, in review.

Mellor, G. L., & Yamada, T. (1974). A hierarchy of turbulence closure models for planetary boundary layers. *Journal of the Atmospheric Sciences*, *31*(7), 1791-1806.

Mennesson, B., Gaudi, S., Seager, S., Cahoy, K., Domagal-Goldman, S., Feinberg, L., Guyon, O., Kasdin, J., Marois, C., Mawet, D., Tamura, M., Mouillet, D., Prusti, T., Quirrenbach, A., Robinson, T., Rogers, L., Scowen, P., Somerville, R., Stapelfeldt, K., et al. (2016) The Habitable Exoplanet (HabEx) Imaging Mission: preliminary science drivers and technical requirements ,

Proc. SPIE 9904, Space Telescopes and Instrumentation 2016: Optical, Infrared, and Millimeter Wave, 99040L.

Miller-Ricci Kempton, E., Zahnle, K., & Fortney, J. J. (2012). The atmospheric chemistry of GJ 1214b: photochemistry and clouds. *The Astrophysical Journal*, *745*(1), 3.

Misra, A., Meadows, V., Claire, M., & Crisp, D. (2014a). Using dimers to measure biosignatures and atmospheric pressure for terrestrial exoplanets. *Astrobiology*, *14*(2), 67–86.

Misra, A., Meadows, V., Crisp, D. (2014b), The Effects of Refraction on Transit Transmission Spectroscopy: Application to Earth-like Exoplanets, *Ap. J.*, 792(1), 61-72.

Moore, J. F. (1971), Infrared absorption of carbon dioxide at high densities with application to the atmosphere of Venus, Ph.D. Thesis, Columbia University, pp 31.

Moroz, V.I., Ekonomov, a. P., Moshkin, B.E., Revercomb, H.E., Sromovsky, L. A. & Schofield, J.T. (1985) Solar and thermal radiation in the Venus atmosphere. *Advances in Space Research* 5:197–232.

Muinonen, K., Lumme, K., Peltoniemi, J. & Irvine, W.M. (1989) Light scattering by randomly oriented crystals. *Applied optics* 28:3051–60.

Owen, J. E. & Mohanty, S. (2016) Habitability of terrestrial-mass planets in the HZ of < dwarfs I. H/He-dominated atmospheres. *Monthly Notices of the Royal Astronomical Society* 459, 4088–4108.

Palmer, K.F. & Williams, D. (1975) Optical constants of Sulfuric Acid; application to the clouds of Venus? *Applied optics* 14:208–219.

Pallé, E., Goode, P. R., Yurchyshyn, V., Qiu, J., Hickey, J., Montañés Rodriguez, P., & Koonin, S. E. (2003). Earthshine and the Earth's albedo: 2. Observations and simulations over 3 years. *Journal of Geophysical Research: Atmospheres*, *108*(D22).

Pallé, E., Osorio, M.R.Z., Barrena, R., Montañés-Rodríguez, P. and Martín, E.L., (2009). Earth's transmission spectrum from lunar eclipse observations. *Nature*, *459*(7248), pp.814-816.

Parkinson, C.D., Gao, P., Schulte, R., Bougher, S.W., Yung, Y.L., Bardeen, C.G., Wilquet, V., Vandaele, A.C., Mahieux, A., Tellmann, S. and Pätzold, M., (2015). Distribution of sulphuric acid aerosols in the clouds and upper haze of Venus using Venus Express VAST and VeRa temperature profiles. *Planetary and Space Science*, 113, pp.205-218.

Pavlov, A. A., Brown, L. L., & Kasting, J. F. (2001). UV shielding of $NH_3$ and $O_2$ by organic hazes in the archean atmosphere. *Journal of Geophysical Research: Planets, 106*(E10), 23267-23287.

Pollack, J., Toon, O., Whitten, R., Boese, R., Ragent, B. & M, T. (1980) Distribution and source of the UV absorption in Venus' atmosphere. *Journal of Geophysical Research* 85:8141–8150.

Pollack, J., Dalton, J., Grinspoon, D., Wattson, R., Freedman, R., Crisp, D., Allen, D., Bezard, B., DeBergh, C., Giver, L., Ma, Q. & Tipping, R. (1993) Near-Infrared Light from Venus' Nightside: A Spectroscopic Analysis. *Icarus* 103:1–42.

Quanz, S. P., Crossfield, I., Meyer, M. R., Schmalzl, E., & Held, J. (2015). Direct detection of exoplanets in the 3–10 μm range with E-ELT/METIS. *International Journal of Astrobiology*, *14*(02), 279-289.

Rannou, P., Cabane, M., Botet, R. & Chassèfiere, E. (1997) A new interpretation of scattered light measurements at Titan's limb. *Journal of Geophysical Research* 102:10997–11013.

Raymond, S. N., Scalo, J., Meadows, V. S. (2007) A Decreased Probability of Habitable Planet Formation around Low‑Mass Stars, *Ap. J.,* 669(1) 606–14. doi:10.1086/521587.

Reiners, A. & Basri, G. (2008). The moderate magnetic field of the flare star Proxima Centauri, *A&A,* 489(3), pp. L45-L48.

Robinson, T. D., Meadows, V. S., & Crisp, D. (2010). Detecting Oceans on Extrasolar Planets Using the Glint Effect. *The Astrophysical Journal., 721*(1), L67–L71.

Robinson, T. D., Meadows, V. S., Crisp, D., Deming, D., A'Hearn, M. F., Charbonneau, D., & Hewagama, T. (2011). Earth as an extrasolar planet: Earth model validation using EPOXI Earth observations. *Astrobiology*, *11*(5), 393-408.

Robinson, T. D., Ennico, K., Meadows, V. S., Sparks, W., Bussey, D. B. J., Schwieterman, E. W., and Breiner, J., (2014) Detection Of Ocean Glint And Ozone Absorption Using *LCROSS* Earth Observations, *Ap. J.*, 787, 2.

Robinson, T. D., Stapelfeldt, K. R., & Marley, M. S. (2016). Characterizing Rocky and Gaseous Exoplanets with 2 m Class Space-based Coronagraphs. *Publications of the Astronomical Society of the Pacific*, *128*(960), 025003.

Robinson, T.D. & Crisp, D. (in prep). Linearized Flux Evolution (LiFE): A Technique for Rapidly Adapting Fluxes from Full-Physics Radiative Transfer Models.

Rothman, L.S., Gordon, I.E., Babikov, Y., Barbe, A., Chris Benner, D., Bernath, P.F., Birk, M., Bizzocchi, L., Boudon, V., Brown, L.R., Campargue, A., Chance, K., Cohen, E.A., Coudert, L.H., Devi, V.M., Drouin, B.J., Fayt, A., Flaud, J.-., Gamache, R.R., Harrison, J.J., Hartmann, J.-., Hill, C., Hodges, J.T., Jacquemart, D., Jolly, A., Lamouroux, J., Le Roy, R.J., Li, G., Long, D.A., Lyulin, O.M., Mackie, C.J., Massie, S.T., Mikhailenko, S., Müller, H.S.P., Naumenko, O.V., Nikitin, A.V., Orphal, J., Perevalov, V., Perrin, A., Polovtseva, E.R., Richard, C., Smith, M.A.H., Starikova, E., Sung, K., Tashkun, S., Tennyson, J., Toon, G.C., Tyuterev, V.G. & Wagner, G. (2013). The HITRAN2012 molecular spectroscopic database, *Journal of Quantitative Spectroscopy and Radiative Transfer,* 130, 4-50.

Rothman, L.S., Gordon, I.E., Barber, R.J., Dothe, H., Gamache, R.R., Goldman, A., Perevalov, V.I., Tashkun, S.A. & Tennyson, J. (2010). HITEMP, the high-temperature molecular spectroscopic database, *Journal of Quantitative Spectroscopy and Radiative Transfer,* 111(15), 2139-2150.

Ribas, I., Bolmont, E., Selsis, F., Reiners, A., Leconte, J., Raymond, S., Engle, S., Guinan, E., Morin, M., Forget, F., Anglada-Escude, G. (2016) The habitability of Proxima Centauri b. I. Irradiation, rotation, and volatile inventory from formation to the present. Astronomy & Astrophysics, submitted.

Schaefer, L., Wordsworth, R., Berta-Thompson, Z., & Sasselov, D. (2016). Predictions of the atmospheric composition of GJ 1132b. *arXiv preprint arXiv:1607.03906.*

Schofield, J. T., Taylor, F. W., & McCleese, D. J. (1982). The global distribution of water vapor in the middle atmosphere of Venus. *Icarus, 52*(2), 263-278.

Schwieterman, E.W., Cockell, C.S. & Meadows, V.S. (2015a) Nonphotosynthetic Pigments as Potential Biosignatures. *Astrobiology* 15:341–361.

Schwieterman, E. W., Robinson, T. D., Meadows, V. S., Misra, A., & Domagal-Goldman, S. (2015). Detecting and Constraining $N_2$ Abundances In Planetary Atmospheres Using Collisional Pairs. *The Astrophysical Journal*, *810*(1), 57.

Schwieterman, E.W., Meadows, V.S., Domagal-Goldman, Deming, L. D., Arney, G. N., Luger, R., Harman, C. E., Misra, A., Barnes, R., (2016) Identifying planetary biosignature impostors: spectral features of CO and $O_4$ resulting from abiotic $O_2/O_3$ Production, *The Astrophysical Journal Letters*, 819:L13.

Segura, A., Krelove, K., Kasting, J.F., Sommerlatt, D., Meadows, V., Crisp, D., Cohen, M. and Mlawer, E. (2003) Ozone concentrations and ultraviolet fluxes on Earth-like planets around other stars. *Astrobiology*, *3*(4), pp.689-708.

Segura, A., Kasting, J.F., Meadows, V., Cohen, M., Scalo, J., Crisp, D., Butler, R.A. and Tinetti, G. (2005) Biosignatures from Earth-like planets around M dwarfs. *Astrobiology*, *5*(6), pp.706-725.

Segura, A., Meadows, V.S., Kasting, J.F., Crisp, D. & Cohen, M. (2007) Abiotic formation of O2 and O3 in high-CO2 terrestrial atmospheres. Astronomy and Astrophysics 472:665–679.

Segura, A., Walkowicz, L.M., Meadows, V., Kasting, J. & Hawley, S. (2010) The effect of a strong stellar flare on the atmospheric chemistry of an earth-like planet orbiting an M dwarf. Astrobiology 10:751–71.

Selsis, F., Wordsworth, R. D., & Forget, F. (2011). Thermal phase curves of nontransiting terrestrial exoplanets-I. Characterizing atmospheres. *Astronomy & Astrophysics*, 532, A1.

Shields, A. L., Meadows, V. S., Bitz, C. M., Pierrehumbert, R. T., Joshi, M. M., & Robinson, T. D. (2013). The effect of host star spectral energy distribution and ice-albedo feedback on the climate of extrasolar planets. *Astrobiology*, *13*(8), 715-739.

Sidis, O. & Sari, R. (2010) Transits of Transparent Planets—Atmospheric Lensing Effects. *The Astrophysical Journal* 720:904–911.

Smith, T. & Guild, J. (1931) The C.I.E. colorimetric standards and their use. *Transactions of the Optical Society* 33:73–134.

Snellen, I., de Kok, R., Birkby, J. L., Brandl, B., Brogi, M., Keller, C., Kenworthy, M., Schwarz, H., & Stuik, R. (2015). Combining high-dispersion spectroscopy with high contrast imaging: Probing rocky planets around our nearest neighbors. *Astronomy & Astrophysics*, 576, A59.

Solomon, S., Qin, D., Manning, M. & Al, E. (2007) Contribution of Working Group I to the Fourth Assessment Report of the Intergovernmental Panel on Climate Change.

Sotin, C., Grasset, O. & Mocquet, A. (2007) Mass-radius curve for extrasolar Earth-like planets and ocean planets. *Icarus* 191:337–351.

Spergel, D., Gehrels, N., Breckinridge, J., Donahue, M., Dressler, A., Gaudi, B.S., Greene, T., Guyon, O., Hirata, C., Kalirai, J. and Kasdin, N.J., 2013. Wide-field infrared survey telescope-astrophysics focused telescope assets WFIRST-AFTA final report. *arXiv preprint arXiv:1305.5422.*

Stevenson, K. B., Désert, J. M., Line, M. R., Bean, J. L., Fortney, J. J., Showman, A. P., Kataria, T., Kreidberg, L., McCullough, P.R., Henry, G.W. & Charbonneau, D. (2014). Thermal structure of an exoplanet atmosphere from phase-resolved emission spectroscopy. *Science,346*(6211), 838-841.

Tian, F. (2009). Thermal escape from super Earth atmospheres in the habitable zones of M stars. *The Astrophysical Journal*, *703*(1), 905.

Tian, F., France, K., Linsky, J. L., Mauas, P. J., & Vieytes, M. C. (2014). High stellar FUV/NUV ratio and oxygen contents in the atmospheres of potentially habitable planets. *Earth and Planetary Science Letters*, 385, 22-27.

Tinetti, G., Meadows, V. S., Crisp, D., Fong, W., Velusamy, T., & Snively, H. (2005). Disk-averaged synthetic spectra of Mars. *Astrobiology*, *5*(4), 461-482.

Tinetti, G., Meadows, V. S., Crisp, D., Kiang, N. Y., Kahn, B. H., Bosc, E., Fishbein, E., Velusamy, T., & Turnbull, M. (2006b). Detectability of planetary characteristics in disk-averaged spectra II: synthetic spectra and light-curves of earth. *Astrobiology*, *6*(6), 881-900.

Tomasko, M.G., Doose, L., Engel, S., Dafoe, L., West, R., Lemmon, M., Karkoschka, E. & See, C. (2008) A model of Titan's aerosols based on measurements made inside the atmosphere. *Planetary and Space Science* 56:669–707.

Toon, O. B., McKay, C. P., Ackerman, T. P., & Santhanam, K. (1989). Rapid calculation of radiative heating rates and photodissociation rates in inhomogeneous multiple scattering atmospheres. *Journal of Geophysical Research: Atmospheres*, *94*(D13), 16287-16301.

Trainer, M. G., Pavlov, A. A., DeWitt, H. L., Jimenez, J. L., McKay, C. P., Toon, O. B., & Tolbert, M. A. (2006). Organic haze on Titan and the early Earth. *Proceedings of the National Academy of Sciences*, *103*(48), 18035-18042.

Turbet, M., Leconte, J., Selsis, F., Bolmont, E., Forget, F., Ribas, I., Raymond, S. & Anglada-Escude, G. (2016) The Habitability of Proxima Centauri b II. Possible climates and observability. *Astronomy & Astrophysics*, submitted.

Ueno, Y., Yamada, K., Yoshida, N., Maruyama, S. & Isozaki, Y. (2006) Evidence from fluid inclusions for microbial methanogenesis in the early Archaean era. *Nature* 440:516–9.

van der Werf, S. Y. (2008). Comment on "Improved ray tracing air mass numbers model". *Applied optics*, 47(2), 153-156.

Walker, J.C.G., Hays, P.B. & Kasting, J.F. (1981) A negative feedback mechanism for the long-term stabilization of Earth's surface temperature. *Journal of Geophysical Research* 86:9776.

Wertheimer, J.G. & Laughlin, G. (2006). Are Proxima and α Centauri Gravitationally Bound?, *The Astrophysical Journal,* 132(5), pp. 1995-1997.

Williams, D. M., & Gaidos, E. (2008). Detecting the glint of starlight on the oceans of distant planets. *Icarus*, *195*(2), 927-937.

Williams, D. M., & Pollard, D. (2002). Habitable planets on eccentric orbits. In *The Evolving Sun and its Influence on Planetary Environments* (Vol. 269, p. 201).

Williams, D. M., & Kasting, J. F. (1997). Habitable planets with high obliquities. *Icarus*, *129*(1), 254-267.

Wiscombe, W. J. (1980). Improved Mie scattering algorithms. *Applied optics*, *19*(9), 1505-1509.

Woolf, N. J., Smith, P. S., Traub, W. A., & Jucks, K. W. (2002). The spectrum of Earthshine: a pale blue dot observed from the ground. *The Astrophysical Journal*, *574*(1), 430.

Wolf, E.T. & Toon, O.B. (2010) Fractal organic hazes provided an ultraviolet shield for early Earth. *Science* 328:1266–8.

Wolf, E.T. & Toon, O.B. (2014) Delayed onset of runaway and moist greenhouse climates for Earth. *Geophysical Research Letters* 41:167–172.

Wordsworth, R., Forget, F., & Eymet, V. (2010). Infrared collision-induced and far-line absorption in dense $CO_2$ atmospheres. *Icarus*, *210*(2), 992-997.

Wordsworth, R., & Pierrehumbert, R. (2014). Abiotic oxygen-dominated atmospheres on terrestrial habitable zone planets. *The Astrophysical Journal Letters*, *785*(2), L20.

Yung, Y. L., & DeMore, W. B. (1982). Photochemistry of the stratosphere of Venus: Implications for atmospheric evolution. *Icarus*, *51*(2), 199-247.

Zahnle, K., Claire, M., & Catling, D. (2006). The loss of mass-independent fractionation in sulfur due to a palaeoproterozoic collapse of atmospheric methane. *Geobiology, 4*(4), 271-283.

Zellem, R. T., Lewis, N. K., Knutson, H. A., Griffith, C. A., Showman, A. P., Fortney, J. J., Cowan, N.B., Agol, E., Burrows, A., Charbonneau, D. and Deming, D. (2014). The 4.5 μm full-orbit phase curve of the hot Jupiter HD 209458b. *The Astrophysical Journal*, *790*(1), 53.

Zerkle, A. L., Claire, M. W., Domagal-Goldman, S. D., Farquhar, J., & Poulton, S. W. (2012). A bistable organic-rich atmosphere on the Neoarchaean Earth. *Nature Geoscience*, *5*(5), 359-363.

Zugger, M. E., Kasting, J. F., Williams, D. M., Kane, T. J., & Philbrick, C. R. (2010). Light scattering from exoplanet oceans and atmospheres. *The Astrophysical Journal*, *723*(2), 1168.

Zugger, M. E., Kasting, J. F., Williams, D. M., Kane, T. J., & Philbrick, C. R. (2011). Searching for Water Earths in the Near-infrared. *The Astrophysical Journal*, *739*(1), 12.